\documentclass[traditabstract]{aa} 

\usepackage{graphicx, times, txfonts, float, rotating, color, url, lscape, bm}

\newcommand{\msol}{\mbox{M$_{\odot}$}} 
\newcommand{\rsol}{\mbox{R$_{\odot}$}}

\newcommand{\lsol}{\mbox{L$_{\odot}$}}

\newcommand{\ks}{km s$^{-1}$}

\newcommand{\muas}{$\mu$as} 

\newcommand{\G}{{\it Gaia}}

\begin{document} 

\title{Spectral energy distributions of classical cepheids in the Magellanic Clouds
\thanks{
Tables~\ref{Tab-Sam}, \ref{Tab:WalPhot}, \ref{Tab:WalPhotMean}, and  \ref{Tab:AppMass} are available in electronic form at the CDS via 
anonymous ftp to cdsarc.u-strasbg.fr (130.79.128.5) or via  http://cdsarc.u-strasbg.fr/viz-bin/qcat?J/A+A/.
The fits to the SEDs are available at \protect\url{https://doi.org/10.5281/zenodo.8032168}.
}
}  
 
\author{
  M.~A.~T.~Groenewegen\inst{1}
  \and
  J.~Lub\inst{2}
}

\institute{ 
Koninklijke Sterrenwacht van Belgi\"e, Ringlaan 3, B--1180 Brussels, Belgium \\ \email{martin.groenewegen@oma.be}
\and
Leiden Observatory, Leiden University, P.O. Box 9513, 2300~RA Leiden, The Netherlands 
} 
 
\date{received: ** 2023, accepted: * 2023} 

\offprints{Martin Groenewegen}

\abstract
{
%
In this study, we constructed spectral energy distributions (SEDs) fot a sample of 142  Large Magellanic Cloud (LMC)
  and 77  Small Magellanic Cloud (SMC) fundamental-mode classical Cepheids (CCs)  using
  photometric data from the literature. When possible, the data were taken to be representative of mean light
  or averaged over the light curve.
The sample was built from stars that either have a metallicity determination from high-resolution spectroscopy
or have been used in Baade-Wesselink types of analyses, or have a radial velocity curve published in {\it Gaia} DR3
or have Walraven photometry, 
or have their light- and radial-velocity curves modelled by pulsation codes.
The SEDs were fitted with stellar photosphere models to derive the best-fitting luminosity and effective temperature.
Distance and reddening were taken from the literature.
Only one star with a significant infrared (IR) excess was found in the LMC and none in the SMC.IR excess in MW CCs is not uncommon
suggesting that IR excess may be more prominent in MW cepheids than in the Magellanic Clouds.

The stars were plotted in a Hertzsprung-Russell diagram (HRD) and compared
to evolutionary tracks for CCs and to theoretical instability strips.
For the large majority of stars, the position in the HRD is consistent with the instability strip.
Period-luminosity ($PL$) and period-radius relations were derived and compared to these relations in the MW.
For a fixed slope, the zero point of the bolometric $PL$ relation does not depend on metallicity, contrary to recent findings of a significant
metallicity term when considering the $PL$ relation in different photometric bands.
The mass-luminosity (ML) relation is derived and it points to an over luminosity of about +0.3~dex with respect to a canonical $ML$ relation.

The most intriguing result concerns the flux-weighted gravity (FWG, a quantity derived from gravity and effective temperature) and
its relation to period and luminosity. Both relations agree with theory, with the results for the MW and with the independent estimates from
the six known LMC eclipsing binaries that contain CCs.
However, the FWG (as determined from dedicated high-resolution spectroscopy for the sample) is too low by about 0.8~dex in 90\% of the cases.
Recent works on time-series data on 20 CCs in the MW were analysed finding a similar (but less extreme) offset in gravity
and the FWG.
Most importantly, other time-series data on the same 20 CCs are in full agreement with the FWG-period relation.

The observed time-series of spectroscopic data and from a two-dimensional hydrodynamical cepheid model was used to investigate the
so-called effective gravity, that is, the gravity corrected for a dynamical term related to the time derivative of the radial velocity.
There is a reasonable good correspondence between the predicted effective gravity and the observed gravity as a function of pulsation phase,
which would potentially allow for an independent estimate of the projection factor, but the dynamical term is too small to explain the
overall difference between the observed (flux-weighted) gravity and the (flux-weighted) gravity derived from the SED modelling and stellar mass estimates.
}


\maketitle

\section{Introduction}
\label{S:Int}

Classical cepheids (CCs) serve as an important standard candle because they are bright and provide a link
between the distance scale in the nearby Universe and that further out, via galaxies that contain both Cepheids and SNIa 
(see \citealt{Riess2022} for a determination of the Hubble constant to 1.0~\ks\ precision).
Typically, the period-luminosity (PL) relations of CCs that are at the core of the distance determinations are derived, namely:
the photometric filters ($V, I, K$) or combination of filters that are designed to be
independent of reddening, the so-called Wesenheit functions \citep{Madore82}; for example, using combinations of ($V,I$), ($J,K$) 
or the combination used by the SH0ES team (F555W, F814W, and F160W HST filters, see \citealt{Riess2022}).

On the other hand, the bolometric magnitude or luminosity is a fundamental quantity of CCs as well as stars in general,
as it is the output of stellar evolution models and the input to CC pulsation models.
In \citet{GrSED} (hereafter G20)  the spectral energy distributions (SEDs) of 477 Galactic CCs were constructed and
fitted with model atmospheres (and a dust component when required). For an adopted distance and reddening these fits resulted in a
best-fitting bolometric luminosity ($L$) and the photometrically derived effective temperature ($T_{\rm eff}$).
This allowed for the derivation of
period-radius ($PR$) and $PL$ relations, the construction of the Hertzsprung-Russell diagram (HRD), and a comparison to
theoretical instability strips (ISs). The position of most stars in the HRD was consistent with theoretical predictions.
Outliers were often associated with sources where the spectroscopically and photometrically  determined
effective temperatures differed, or in cases where the sources exhibit a high degree of reddening with large uncertainties as well.

This sample was further studied in \citet{GrFWG}, where the relation between
bolometric absolute magnitude and the flux-weighted gravity (FWG),
defined as
$\log g_{\rm F} = \log g - 4 \cdot \log \left(T_{\rm eff} / 10^4\right)$ \citep{K2003}, 
was investigated: the so-called flux-weighted gravity-luminosity relation (FWGLR).
The tight correlation between $g_{\rm F}$ and luminosity was first
demonstrated by \citet{K2003,K2008} for blue supergiants (BSGs) and was later used for extra galactic distance determinations in \citet{K16}. 
\citet{Anderson16} then demonstrated that theoretical pulsation models for CCs also followed a tight FWGLR, one that is,
in fact, tighter than the $PL$ relation, finding that there was a good correspondence between observed $g_{\rm F}$ and period for a sample of CCs.
\citet{GrFWG} presented the currently best observationally determined FWGLR for Milky Way (MW) CCs, based on the luminosities
derived in \citet{GrSED} and gravity and effective temperatures from the literature.

In \citet{GrSED} the adopted distances were based as much as possible on {\it Gaia} parallax data (from DR2 in that case).
However, we need to correct the catalogued parallaxes for the parallax zero-point offset (PZPO). 
In GDR2, this value was $-0.029$ mas for quasars \citep{Lindegren18} and for CCs $-0.046 \pm 0.013$ \citep{RiessGDR2} or
$-0.049 \pm 0.018$~mas (\citealt{Gr_GDR2}, hereafter G18), and was also a limiting factor in improving upon the local distance scale.
In GEDR3 the {\it Gaia} team provided a Python script to the community (\citealt{GEDR3_LindegrenZP}, hereafter L21), which
\cite{RiessGEDR3} applied to a sample of 75 CCs in the MW, concluding that a counter correction of $-14 \pm 6$ \muas\ is required.
The advantage of using the Magellanic Clouds (MCs) is that accurate and independently derived mean distances are available based on
the analysis of samples of eclipsing binaries \citep{Pietrzynski19, Graczyk20}

The present paper performs a study similar to \citet{GrSED} and \citet{GrFWG} for a sample of CCs in the Small (SMC) and
Large (LMC) Magellanic Clouds.
The paper is structured as follows.
In Section~\ref{S-Sam}, the sample of 219 MC CCs is introduced, while 
Section~\ref{S:PDM} introduces the photometry that is used, the distances used, how the stellar mass is estimated and how the
modelling of the SED is done.
Section~\ref{S:Res} offers a discussion of key results, in particular, the location of the CCs in the Hertzsprung-Russell diagram, the
presence of any infrared excess, the period-radius and period-luminosity relations, the mass-luminosity relation, and the relation
between the FWG and period and luminosity.
A brief discussion and summary is given in Section~\ref{S:Dis}.

\section{Sample} 
\label{S-Sam}

In this paper, we study a sample of 142 LMC and 77 SMC CCs.
Although this is a small subset of the about 4700 LMC and 4900 SMC CCs known in the MCs (see e.g. \citealt{Ripepi17,Ripepi22}), 
the stars in this sample are of special interest as they have been studied in other respects.
Specifically, the sample is composed of:
\begin{itemize}
\item
 89 LMC CCs for which \citet{Romaniello22} derived iron and oxygen abundances
  (as well as effective temperatures and gravities) from high-resolution (HR) spectroscopy.
This sample is composed of 68 CCs used to define the PL-relation in the LMC in the SH0ES program \citep{Riess19} and
21 for which archival spectra, first presented in \citet{Romaniello08}, were re-analysed.
\item 14 SMC CCs for which \citet{Romaniello08} performed an abundance analysis.
We note that for the LMC CCs in overlap \citet{Romaniello22} derived an iron abundance that was (on average) 0.1 dex smaller
compared to \citet{Romaniello08}.
\item 7 CCs in the LMC cluster NGC 1866 \citep{Molinaro12,Lemasle17} and four field SMC objects \citep{Lemasle17} with
  iron abundances from HR spectroscopy. 
\item CCs for which a Baade-Wesselink analysis has been carried out, in particular: 
  36 LMC and five SMC stars considered in \citet{Storm11b}, and the almost identical sample of
  36 LMC and six  SMC stars analysed in \citet{Gr2013}.  
Similarly, 27 LMC and eight SMC stars that have been analysed with the SPIPS code \citep{Merand2015} in \citet{Gallenne17} to
  derive the pulsation ($p$) factor. 
\item CCs for which light-curves (and sometimes radial-velocity curves) have been fitted with theoretical pulsation models.
In such a modelling \citep{Marconi13} the stellar mass, luminosity, and (mean) effective temperature are derived by
fitting the light-curves
(typical $V$, $I,$ and $K$). The apparent distance moduli (DM) are derived from which the true DM and reddening are found.
If RV curves are fitted, the projection-factor ($p$-factor) is also derived.
Here, we consider the 11 LMC and 9 SMC fundamental-mode (FM) CCs studied in \citet{Ragosta19} and \citet{Marconi17}, respectively.
\item CCs with (previously unpublished) photometry in the Walraven \citep{Walraven60} system.
This system is very usefull in constraining effective temperature and reddening as the photometric bands extend into the blue. 
\citet{vanGenderen1983} published VBLUW photometry for 21 SMC and 20 LMC CCs using data taken between 1971 and 1978 in South Africa
(also see \citealt{vanGenderen81,PelvGL81,vanGenderen1989}).
However, data collection continued from 1979 onwards from Chile and in Appendix~\ref{WApp} we report on these observations.

\item CCs in the MCs with RV curves published in \G\ DR3 \citep{RipepiDR3Cep22,GaiaDR3Vallenari22,GC2016a}. 
\item SMC FM CCs for which UVES spectra and in part {\it HST} photometry will be obtained in the near
  future\footnote{See the publically available information on ESO program 0109.D-0846(A) (P.I. M.~Romaniello) as
    per October 1st, 2022 and \url{https://www.stsci.edu/hst/phase2-public/17097.pro} (P.I. A.~Riess), respectively.}.

\end{itemize}

\noindent
There are stars contained in the overlap between the different subsamples and the final sample
consists of 142 LMC and 77 SMC CCs, all of which are FM pulsators.
The basic parameters of the stars are compiled in Table~\ref{Tab-Sam}.

\begin{sidewaystable*}
    \small
\setlength{\tabcolsep}{1.5mm}
  \caption{\label{Tab-Sam} Sample of stars (selected entries) } 
\begin{tabular}{crrccccccccrrrrlllll}
  \hline  \hline
  Identifier & HV    &  Period    &  $d$      &  $A_{\rm V}$      & $T_{\rm eff}$ & $\log g$ &  Ref & Luminosity  &  $T_{\rm eff}$  & $R$       & Mass & $\log g$ & $\log g_{\rm F}$ & $\chi^2_{\rm r}$  \\ 
           &       &   (d)      &    (kpc)  &  (mag)          &     (K)      &  (cgs)   &      &  (\lsol)    &    (K)        &    (\rsol)  &  (\msol) & (cgs) & (cgs)      \\ 
\hline
%
LMC0046 & 12717 &   8.844 & 50.62 & 0.39 & 5224  $\pm$ 134  & 2.45 $\pm$ 0.19 & (5)  &   3777.1 $\pm$    88.3 & 5750  $\pm$  125  &  62.0 $\pm$   2.7 &   5.50 $\pm$   0.22 & 1.59 $\pm$ 0.04 & 2.55 $\pm$ 0.05 &    9.5 \\ 
LMC0079 &    -1 &  22.544 & 50.52 & 0.36 & 6150  $\pm$  97  & 1.50 $\pm$ 0.10 & (4)  &   7823.3 $\pm$   102.0 & 5125  $\pm$   88  & 112.4 $\pm$   3.8 &   6.58 $\pm$   0.44 & 1.16 $\pm$ 0.04 & 2.32 $\pm$ 0.05 &    6.1 \\ 
LMC0107 & 12452 &   8.739 & 50.48 & 0.43 & 5390  $\pm$  42  & 0.80 $\pm$ 0.07 & (1)  &   3834.1 $\pm$    91.3 & 5750  $\pm$  125  &  62.5 $\pm$   2.7 &   5.68 $\pm$   0.25 & 1.60 $\pm$ 0.05 & 2.57 $\pm$ 0.06 &    6.7 \\ 
LMC0328 &   873 &  34.449 & 50.75 & 0.24 & 5222  $\pm$  27  & 1.25 $\pm$ 0.06 & (5)  &  17626.4 $\pm$   335.4 & 5250  $\pm$  144  & 160.8 $\pm$   8.6 &   8.80 $\pm$   0.40 & 0.97 $\pm$ 0.05 & 2.09 $\pm$ 0.07 &    7.0 \\ 
LMC0367 &   872 &  29.822 & 49.79 & 0.24 & 5675  $\pm$ 120  &                 & (5)  &   9577.3 $\pm$   137.1 & 5000  $\pm$  125  & 130.6 $\pm$   6.4 &   6.38 $\pm$   0.56 & 1.01 $\pm$ 0.06 & 2.22 $\pm$ 0.07 &   16.7 \\ 
LMC0434 &   875 &  30.349 & 50.25 & 0.26 & 5660  $\pm$ 100  & 0.30 $\pm$ 0.13 & (2)  &  17254.3 $\pm$   617.9 & 5625  $\pm$  144  & 138.6 $\pm$   7.3 &   7.22 $\pm$   0.36 & 1.02 $\pm$ 0.05 & 2.02 $\pm$ 0.07 &    5.2 \\ 
LMC0461 &   877 &  45.166 & 49.86 & 0.26 & 4890  $\pm$ 109  & 0.70 $\pm$ 0.08 & (1)  &  14891.6 $\pm$   257.9 & 4750  $\pm$  125  & 180.5 $\pm$   9.3 &   7.85 $\pm$   0.70 & 0.82 $\pm$ 0.06 & 2.12 $\pm$ 0.08 &   13.6 \\ 
LMC0467 &   876 &  22.720 & 49.70 & 0.23 & 5391  $\pm$  74  & 1.63 $\pm$ 0.16 & (5)  &   9500.7 $\pm$   251.0 & 5375  $\pm$  125  & 112.6 $\pm$   5.3 &   6.54 $\pm$   0.22 & 1.16 $\pm$ 0.05 & 2.23 $\pm$ 0.06 &    4.4 \\ 
LMC0501 &   878 &  23.311 & 50.37 & 0.24 & 5130  $\pm$  77  & 0.30 $\pm$ 0.05 & (2)  &  10708.0 $\pm$   279.0 & 5500  $\pm$  125  & 114.2 $\pm$   5.2 &   6.62 $\pm$   0.12 & 1.14 $\pm$ 0.04 & 2.18 $\pm$ 0.05 &    4.6 \\ 
LMC0504 & 12505 &  14.393 & 50.05 & 0.33 &               &                    &      &   3867.4 $\pm$   132.1 & 5000  $\pm$  189  &  83.0 $\pm$   6.1 &   5.63 $\pm$   0.32 & 1.35 $\pm$ 0.07 & 2.55 $\pm$ 0.09 &   20.6 \\ 
LMC0510 &   879 &  36.831 & 50.22 & 0.34 & 5530  $\pm$  64  & 0.10 $\pm$ 0.15 & (2)  &  13853.0 $\pm$   335.2 & 4875  $\pm$  153  & 165.3 $\pm$  10.1 &   8.48 $\pm$   0.53 & 0.93 $\pm$ 0.06 & 2.18 $\pm$ 0.08 &   33.5 \\ 
LMC0512 &  2257 &  39.398 & 50.36 & 0.22 & 5200  $\pm$  79  & 0.00 $\pm$ 0.09 & (2)  &  17977.8 $\pm$   152.2 & 5125  $\pm$   88  & 170.4 $\pm$   5.8 &   8.40 $\pm$   0.48 & 0.90 $\pm$ 0.04 & 2.07 $\pm$ 0.05 &    6.2 \\ 
LMC0528 &   881 &  35.731 & 50.47 & 0.15 & 5200  $\pm$  64  & 0.10 $\pm$ 0.08 & (2)  &  15023.5 $\pm$   494.6 & 5125  $\pm$  189  & 155.7 $\pm$  11.2 &   7.70 $\pm$   0.46 & 0.95 $\pm$ 0.07 & 2.11 $\pm$ 0.10 &   21.2 \\ 
LMC0545 &  2262 &  15.832 & 50.38 & 0.25 & 5420  $\pm$  85  & 0.80 $\pm$ 0.05 & (2)  &   5898.8 $\pm$   124.2 & 5250  $\pm$  125  &  93.0 $\pm$   4.4 &   6.52 $\pm$   0.38 & 1.31 $\pm$ 0.05 & 2.43 $\pm$ 0.06 &    7.3 \\ 
LMC0546 &  2249 &  15.216 & 49.64 & 0.21 & 6730  $\pm$ 285  & 1.40 $\pm$ 0.29 & (5)  &   6215.9 $\pm$   160.9 & 5500  $\pm$  189  &  87.0 $\pm$   5.8 &   5.99 $\pm$   0.25 & 1.34 $\pm$ 0.06 & 2.38 $\pm$ 0.08 &   16.2 \\ 
LMC0561 &   880 &  11.670 & 49.73 & 0.22 & 5383  $\pm$ 202  & 2.18 $\pm$ 0.26 & (5)  &   4836.7 $\pm$    88.8 & 5875  $\pm$  168  &  67.2 $\pm$   3.7 &   4.72 $\pm$   0.15 & 1.45 $\pm$ 0.05 & 2.38 $\pm$ 0.07 &    7.1 \\ 
LMC0590 &   882 &  31.787 & 50.18 & 0.31 & 5880  $\pm$ 322  & 0.00 $\pm$ 0.10 & (2)  &  14075.5 $\pm$   307.3 & 5250  $\pm$  153  & 143.7 $\pm$   8.2 &   7.48 $\pm$   0.29 & 1.00 $\pm$ 0.06 & 2.12 $\pm$ 0.08 &   16.6 \\ 
LMC0594 &    -1 &   6.733 & 50.19 & 0.27 & 5520  $\pm$ 194  & 0.90 $\pm$ 0.05 & (2)  &   2245.8 $\pm$    44.7 & 5625  $\pm$  189  &  50.0 $\pm$   3.2 &   4.90 $\pm$   0.17 & 1.73 $\pm$ 0.06 & 2.73 $\pm$ 0.08 &   17.2 \\ 
LMC0619 &   883 & 133.779 & 49.89 & 0.28 & 4754  $\pm$  11  & 1.65 $\pm$ 0.02 & (5)  &  48057.0 $\pm$  2103.3 & 4625  $\pm$  208  & 342.0 $\pm$  29.8 &   (8.03 $\pm$   0.77) & (0.27 $\pm$ 0.08) & (1.61 $\pm$ 0.11) &   86.4 \\ 
LMC0648 &  2270 &  13.626 & 50.17 & 0.28 & 5300  $\pm$  90  & 0.50 $\pm$ 0.07 & (2)  &   4192.2 $\pm$   109.1 & 5250  $\pm$  125  &  78.4 $\pm$   3.7 &   5.53 $\pm$   0.31 & 1.39 $\pm$ 0.05 & 2.51 $\pm$ 0.06 &    7.4 \\ 
$\cdots$ \\
SMC3533 &  1950 &   7.990 & 62.44 & 0.13 & 5900  $\pm$ 100  & 1.64 $\pm$ 0.10 & (6)  &   3226.2 $\pm$   123.4 & 5375  $\pm$  189  &  65.6 $\pm$   4.6 &   6.86 $\pm$   0.89 & 1.65 $\pm$ 0.08 & 2.72 $\pm$ 0.10 &   24.8 \\ 
SMC3565 &  1954 &  16.694 & 62.44 & 0.15 & 5890  $\pm$ 100  & 1.00 $\pm$ 0.10 & (3)  &   9712.3 $\pm$   285.1 & 5500  $\pm$  153  & 108.7 $\pm$   6.0 &   8.97 $\pm$   0.66 & 1.32 $\pm$ 0.06 & 2.36 $\pm$ 0.08 &   13.8 \\ 
SMC3588 &  1957 &   5.319 & 62.44 & 0.12 & 5975  $\pm$ 100  & 1.84 $\pm$ 0.10 & (6)  &   1964.1 $\pm$    75.0 & 5750  $\pm$  189  &  44.7 $\pm$   2.9 &   4.85 $\pm$   0.27 & 1.83 $\pm$ 0.06 & 2.79 $\pm$ 0.08 &   16.0 \\ 
SMC3611 &  1956 & 208.799 & 62.44 & 0.10 & 5677  $\pm$  63  & 1.94 $\pm$ 0.19 & (5)  &  69071.1 $\pm$  2229.5 & 4375  $\pm$  168  & 458.2 $\pm$  34.0 &   (9.60 $\pm$   0.79) & (0.10 $\pm$ 0.07) & (1.54 $\pm$ 0.10) &   37.4 \\ 
SMC4555 &  2209 &  22.642 & 62.44 & 0.17 & 6130  $\pm$ 100  & 1.10 $\pm$ 0.10 & (3)  &  14112.7 $\pm$   430.0 & 5625  $\pm$  153  & 125.3 $\pm$   6.8 &   8.64 $\pm$   0.08 & 1.18 $\pm$ 0.04 & 2.18 $\pm$ 0.06 &   18.0 \\ 
SMC4697 &   817 &  18.901 & 62.44 & 0.12 & 5850  $\pm$ 100  & 1.00 $\pm$ 0.10 & (3)  &   9953.4 $\pm$   533.2 & 5500  $\pm$  237  & 110.1 $\pm$   9.4 &   7.91 $\pm$   0.35 & 1.26 $\pm$ 0.08 & 2.30 $\pm$ 0.11 &   59.7 \\ 
SMC4919 &  6357 &  33.338 & 62.44 & 0.18 & 6130  $\pm$ 100  & 0.50 $\pm$ 0.10 & (3)  &  21581.7 $\pm$   924.8 & 5250  $\pm$  204  & 177.9 $\pm$  13.6 &  11.61 $\pm$   0.24 & 1.00 $\pm$ 0.07 & 2.12 $\pm$ 0.09 &   13.5 \\ 
SMC4953 & 11211 &  21.386 & 62.44 & 0.09 & 4830  $\pm$ 100  & 0.00 $\pm$ 0.10 & (3)  &  11665.8 $\pm$   387.5 & 5250  $\pm$  153  & 130.8 $\pm$   7.6 &  10.08 $\pm$   0.59 & 1.21 $\pm$ 0.06 & 2.33 $\pm$ 0.08 &   15.8 \\ 
SMC4976 &  2231 &  36.737 & 62.44 & 0.17 &               &                    &      &  16641.3 $\pm$  1921.0 & 5125  $\pm$  312  & 163.9 $\pm$  20.5 &   8.38 $\pm$   0.55 & 0.93 $\pm$ 0.11 & 2.09 $\pm$ 0.15 &   47.9 \\ 
\hline
\end{tabular}
\tablefoot{
Column~1. The identifier used in this paper, which is related to the identifier used by OGLE.
The first entry, LMC0046, for example, would be known as OGLE-LMC-CEP-0046.
Column~2. Harvard variable (HV) identifier, when available.
Column~3. Pulsation period.
Column~4. The adopted distance.
For the LMC objects this includes the geometric correction, see Sect.~\ref{SS-dist}.
Column~5. The adopted reddening value $A_{\rm V}$ based on \citet{Skowron21}, see main text.
Column~6. Effective temperature in the literature. For references 3 and 6 an uncertainty of 100~K has been adopted.
Column~7. log gravity in the literature. For references 3, 4 and 6 an uncertainty of 0.1~dex has been adopted.
Column~8. Reference for the data in Columns~6 and 7,
(1) \citet{Romaniello22}, re-analysed data from \citet{Romaniello08} (Tables 4 and 6),
(2) \citet{Romaniello22} (new spectra, Tables 3 and 5),
(3) \citet{Romaniello08},
(4) \citet{Lemasle17},
(5) \citet{Sprague22}.
(6) \citet{Ragosta19} (LMC) and \citet{Marconi17} (SMC).
For reference (6) the values are not determined from high-resolution spectroscopy but from their best-fitting LC fitting models.
Column~10. Luminosity with error bar from the SED fitting.
Column~11. (photometric) effective temperature with error bar from the SED fitting.
Column~12. Radius with error bar, derived from $L$ and $T_{\rm eff}$.
Column~13. Adopted stellar mass, see Appendix~\ref{MApp}.
Column~14. log gravity determined from mass and radius.
Column~13. Flux-weighted gravity calculated from from the gravity and $T_{\rm eff}$.
For a few stars the mass estimate is clearly too low given their period and the value for the mass
and (flux-weighted) gravity are not used and listed between parenthesis.
Column~14. The reduced chi-square of the fit to the SED.

}
\end{sidewaystable*}

\section{Photometry, distance, masses, and modelling}
\label{S:PDM}

\subsection{Photometry}
\label{SS-phot}

The SEDs were constructed using photometry retrieved mainly (but not exclusively) 
via the VizieR web-interface\footnote{\url{http://vizier.u-strasbg.fr/viz-bin/VizieR}}.
Data were considered (in increasing wavelength) in the UV
from GALEX \citep{Bianchi17},
in the optical from a variety of sources, namely,
OGLE ($B, V,I$) photometry from \citet{Udalski1998} (SMC) and \citet{Udalski2000} (LMC),
($V,I$) photometry from the OGLE Shallow Survey in the LMC \citep{Ulaczyk12,Ulaczyk13}, and OGLE-IV data
for both Clouds \citet{Soszynski17}, 
{\it Gaia}  $B_{\rm p}$, $G$, and $R_{\rm p}$ photometry from DR3  \citep{GaiaDR3Vallenari22},
($B,R$) photometry from the EROS survey in the LMC \citep{Kim2014},
($u,v,g,r,i,z$) data from Skymapper DR2 \citep{Onken19},
($B,V,R,I$) photometry for LMC CCs from \citet{Sebo02},
(U,B,V) photometry from \citet{Madore1975}, \citet{Eggen1977}, and \citet{MW1979},
Walraven (VBLUW) photometrey (see Appendix~\ref{WApp}),
($U,B,V,I,K$) for CCs in NGC 1866 from \citet{Musella16},
HST F555W, F814W, and F160W photometry from \cite{Riess19} for LMC cepheids, and further to the near- and mid-infrared,
($Y,J,K$) photometry from the VMC survey \citep{Cioni11} for the SMC \citep{Ripepi16} and LMC \citep{Ripepi22}, and
from the public DR6 for a few remaining stars, 
($J,H,K$) photometry for LMC CCs from \citet{Persson04} and \citet{Macri15},
Akari photometry for the SMC \citep{Ita2010} and LMC \citep{Kato12}.
We note that for the S7, S11, and L15 filters, only errors in the magnitudes of
 $<$0.15, $<$0.20, and $<$0.20~mag, respectively, were accepted.
Then, we also used AllWISE photometry (\citealt{Cutri_Allwise}; in the W3 and W4 filters only errors in the magnitudes of
$<$0.30, and $<$0.25~mag, respectively, were accepted),
{\it Spitzer} IRAC photometry from \citet{Chown21} (mean magnitudes from template fitting in the 3.6 and 4.5~$\mu$m bands) and
from VizieR catalogue II/305/catalog in the 5.8 and 8.5~$\mu$m bands.  Finally, we used
MIPS photometry at 24~$\mu$m available from the IRSA\footnote{\url{https://irsa.ipac.caltech.edu/applications/Gator/}, 
"SAGE MIPS 24 um Epoch 1 and Epoch 2 Full List."}. No MIPS data at 70~$\mu$m were available.

The number of available photometric data points for the LMC CCs ranges from 16 to 39, with a median of 30,
and for the SMC objects from 15 to 29, with a median of 20 photometric points.
The data contain single-epoch observations (typically from GALEX and Akari) but whenever possible values
at mean light were taken or multiple datapoints were averaged.

\subsection{Distance and geometric correction}
\label{SS-dist}

The mean distance to the LMC is adopted to be $d_{\rm LMC}= 49.59 \pm 0.09~{\rm (stat)} \pm 0.54$ (syst) kpc \citep{Pietrzynski19}),
and to the SMC of $d_{\rm SMC}= 62.44 \pm 0.47~{\rm (stat.)} \pm 0.81$ (syst.) kpc \citep{Graczyk20}, based, in both cases, on
the analysis of samples of eclipsing binaries.
The depth effect in the SMC is considerable, for example \citet{Ripepi17}, but all SMC sources have been adopted to be at the mean distance.
For the LMC the first order approximation of an inclined disk is adopted to compute the geometric correction and the procedure 
in \citet{Riess19} is followed, taking the inclination 
and position angle of the line of nodes 
of the disk from \citet{Pietrzynski19}
and the LMC center-of-mass coordinates from \citet{vdMarel14}. 
The adopted distances are listed in column~4 in Table~\ref{Tab-Sam}.

\subsection{Stellar masses}
\label{SS-mass}

An estimate for the stellar mass is required when computing the stellar gravity, with the stellar radius available from the
best-fitting luminosity and effective temperature combination, as detailed below.
Several methods have been employed
\begin{itemize}

\item
the period-luminosity-mass-effective temperature-metallicity relation derived in \citet{GrJu17b} based on the
  models of \cite{Bono_2000}.

\item
Similarly, such a relation was derived here based on the models of \citet{Anderson16}.
That paper gives, for a given mass, metallicity and rotational speed -- the period, luminosity, and effective temperature
at the start and end of the IS, and for different crossings of the IS a star may undergo.
As the first-crossing is very short compared to the other crossings these models were not considered.
An initial fit showed that the metallicity term is not significant.
The best linear fit for average rotation, all metallicities, and the second and third crossing models for FM pulsators is:

\begin{align}
\log P = & \;  (13.095 \pm 0.114) +(0.857 \pm 0.010) \log L               \nonumber \\ 
         & \; -(0.669  \pm 0.032)  \log M -(3.912 \pm 0.030) \log T_{\rm eff} \nonumber \\
         & \;  {\rm (N= 46, \sigma = 0.0076)}
\label{EqPCEP}
\end{align}

\item

Based on the model fitting of LMC Cepheid light curves \citet{Ragosta19} presented a period-mass-radius relation of
\begin{align}
  \log P =  & \; (-1.618 \pm 0.007) + (-0.68 \pm 0.02) \log M  \nonumber \\
            & \;  + (1.72 \pm 0.01) \log R
\end{align}
with a dispersion of only 0.005~dex. This relation will be applied to the SMC ones as well.

\item

\citet{Pilecki2018} analysed the light- and radial-velocity curves of the six known cepheid containing
eclipsing binaries (containing seven cepheids, including one type-II cepheid (T2C), all in the LMC).
For convenience, the stellar parameters they derived for the
cepheids are compiled in Table~\ref{Tab:DEB}, as they appear in several plots.
\citet{Pilecki2018} derived the following period-mass-radius relation
\begin{align}
  \log P =  & \;  (-1.555 \pm 0.035) - (0.795 \pm  0.044) \log M  \nonumber \\
            & \; + (1.703  \pm 0.023) \log R 
\end{align}
with a dispersion of 0.037~dex. This relation will be applied to the SMC ones as well.
The mass range on which this relation is based is smaller than that of the other relations.

\begin{table*}
  \centering

\caption{Stellar parameters of the Cepheid in known LMC EBs (from \citealt{Pilecki2018}) }
\begin{tabular}{rrccccccccccccccccccr} \hline \hline 
  Name   & Period  &    $M_1$        & $\log g_1$        & $T_{\rm eff1}$    &  $\log L_1$      &     $R_1$       & $p_{\rm f}$     & Remarks \\
         &  (days) &   (\msol)      &   (cgs)            &    (K)          &  (\lsol)        &     (\rsol)     &                &         \\
  \hline                                                                                        
CEP0227 & 3.79708 & 4.15 $\pm$ 0.03 & 1.970 $\pm$ 0.004 & 6000 $\pm$ 160 & 3.15 $\pm$ 0.05 & 34.87 $\pm$ 0.12 & 1.21 $\pm$ 0.05 &  \\
CEP4506 & 2.98785 & 3.61 $\pm$ 0.03 & 2.087 $\pm$ 0.007 & 6120 $\pm$ 160 & 3.01 $\pm$ 0.05 & 28.5  $\pm$ 0.2  & 1.35 $\pm$ 0.09 &  \\
CEP2533 & 2.833   & 3.98 $\pm$ 0.10 & 2.10  $\pm$ 0.04  & 6350 $\pm$ 150 & 3.10 $\pm$ 0.06 & 29.2  $\pm$ 1.4  &                 & FO \\
CEP1812 & 1.31290 & 3.76 $\pm$ 0.03 & 2.509 $\pm$ 0.007 & 6120 $\pm$ 150 & 2.61 $\pm$ 0.04 & 17.85 $\pm$ 0.13 & 1.26 $\pm$ 0.08 &  \\
CEP1718 & 2.732   & 4.27 $\pm$ 0.04 & 2.18  $\pm$ 0.04  & 6310 $\pm$ 150 & 3.04 $\pm$ 0.06 & 27.8  $\pm$ 1.2  &                 & FO  \\
CEP1718 & 3.460   & 4.22 $\pm$ 0.04 & 2.02  $\pm$ 0.03  & 6270 $\pm$ 160 & 3.18 $\pm$ 0.06 & 33.1  $\pm$ 1.3  &                 & FO \\
T2C098  & 4.97372 & 1.51 $\pm$ 0.09 & 1.81  $\pm$ 0.03  & 5300 $\pm$ 120 & 2.66 $\pm$ 0.04 & 25.3  $\pm$ 0.2  & 1.30 $\pm$ 0.05 &  \\
\hline
\end{tabular} 
\tablefoot{
  Column~1: Name/identifiers. CEP0227 refers to OGLE-LMC-CEP-0227, etc. T2C098 refers to LMC-T2CEP-098. CEP1718 appears twice as both components of the EB are cepheids.
  Column~2: FM Period. For the first overtone (FO) pulsators the conversion is outlined in \citet{Pilecki2018}.
  Column~3-7: Stellar parameters of the cepheid: mass, gravity, effective temperature, luminosity and radius.
  Column~8: projection factor when available. 
  Column~9: Remarks, indicating which cepheids are FO pulsators. All data from \citet{Pilecki2018}.
}
\label{Tab:DEB}
\end{table*}

\item

\citet{Marconi20} used nonlinear convective pulsation models  to link period and mass to a
Wesenheit index based on {\it Gaia} magnitudes.
For a mixing length parameter of 1.7 and FM pulsators \citet{Marconi20} give:
\begin{equation}
  WG - {\rm DM} = -1.686 -2.496 \log P -2.285 \log M
\end{equation}
with a dispersion of 0.058~mag, and where $WG = G - 1.90 \cdot (B_{\rm p} - R_{\rm p})$  \citep{Ripepi_GDR2}
and DM is the distance modulus.  The pulsation models were calculated at solar metallicity but will be used for the MC CCs.

\end{itemize}

\noindent
The mass estimates from these five methods are listed in Table~\ref{Tab:AppMass}.
The adopted mass is the median among the five estimates, and is listed in Table~\ref{Tab:AppMass} and in column~12 in Table~\ref{Tab-Sam}.
To estimate the error bar, the error in the mass estimate of the median value is added in quadrature to
the median-absolute-deviation times 1.48 (to get the equivalent of one sigma in a Gaussian distribution)
among the five estimates.

The different estimates are in good agreement in most cases, except for some of the longest period cepheids where
some individual estimates give unrealistically low masses, leading to the median value becoming also unrealistically low.
These values are set between parenthesis and have not been used in the analysis.
The origin of the discrepancy is probably related to the fact that the different mass-estimate formalisms are not derived from such long-period cepheids.

\begin{figure*}

\begin{minipage}{0.45\textwidth}
\resizebox{\hsize}{!}{\includegraphics{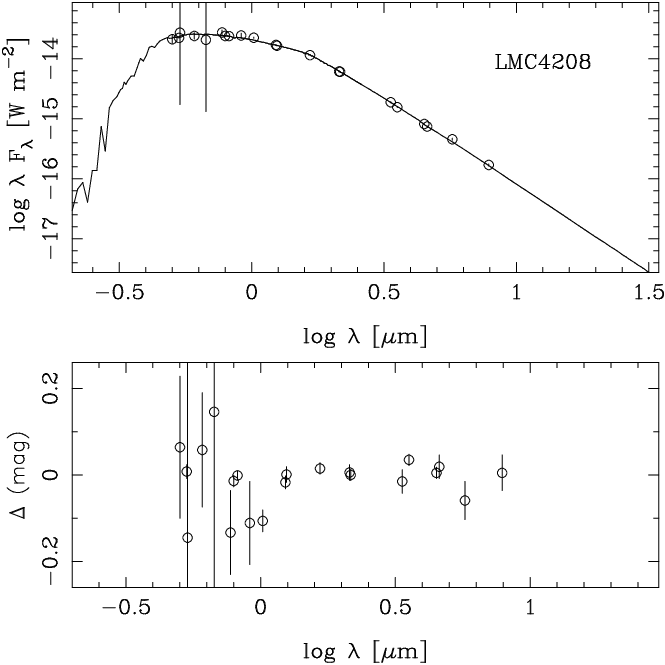}}
\end{minipage}
\begin{minipage}{0.45\textwidth}
\resizebox{\hsize}{!}{\includegraphics{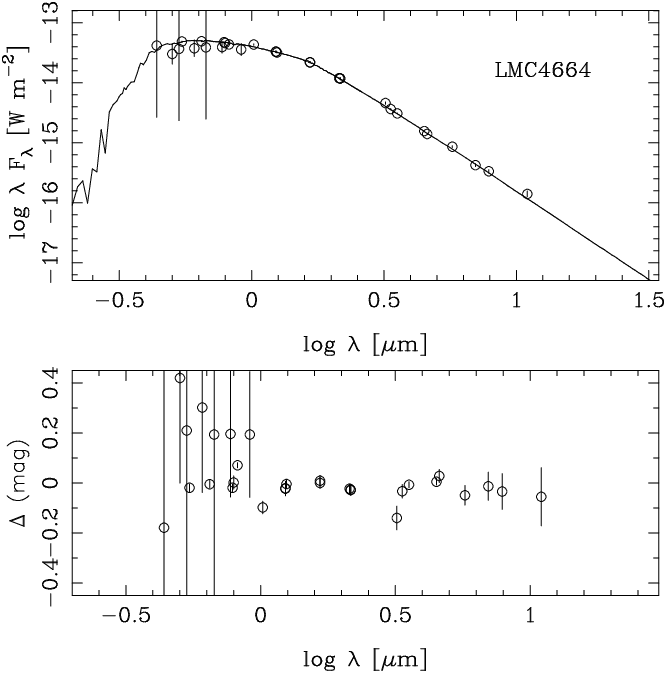}}
\end{minipage}

\begin{minipage}{0.45\textwidth}
\resizebox{\hsize}{!}{\includegraphics{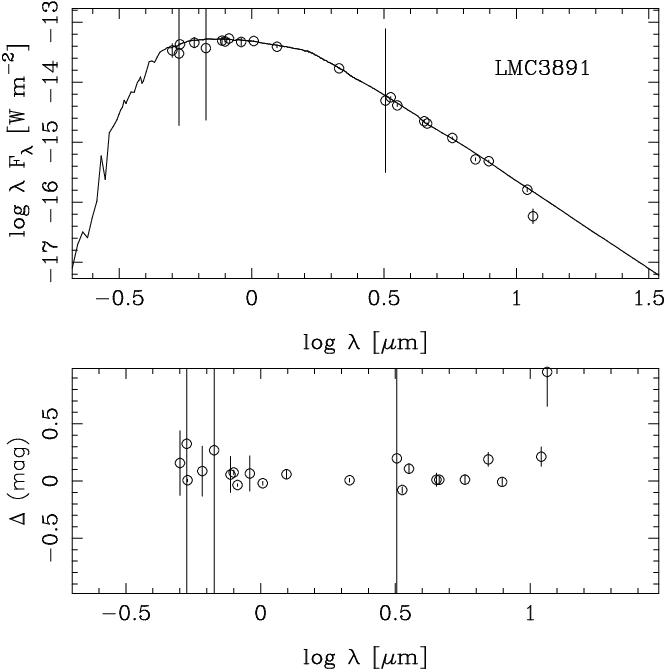}}
\end{minipage}
\begin{minipage}{0.45\textwidth}
\resizebox{\hsize}{!}{\includegraphics{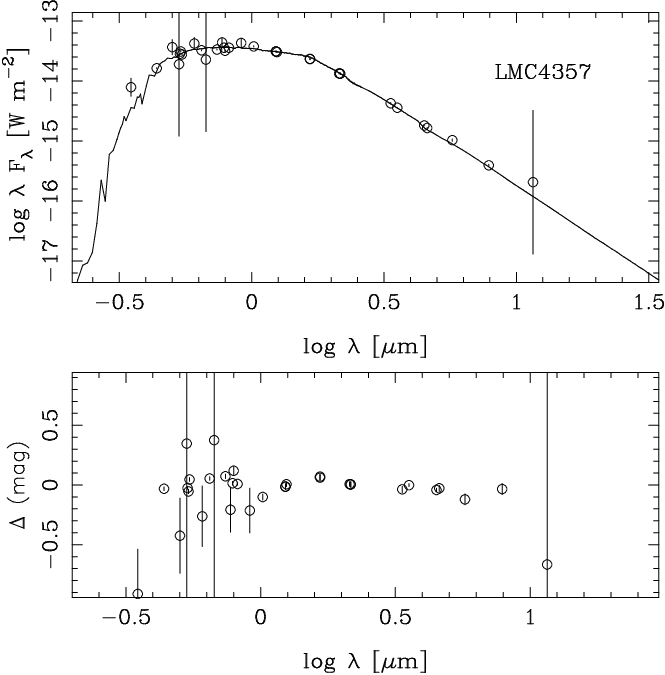}}
\end{minipage}

\caption{Four best-fitting models. The upper panels: Observations (with error bars) and the model.
  The lower panels: Residuals. Outliers that have been clipped are plotted with an (arbitrary) error bar of 3.0~mag.
}
\label{Fig:sed1}
\end{figure*}

\subsection{Modelling}
\label{SS-model}

The SEDs are fitted with the code
More of DUSTY (MoD, \cite{Gr_MOD})\footnote{\url{http://homepage.oma.be/marting/codes.html}} %
which uses a slightly updated and modified version of the DUSTY dust radiative transfer (RT) code \citep{Ivezic_D} as
a subroutine within a minimisation code. As we are not interested in any dust component
the dust optical depth is initially set to zero.
In that case, the input to the model are the distance, reddening, and a model atmosphere.
The few cases where an infrared (IR) excess may be present are discussed in Sect.~\ref{S:dust}.

The model atmosphere fluxes are reddened to be compared to the observations.
The reddening map of \citet{Skowron21} for the MCs is adopted and the $E(V-I)$ value in the map closest to the source is taken.
The visual extinction is then assumed to be $A_{\rm V}= 3.1 \cdot E(V-I) /1.318$, and is listed in column~5 in Table~\ref{Tab-Sam}.

MARCS model atmospheres are used as input \citep{Gustafsson_MARCS} for $\log g= 1.5$ and metallicity $-0.50$ and $-0.75$ dex
for the LMC and SMC stars, respectively.
The model grid is available at 250~K intervals for the effective temperature 
range of interest, and adjacent model atmospheres are used to interpolate models at 125~K intervals, 
which better reflects the accuracy in $T_{\rm eff}$ that can be achieved.
For every model atmosphere (that is, $T_{\rm eff}$) a best-fitting luminosity (with its [internal] error bar, based on the covariance
matrix) is derived with the corresponding reduced $\chi^2$ ($\chi_{\rm r}^2$) of the fit. 
The model with the lowest $\chi_{\rm r}^2$ then gives the best-fitting effective temperature. 
Considering models within a certain range above this minimum $\chi_{\rm r}^2$ then gives the estimated error in the 
effective temperature and luminosity. For the luminosity, this error is added in quadrature to the internal error in luminosity.

The best fitting effective temperature and luminosity with error bar are listed in columns 9 and 10, and the resulting radius in column 11
of Table~\ref{Tab-Sam}. Combined with the mass this gives the gravity (column 13), and gravity combined with $T_{\rm eff}$ the FWG (column 14).

\section{Results and discussion}
\label{S:Res}

\subsection{General}

Figure~\ref{Fig:sed1} and ~\ref{Fig:sed2} present the four best, respectively, the four poorest fits to the SEDs (according to the
$\chi_{\rm r}^2$), respectively, with the residual (model minus observations) in the bottom part
of each panel\footnote{The complete set of SEDs is are available at \protect\url{https://doi.org/10.5281/zenodo.8032168}}.
In the model fitting procedure photometric outliers were excluded in the following way.
The rms in the residuals was determined and added in quadrature to the photometric error bar for each data point.
If the absolute difference between model and observations was larger than 4$\sigma$ the point is flagged and
plotted with an error bar of 3.0~mag to still identify it but to have no influence on the fitting.

The fits are quite acceptable. In the poorest fits the scatter among the various photometric points is larger overall.
In the case of SMC0417 and SMC0921 this leads to the result that the most visually discrepant points (the VMC $JHK$ points)
are not marked as 4$\sigma$ outliers and that therefore the reduced $\chi^2$ is large.

A limitation of our procedure is that time variability of the photometry is not taken into account.
Values at mean light have been considered whenever possible, but the SEDs also contain single-epoch data.
Pulsation amplitudes decrease with wavelength so the effect should not play a major role in the mid and far-IR and in the
NIR where the SEDs peak mean-light magnitudes are typically available.
The construction of the SED at mean light also ignores possible phase shifts between photometric bands.

We have compared our results to the fully independent modelling  by \citet{Gallenne17} using the SPIPS code \citep{Merand2015} for
35 stars in overlap. The SPIPS code takes light curves as input and therefore considers the time variability.
It is also independent in the sense that it fits ATLAS9 model atmospheres \citep{ATLAS9}.

\begin{figure}
  \centering

\begin{minipage}{0.24\textwidth}
\resizebox{\hsize}{!}{\includegraphics{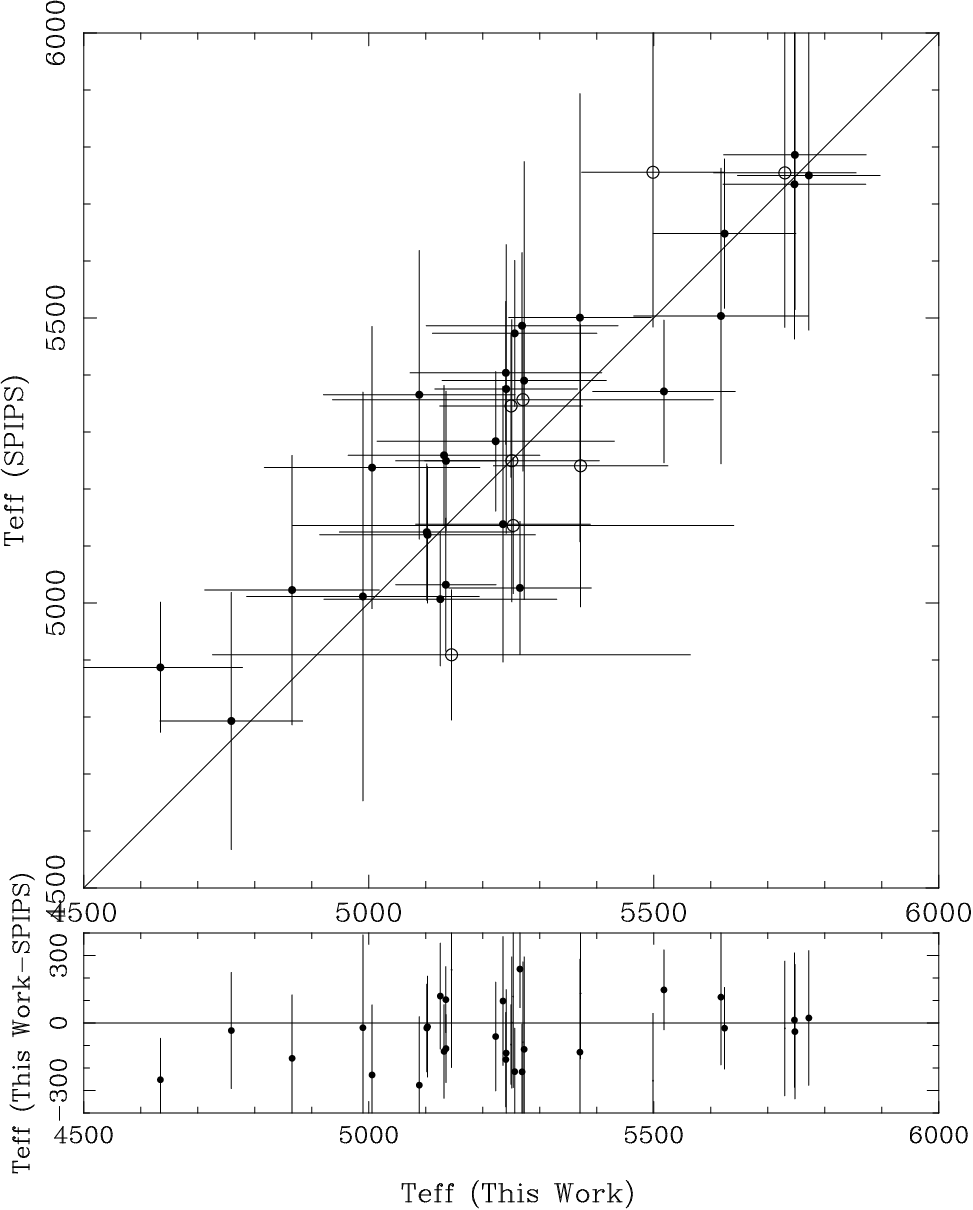}}
\end{minipage}
\begin{minipage}{0.24\textwidth}
\resizebox{\hsize}{!}{\includegraphics{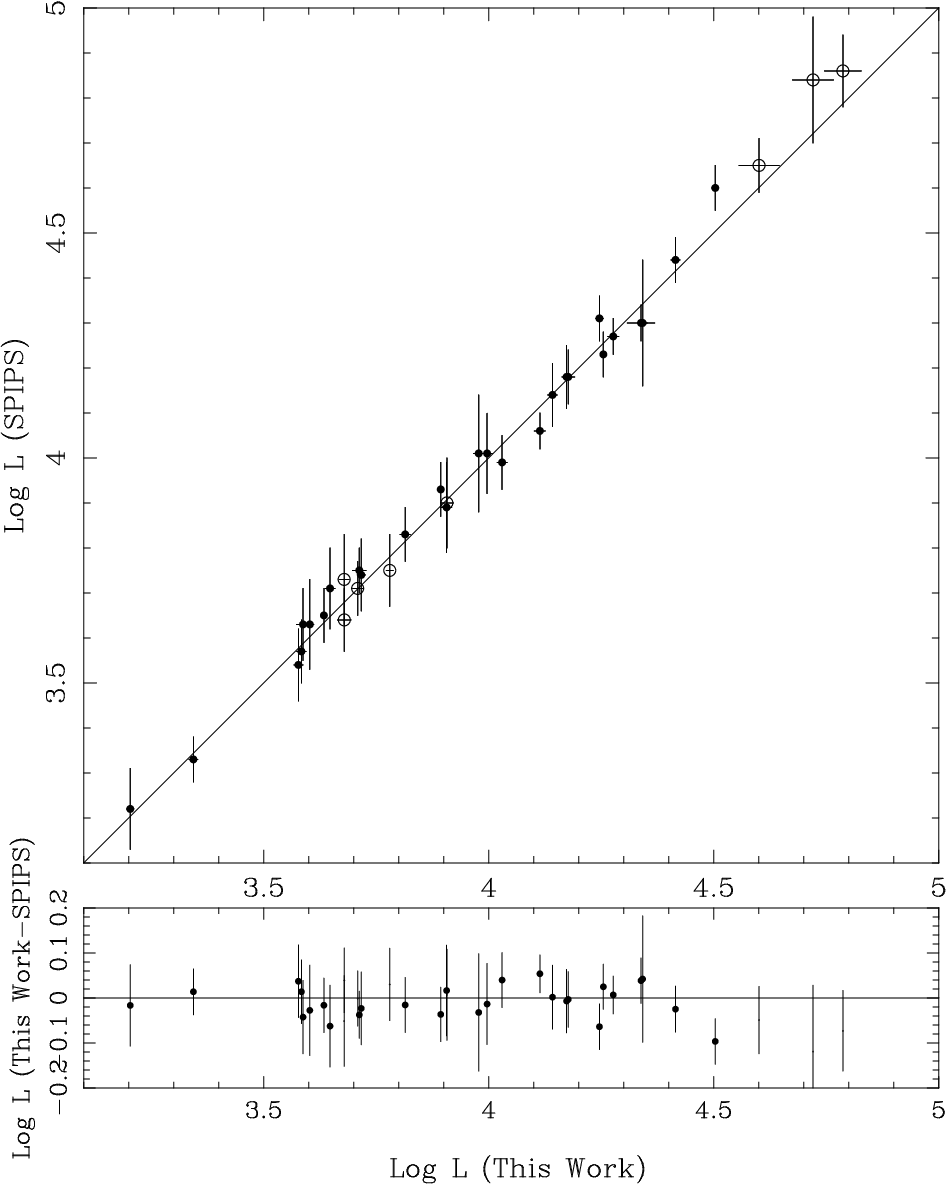}}
\end{minipage}

\caption{Comparison of effective temperature and luminosity between the present work and the
  27 LMC (filled circles) and 8 SMC (open circles) stars in overlap.
  The bottom panels show the residuals.
}
\label{Fig:SPIPS}
\end{figure}

Figure~\ref{Fig:SPIPS} compares the result and the agreement is excellent. The rms in the residuals is about 160~K in
$T_{\rm eff}$ and 0.05~dex in $\log L$. The effective temperature plot suggest that the errors in effective temperature
may have been overestimated by about $\sim 40\%$ in both studies.

\subsection{Hertzsprung-Russell diagram}

\begin{figure}
  \centering

\begin{minipage}{0.49\textwidth}
\resizebox{\hsize}{!}{\includegraphics{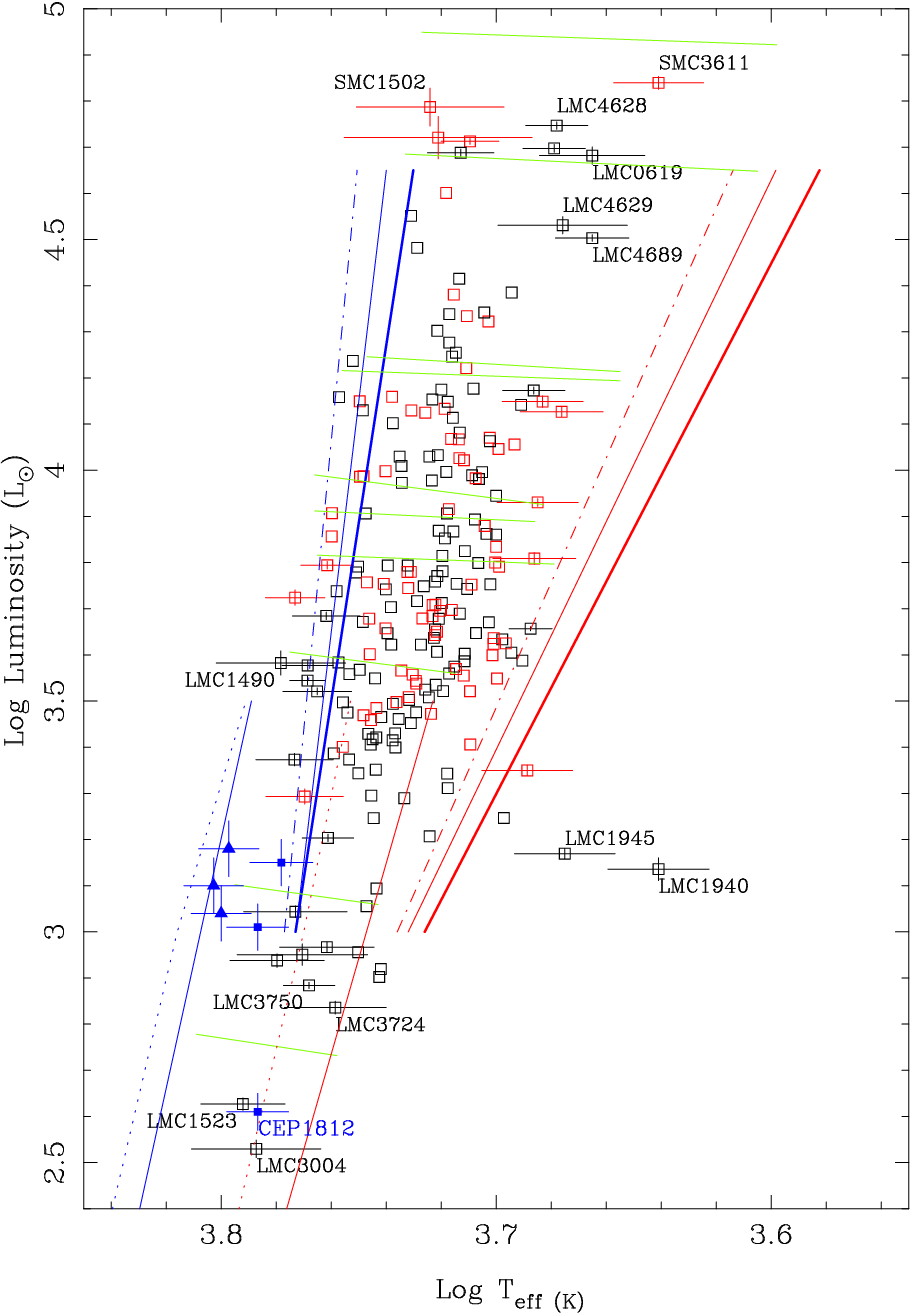}}
\end{minipage}

\caption{Hertzsprung-Russell diagram.
  Black and red open squares indicate LMC and SMC CCs from the sample, respectively.
  Stars located outside the bulk of objects are plotted with error bars, and some are labelled as well.
  Blue symbols with error bars indicate the six CCs in EBs (three FM as filled squares, three FO as filled triangles, see Table~\ref{Tab:DEB}).
  Blue and red lines indicate the blue and red edge of the IS.
  In the upper part (brighter than $\log L = 3$), the results from \citet{DeSomma21} are plotted.
  The thinner solid and dot-dashed line are for $Z = 0.008$ and $Z = 0.004$ models,
  respectively, for their type A mass-luminosity relation. The tick line is for $Z = 0.008$ for their
  type-B mass-luminosity relation.  
  In the lower part (fainter than $\log L = 3.5$), the results from \citet{Pilecki2018} are plotted for FM (solid lines)
  and FO (dotted lines) model.
  Green lines indicate evolutionary models from \citet{Anderson16}. 
  See the main text for details.
}
\label{Fig:HRD}
\end{figure}

Figure~\ref{Fig:HRD} shows the HRD together with sets of evolutionary tracks and ISs.
Objects from the sample are plotted as black (LMC) and red (SMC) open squares, respectively.
Stars located outside the bulk of objects are plotted with error bars, and some are labelled as well.
Blue symbols with error bars indicate the six CCs in EBs (three FM as filled squares, three FO as filled triangles; see Table~\ref{Tab:DEB}).

Two sets of ISs from \citet{DeSomma21} (at brighter magnitudes) and from \citet{Pilecki2018} (at fainter magnitudes) are plotted.
The near horizontal green lines indicate the evolutionary tracks for $Z = 0.006$ and average initial rotation rate
$\omega_{\rm ini} = 0.5$ from \citet{Anderson16}.  Increasing in luminosity are tracks for initial mass
(number of the crossing through the IS): 4 (1), 5 (1), 7 (1), 7 (2), 7 (3), 9 (1), 9 (2), 9 (3), 12 (1), and 15 \msol\ (1).
The density of stars in the HRD is qualitatively consistent with the fact that the first crossing is much faster than the second and third crossings.

There are only two clear outliers, LMC1940 and LMC1945, and their position in the HRD remains unexplained.
The former object has the fourth poorest fit, but the $\chi^2_{\rm r}$ of the fit of LMC1945 is not extremely poor.
LMC1940 and LMC1945 have large \G\ {\tt astrometric\_gof\_al} (GoF) parameters of about 12 and 9.9, respectively
(and {\tt RUWE} values of 1.56 and 1.43, respectively), which may hint to binarity.
However, 38 stars in the sample have a larger GoF than 9.9 and are not outliers in the HRD diagram.
A cautionary note is that the SED fitting assumes the stars to be single.
Contamination of the photometric points by a companion will have an influence on the results of the fitting procedure.
No spectroscopic temperature determinations is available for LMC1945.
For LMC1940 there is a value of $4909 \pm 126$~K determined from APOGEE data \citep{Sprague22}.
A model with $4875$~K (the closest in the available grid) was run to find that the luminosity is about 13\% less than in the best-fitting model.
This temperature and luminosity would put the star closer to the red edge of the IS, but still too cool and
overly under-luminous, compared to expectations.

The location of the known CCs in EBs in noteworthy. Five of them are clearly hotter than expected from the IS by \citet{DeSomma21}
but are consistent with the IS as calculated by \citet{Pilecki2018}.
It is noted that the effective temperatures in \citet{Pilecki2018} have not been derived from the available (disentangled) spectra
that these authors used to obtain the radial velocities, but from effective temperature--colour relations using
the $V-I$ (sometimes $V-K$) colour of the two components, as derived from the modelling of the light curve.
This may have possibly introduced a systematic effect. If the temperatures derived in this way would end up being too high, the luminosities would also
be too large, as indeed inferred from the $PL$-relation (see below).

\subsection{Infrared excess}
\label{S:dust}

The default assumption in the modelling has been that there is no IR excess and the SEDs can be modelled by a stellar atmosphere.
However, near- and mid-IR excess is known to exist in Galactic CCs, for example, direct interferometric observations in the optical or NIR
(e.g. \citealt{Kervella2006, Merand2006, GalVISIR, Nardetto16}),
modelling with the SPIPS code (e.g. \citealt{Breitfelder16, Trahin19, Trahin21}, and \citealt{Gallenne17} for the LMC) 
and was also found in modelling the SEDs of Galactic CCs (\citealt{Gallenne13b}, G20).

Visual inspection of the SEDs showed five cases where an IR excess could explain the shape of the SEDs\footnote{Cases where the excess consisted
only in a single point, typically in the WISE3 filter are not discussed here. They are probably related to contamination in the larger W3 (and W4) beams.
An a-priori selection on photometric error in the W3 and W4 filters was applied (Sect.~\ref{SS-phot}) but this did not
remove all likely unreliable points.
The stars discussed in the main text appear to have IR excess in multiple filters.}.
Following G20, models were run for these five stars including a dust component and additionally fitting for the dust temperature at the inner radius
and the dust optical depth under the assumption of spherical symmetry. Following G20, a mixture of 3\% silicate, 3\% aluminium oxide and 94\% iron dust
was adopted. The analysis of the MW CCs in G20 with available mid-IR spectra showed that these are near featureless requiring a large fraction of
featureless iron dust, although the nature of the excess is in fact unclear.
The results of the fitting are listed in Table~\ref{Tab:dust} that include the magnitudes in various filters for the best-fitting model excluding and including
a dust component.

\begin{figure*}

\begin{minipage}{0.48\textwidth}
\resizebox{\hsize}{!}{\includegraphics{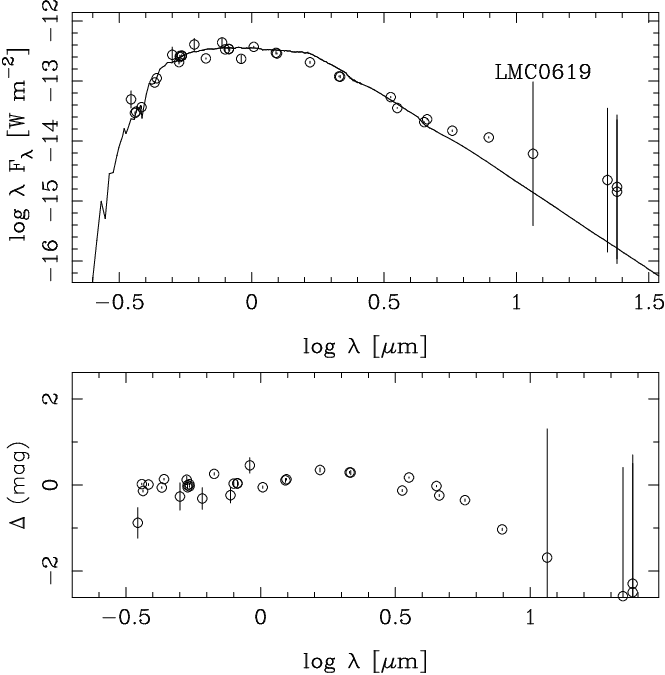}}
\end{minipage}
\begin{minipage}{0.48\textwidth}
\resizebox{\hsize}{!}{\includegraphics{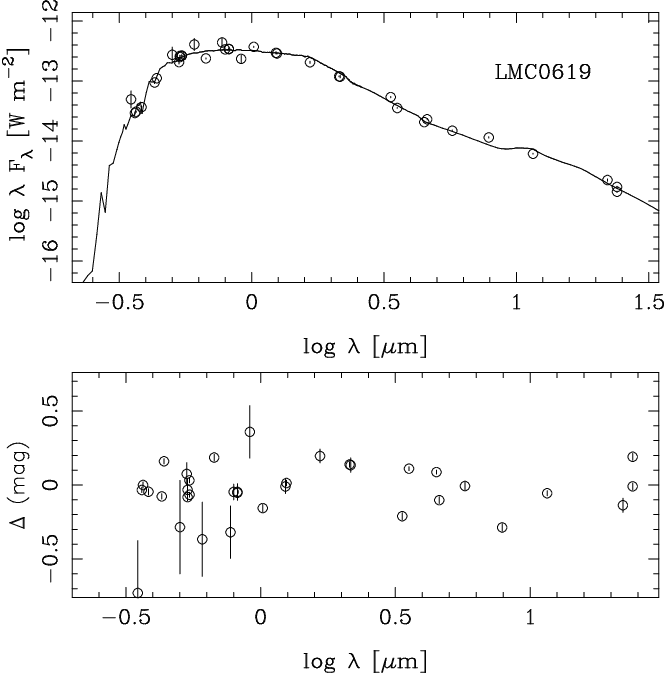}}
\end{minipage}

\caption{Best-fitting model without dust (left side) and with dust (right side) for the star with the most convincing presence of an IR excess.
In the model without dust, the long-wavelength points are considered outliers.
}
\label{Fig:dust}
\end{figure*}

Only the model for LMC0619 is convincing with an excess in four to five filters (see Figure~\ref{Fig:dust}) and an SED comparable in shape to the
SEDs of the MW CCs with IR excess.
In the other four cases, the temperature at the inner radius is very low and based on two filters only (see Figure~\ref{Fig:dust2}).
Also, the reduction in the Bayesian information criteria (BIC) is small.

Interestingly, the best-fitting model with dust predicts fainter magnitudes for LMC0619. For fixed effective temperature and luminosity and for optically thin
cases, one expects some absorption in the optical and emission in the IR. However, in the fitting the effective temperature and luminosity were
allowed to vary, and the best-fitting luminosity is lower when including dust. The difference in the Wesenheit filters is around 0.15~mag, which is significant.

Only one of 142 CCs in the LMC, and 0/77 in the SMC has a convincing IR excess. Coincidence or not, LMC0619 has the longest period of
the LMC objects (133 days) and one of the highest luminosities.
In the MW, IR excess is quite common (see references at the beginning of the section) and G20 lists 16/347 stars as having an IR excess, also based on SED fitting.
It appears that the presence of IR excess is more common in the MW than in LMC and SMC CCs
For comparison, one of the MW stars with IR excess from G20, LS Pup, was refitted and the results are included in Table~\ref{Tab:dust} and Figure~\ref{Fig:dust2}.
A definite conclusion would require a more in debt study beyond the scope of the present paper, as one would have to consider the impact of
the bias due to the fact that for MW CCs more data is available (in some cases even mid-IR spectra) especially at longer wavelengths.
The SED modelling is therefore more likely to find an IR excess in MW stars than in the MCs.

\begin{table*}
  \centering

\footnotesize
\setlength{\tabcolsep}{1.3mm}
\caption{Effect of dust on the colours of CCs }
\begin{tabular}{ccrrcccccccccccccccc} \hline \hline 
  Name   & $T_{\rm eff}$ &    $L$            &  BIC     & Bp     &  G     &  Rp    &   $V$  & $I$   & $J$    &  $K$     & F160W   &  $WG$  & $WVI$    & $WJK$   & $WH$ \\
         &  (K)        &    (\lsol)        &          & (mag)  &  (mag) &  (mag) &  (mag) &  (mag) &  (mag) &  (mag)   &   (mag)  &  (mag) &  (mag)  & (mag)  &  (mag) \\
  \hline   
LMC0619 &   4625 &    48057 $\pm$   1645 &    2809 &  12.37 &  11.71 &  11.12 &  12.27 &  10.97 &  10.18 &   9.45 &   9.67 &   9.34 &   8.96 &   8.95 &   9.18 \\
        &   4750 &    45036 $\pm$ \phantom{1}946 &     713 &  12.42 &  11.79 &  11.20 &  12.30 &  11.06 &  10.30 &   9.60 &   9.81 &   9.47 &   9.14 &   9.12 &   9.35 \\ 
        &        & $T_{\rm c}$=    535 $\pm$   26~K &         &  -0.05 &  -0.08 &  -0.08 &  -0.03 &  -0.09 &  -0.12 &  -0.15 &  -0.14 &  -0.13 &  -0.18 &  -0.17 &  -0.17 \\
LMC1616 &   4875 &     4026 $\pm$     85 &     752 &  14.88 &  14.30 &  13.75 &  14.78 &  13.62 &  12.92 &  12.28 &  12.47 &  12.15 &  11.82 &  11.84 &  12.03 \\
        &   4875 &     4185 $\pm$     91 &     614 &  14.93 &  14.34 &  13.77 &  14.81 &  13.64 &  12.91 &  12.25 &  12.45 &  12.14 &  11.83 &  11.79 &  12.01 \\ 
        &        & $T_{\rm c}$=    329 $\pm$   40~K &         &  -0.05 &  -0.04 &  -0.02 &  -0.03 &  -0.02 &  +0.01 &  +0.03 &  +0.02 &  +0.01 &  -0.01 &  +0.05 &  +0.02 \\
LMC0966 &   5000 &    21971 $\pm$    379 &     971 &  12.99 &  12.44 &  11.92 &  12.89 &  11.79 &  11.13 &  10.54 &  10.71 &  10.41 &  10.08 &  10.13 &  10.29 \\
        &   5000 &    21876 $\pm$    387 &     885 &  13.02 &  12.46 &  11.92 &  12.90 &  11.80 &  11.14 &  10.55 &  10.71 &  10.37 &  10.10 &  10.14 &  10.29 \\ 
        &        & $T_{\rm c}$=    417 $\pm$  125~K &         &  -0.03 &  -0.02 &  +0.00 &  -0.01 &  -0.01 &  -0.01 &  -0.01 &  +0.00 &  +0.04 &  -0.02 &  -0.01 &  +0.00 \\
SMC1172 &   5250 &     4960 $\pm$ \phantom{1}83 &     214 &  14.97 &  14.50 &  14.02 &  14.89 &  13.89 &  13.31 &  12.79 &  12.93 &  12.69 &  12.34 &  12.43 &  12.55 \\
        &   5500 &     5958 $\pm$    182 &     130 &  14.96 &  14.52 &  14.05 &  14.87 &  13.94 &  13.30 &  12.77 &  12.93 &  12.79 &  12.50 &  12.40 &  12.58 \\ 
        &        & $T_{\rm c}$=     90 $\pm$    7~K &         &  +0.01 &  -0.02 &  -0.03 &  +0.02 &  -0.05 &  +0.01 &  +0.02 &  +0.00 &  -0.10 &  -0.16 &  +0.03 &  -0.03 \\
LMC0107 &   5750 &     3834 $\pm$     31 &      55 &  14.89 &  14.46 &  14.01 &  14.81 &  13.89 &  13.35 &  12.92 &  13.04 &  12.79 &  12.46 &  12.62 &  12.69 \\
        &   5875 &     4047 $\pm$     39 &      60 &  14.88 &  14.46 &  14.01 &  14.79 &  13.90 &  13.36 &  12.94 &  13.06 &  12.81 &  12.52 &  12.65 &  12.73 \\ 
        &        & $T_{\rm c}$=    217 $\pm$   28~K &         &  +0.01 &  +0.00 &  +0.00 &  +0.02 &  -0.01 &  -0.01 &  -0.02 &  -0.02 &  -0.02 &  -0.06 &  -0.03 &  -0.04 \\
LS Pup  &   5500 &     4395 $\pm$    152 &    3436 &  10.49 &   9.76 &   9.11 &  10.39 &   8.94 &   8.08 &   7.43 &   7.67 &   7.14 &   6.69 &   6.98 &   7.13 \\
        &   5750 &     4364 $\pm$    153 &    2752 &  10.55 &   9.81 &   9.15 &  10.45 &   8.97 &   8.05 &   7.40 &   7.65 &   7.15 &   6.68 &   6.95 &   7.10 \\ 
        &        & $T_{\rm c}$=  3620 $\pm$  1182~K &         &  -0.06 &  -0.05 &  -0.04 &  -0.06 &  -0.03 &  +0.03 &  +0.03 &  +0.02 &  -0.01 &  +0.01 &  +0.03 &  +0.03 \\
\hline
\end{tabular} 
\tablefoot{
For every star the first line gives the best-fitting parameters based on the model atmosphere, the second line when including a dust component,
and the third line the difference in magnitude. In the column of the luminosity the dust temperature at the inner radius ($T_{\rm c}$) is listed in the third line.
$W$ refers to the Wesenheit magnitudes, defined as follows: 
$WG$ = $G - 1.90 \cdot (Bp-Rp)$,
$WVI$ = $I - 1.55 \cdot (V-I)$,
$WJK$ = $K - 0.69 \cdot (J-K)$,
$WH$  =  F160W $-0.386\, \cdot$ (F555W $-$ F814W).
Note that F555W and F814W magnitudes are not explicitly listed in the table (they are close to $V$ and $I$) but were
calculated and used in calculating the $WH$ magnitude.
}
\label{Tab:dust}
\end{table*}

\subsection{Period-luminosity relations}
\label{SS:PL}

\begin{figure}
  \centering

\begin{minipage}{0.49\textwidth}
\resizebox{\hsize}{!}{\includegraphics{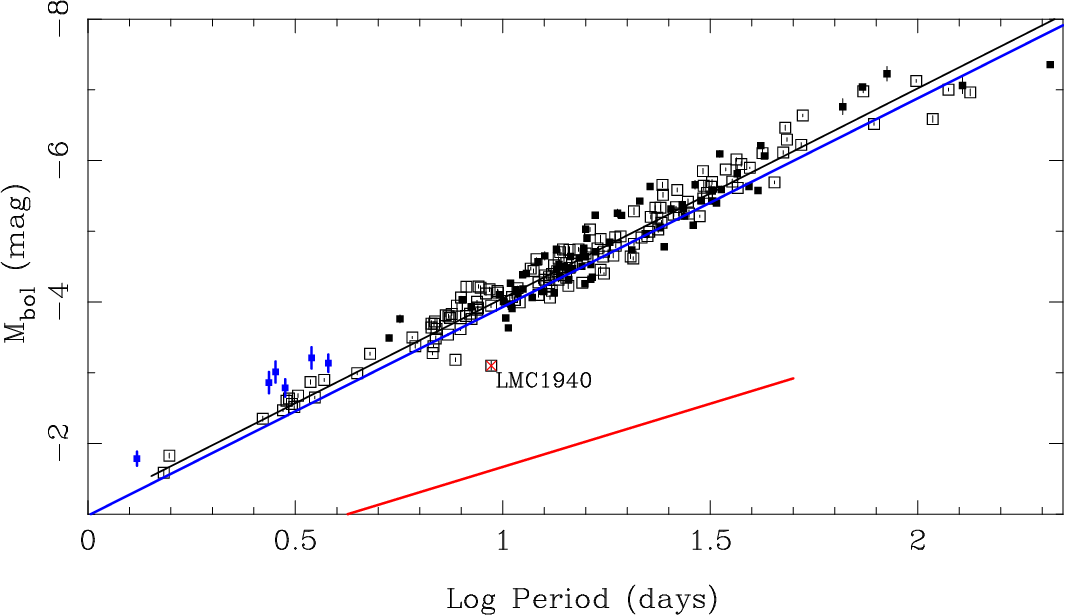}}
\end{minipage}

\caption{Period-$M_{\rm bol}$ relation.
Error bars in $M_{\rm bol}$ are plotted but are typically smaller than the symbol size.
LMC objects are plotted as open squares, SMC objects are filled squares.
Blue symbols indicate the six CCs in EBs.
One outlier is marked by a red cross and is identified.
The black line indicates the best fit (excluding the outlier) to the complete MC sample.
Fits to the LMC and SMC stars separately are given in the text.
The blue line gives the $PL$ relation for MW CCs from G20, while the red line gives the recommended solution
for T2C \citep{GrSED}.
}
\label{Fig:PL}
\end{figure}

Figure~\ref{Fig:PL} shows the $PL$ relation.
A fit to 141 LMC objects (removing one object through 3$\sigma$ clipping) is:
\begin{equation}
  M_{\rm bol} =  (-2.96 \pm 0.05) \log P   + (-1.10 \pm 0.05)  
  \label{Eq:MbolL}
\end{equation}
with an rms of 0.20~mag.
A fit to 77 SMC objects (removing zero outliers) is:
\begin{equation}
  M_{\rm bol} =  (-3.04 \pm 0.11) \log P   + (-0.99 \pm 0.14) 
\label{Eq:MbolS}
\end{equation}
with an rms of 0.25~mag.

LMC1940 is an outlier in the HRD, but also in the $PL$-relation. Its period would suggest $M_{\rm bol} \approx -4.0$,
or $\log L \approx 3.5$.  
In the HRD, this would move the star up but it would still be an outlier and too cool for its expected position.
Although it is underluminous for a CC LMC1940 is too luminous to be a Type-{\sc ii} cepheid (T2C), as indicated by the $PL$-relation
for T2Cs from \citet{GrSED}.
We also note that the CCs in EBs are slightly brighter than the mean relation.

In G20 the slope and the zero point (ZP) based on 380 Galactic CCs were derived to be $-2.95 \pm 0.09$
and $-0.98 \pm 0.07$ (rms of 0.40~mag), respectively.
The slopes derived for the three galaxies agree to within the error bar.
Fixing the slope to the most precise one of that in the LMC ($-2.96$), ZPs of $-4.057 \pm 0.002$ and
$-4.046 \pm 0.004$ are found for LMC and SMC at $\log P = 1$, respectively.

The bolometric $PL$ relation to the Galactic Cepheids presented in G20 did not select based on metallicity.
Figure~\ref{Fig:fehhist} shows the distribution in [Fe/H] of the stars that went into that fitting. The 5-95\% range
is from $-0.31$ to $+0.29$~dex, with a median of $+0.06$.
Selecting stars in the range 0.0 $\le$ [Fe/H] $<$+0.2 to better have a sample of near solar metallicity leads to a median
of +0.09 (with 0.06~dex dispersion\footnote{Calculated as 1.48 times the median absolute deviation.}) and a $PL$-relation of
\begin{equation}
  M_{\rm bol} =  (-2.64 \pm 0.11) \log P   + (-1.34 \pm 0.09)  
\label{Eq:Mbol}
\end{equation}
with an rms of 0.35~mag using 191 stars, and a ZP of $-4.041 \pm 0.002$ at 10 days for a fixed slope of $-2.96$.

\begin{figure}
  \centering

\begin{minipage}{0.49\textwidth}
\resizebox{\hsize}{!}{\includegraphics{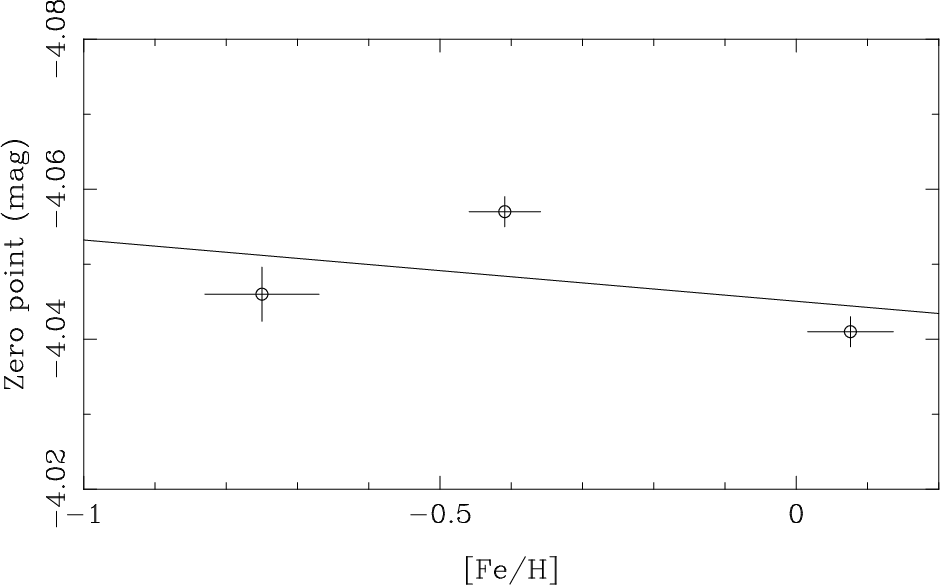}}
\end{minipage}

\caption{Zero point at 10 days of the $PL$-relation for SMC, LMC, and MW.
  The line is a linear fit with a slope $(0.0082 \pm 0.0076)$ mag/dex which is not significant.
}
\label{Fig:metal}
\end{figure}

Figure~\ref{Fig:metal} shows the ZP at 10 days plotted against metallicity. The metallicity for the MW is the one just derived,
for the LMC  $-0.409$~dex with dispersion of $0.076$ \citep{Romaniello22} is adopted and
for the SMC, a value of  $-0.75$~dex with dispersion of $0.08$ \citep{Romaniello08} is assumed.
A least-squares fit taking into account the error bars in both axes gives
\begin{equation}
  {\rm ZP @ 10 d}=  (-4.0451 \pm  0.0036) + (+0.0082 \pm 0.0075) \cdot {\rm [Fe/H]}
\end{equation}
Using slightly different values for the LMC, for example, $-0.35 \pm 0.09$~dex for BSGs \citep{Urbaneja17}, or
$-0.62 \pm 0.14$~dex for SMC cluster giants (\citealt{Trundle07}, as quoted in \citealt{Romaniello08}) all lead
to slight positive but insignificant slopes.
A constant value of $-4.049$~mag with dispersion 0.008~mag would fit the ZPs of all three galaxies.

The result that the bolometric $PL$ relation does not seem to depend on metallicity is in contrast with the most recent
results of \citet{Breuval22} that derive the metallicity term in various filters from \G\ Bp band to IRAC 4.5~$\mu$m and
find little dependence on wavelength and an average of $\gamma= -0.29$~mag/dex.
Our results indicate that the LMC cepheids are indeed brighter than the SMC ones (this is so when $\gamma$ is negative, and fitting only
the SMC and LMC data points gives $\gamma= -0.04$) but it would imply that the ZP for the MW cepheids is too faint
by $(-0.75 - +0.08) \cdot -0.29 \approx 0.24$~mag. 
Further study is required, especially on the MW sample, and parallaxes from \G\ DR4 will be crucial in this regard.
Additionally, the difference between the bolometric magnitude and the magnitude in any photometric band involves a
bolometric correction that should depend on wavelength and makes a direct comparison of the $\gamma$ terms less evident.

\subsection{Period-radius relation}
\label{SS:PR}

\begin{figure}
  \centering

\begin{minipage}{0.49\textwidth}
\resizebox{\hsize}{!}{\includegraphics{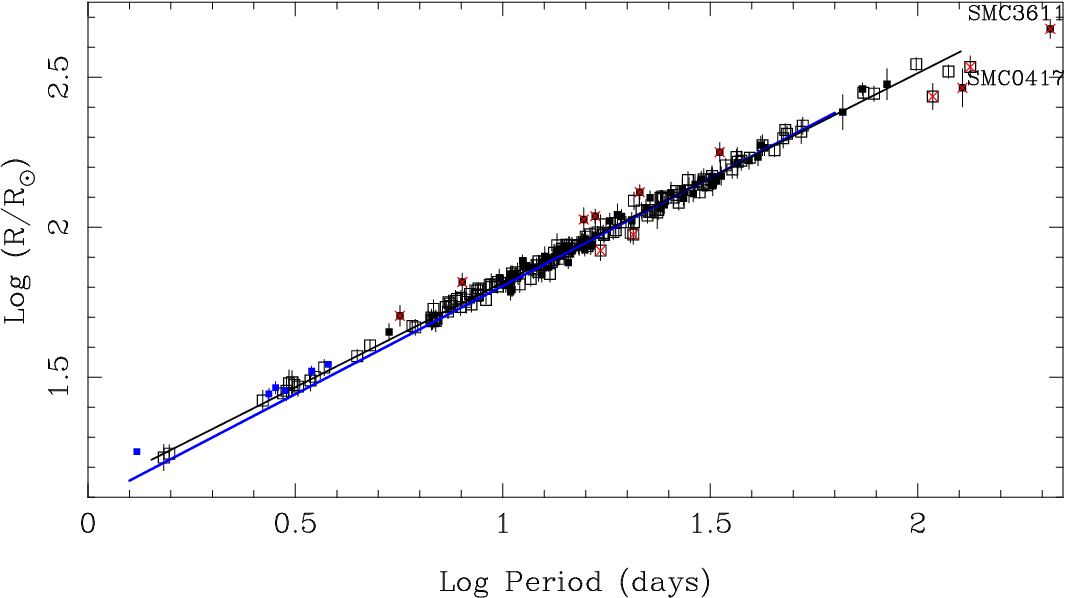}}
\end{minipage}

\caption{
Period-radius relation.
LMC objects are plotted as open squares, SMC objects are filled squares.
Blue symbols indicate the six CCs in EBs.
Outliers removed from the fit are marked by a red cross and some are identified.
The black line indicates the best fit (excluding the outliers) to the complete MC sample.
Fits to the LMC and SMC stars separately are given in the text.
The blue line gives the $PR$ relation for MW CCs from G20.
}
\label{Fig:PR}
\end{figure}

Figure~\ref{Fig:PR} shows the $PR$ relation.
The relation for the LMC is:
\begin{equation}
 \log R =  (0.6966 \pm 0.0043) \log P  + (1.1194 \pm 0.0052),   
\label{Eq:PRL}
\end{equation}
with an rms of 0.017~dex and using 138 stars (removing 4 outliers), 
and then for the SMC
\begin{equation}
 \log R =  (0.697 \pm 0.013) \log P  + (1.134 \pm 0.017),   
\label{Eq:PRS}
\end{equation}
with an rms of 0.027~dex and using 76 stars (removing 1 outlier).
Figure~\ref{Fig:PR} also shows the $PR$ relation for MW CCs from G20.
The slope derived there was $0.721 \pm 0.013$ with a ZP of $1.083 \pm 0.012$.
Refitting the data in G20 restricting the metallicity to 0.0 $\le$ [Fe/H] $<$+0.2 gives:
\begin{equation}
 \log R =  (0.668 \pm 0.020) \log P  + (1.143 \pm 0.017),  
\label{Eq:PRMW}
\end{equation}
with an rms of 0.069~dex and using 190 stars.
The slopes are consistent with each other and at $P= 10$~d the radii in MW, LMC, and SMC CCs are identical to within the error bars.

\subsection{The mass-luminosity relation}

\begin{figure}
  \centering

\begin{minipage}{0.49\textwidth}
\resizebox{\hsize}{!}{\includegraphics{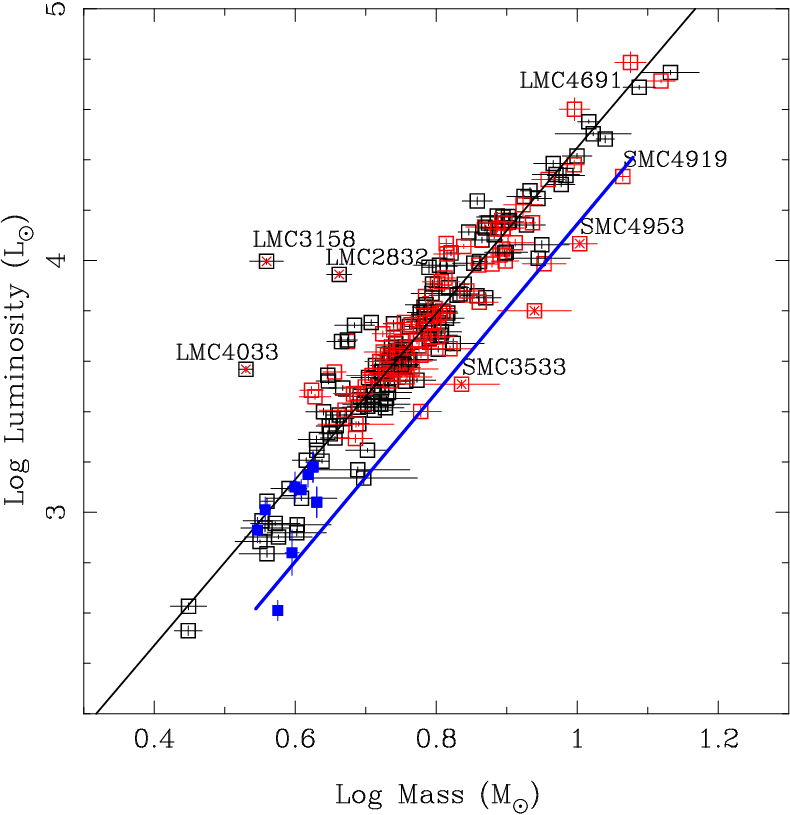}}
\end{minipage}

\caption{Mass-luminosity relation.
Objects in the LMC are plotted as black, those in the SMC as red open squares.
Stars outside the bulk of objects are identified.
Objects in blue are the known CCs and their companions in EBs.
The blue line indicates the canonical ML relation from \citet{Bono00} for $Y= 0.255$ and $Z= 0.008$.
For $Z= 0.004$ the line would be at higher luminosities by about 0.1~dex.
The black line indicates a fit to the LMC stars minus the outliers (see Eq.~\ref{Eq:ML}) and
lies almost exactly +0.3~dex above the canonical ML relation 
}
\label{Fig:ML}
\end{figure}

Figure~\ref{Fig:ML} shows the relation between mass and luminosity.
The best fit is:
\begin{equation}
\log L  = (3.193 \pm 0.060) \log M + (1.237 \pm 0.048) 
\label{Eq:ML}
\end{equation}
based on 208 stars and an rms of 0.12~dex, and lies approximately
0.3~dex above the canonical ML relation from \citet{Bono00} for Helium abundance $Y= 0.255$ and metallicity $Z= 0.008$.
A few stars scatter clearly above the relation.
The few stars that scatter below the best-fit relation are consistent with the canonical ML relation.
The best-fit relation is the intermediate between the case B (+0.2 dex w.r.t. the canonical relation)
and case C (+0.4 dex) ML relations adopted in \citet{Marconi20}.

\subsection{Comparing stellar parameters}

The determination of the photometric effective temperature allows one to compare it to the spectroscopic temperature determined in
the literature.
In addition, via the derived luminosity, and estimated mass, it is possible to compare the gravity to that derived by spectroscopy.
The values for the spectroscopic effective temperature and gravity come from the papers that contribute a large fraction
of the stars in the sample \citep{Romaniello08,Romaniello22,Lemasle17}.
In addition, spectroscopicly derived parameters from the APOGEE survey have been considered \citep{Sprague22}, as well
as the values derived from the LC fitting in \citet{Ragosta19} and \citet{Marconi17}.
The adopted values for temperature and gravity are listed in columns 6 and 7 of Table~\ref{Tab-Sam}.
When data from multiple references were available the order of preferences was \citet{Romaniello22}, \citet{Romaniello08},
\citet{Lemasle17}, \citet{Sprague22}, and \citet{Marconi17} or \citet{Ragosta19}.
Results from GDR3 were not considered in the end. The results from the GSP\_Spec analysis \citep{Recio-Blanco22} were inspected
but only one, respectively two, had an entry from the so called Matisse-Gauguin and ANN pipeline when selecting 
'000000' for the first six values in the {\tt flags\_gspspec} and {\tt flags\_gspspec\_ann} flags.

\begin{figure}
  \centering

\begin{minipage}{0.49\textwidth}
\resizebox{\hsize}{!}{\includegraphics{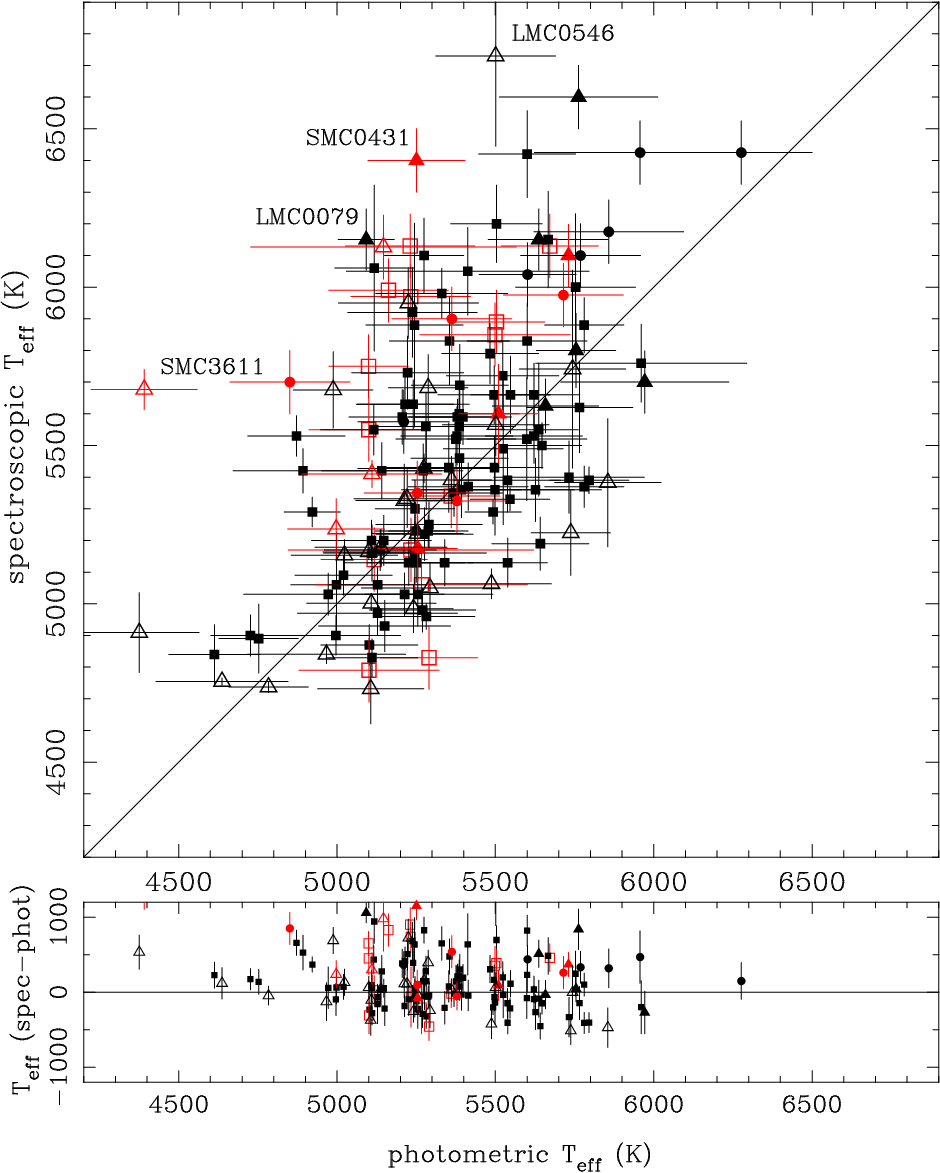}}
\end{minipage}

\caption{Comparison between the effective temperatures determined from the SED fitting and in the literature.
Objects in the LMC are plotted in black and objects in the SMC in red.
The one-to-one line is indicated.
Stars where the absolute difference is larger than 1000~K are identified.
Objects from reference 1 and 2 (see the note to Table~\ref{Tab-Sam}) are plotted as filled squares, from reference 3 as open squares,
from reference 4 as filled triangles, from reference 5 as open triangles, and from reference 6 as filled circles.
In the lower panel, the difference between literature and photometric temperature is plotted.
}
\label{Fig:TeffTeff}
\end{figure}

Figure~\ref{Fig:TeffTeff} shows the comparison of the photometric and literature temperatures.
On average the spectroscopic temperatures seem to be slightly larger, although there is large scatter.
The median offset is  $+91 \pm 330$~K.
The effective temperature changes over the pulsation cycle so an exact agreement is in fact not expected.
The spectra taken from the works  dedicated to determining the metallicity (and $T_{\rm eff}$ and $\log g$ in
the process, \citealt{Romaniello08,Romaniello22,Lemasle17}) of CCs typically try to avoid the phases where
shocks play a role the objects.
On the other hand the APOGEE data were taken at random phases while the photometric temperatures was derived
from the SED that was constructed to be as much as representative of mean light as possible.

\begin{figure}
  \centering
 
\begin{minipage}{0.49\textwidth}
\resizebox{\hsize}{!}{\includegraphics{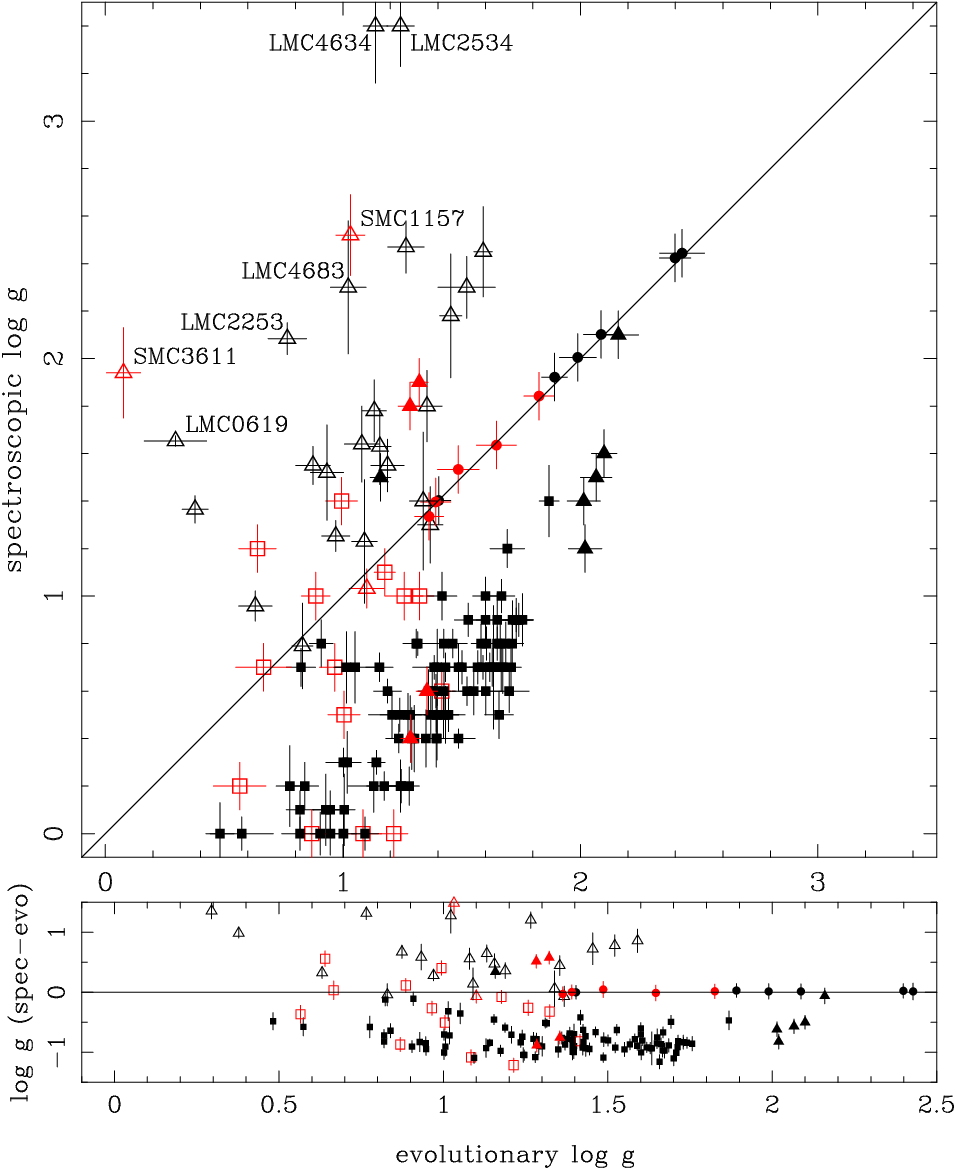}}
\end{minipage}

\caption{Comparison between the $\log g$ determined in the literature, and the evolutionary value
  as determined from $L$ and $T_{\rm eff}$ from
  the SED fitting and the stellar mass based on various methods (see Appendix~\ref{MApp}).
The one-to-one line is indicated.
Stars outside the bulk of objects are identified.
Symbols as in Fig.~\ref{Fig:TeffTeff}.
}
\label{Fig:logg}
\end{figure}

Figure~\ref{Fig:logg} shows the same for the $\log g$ values determined in the literature.
The overall median offset is $-0.70 \pm 0.36$~dex, but it strongly depends on the source.
The $\log g$ values derived from the pulsation models are in very good agreement with the values
determined in the present paper, the values from APOGEE are larger     (median offset $+0.58 \pm 0.46$~dex), while 
those derived from dedicated HR spectroscopy are significantly smaller (median offset $-0.80 \pm 0.21$).           
A similar plot for the FWG is shown in the Appendix (Fig.~\ref{Fig:loggFSPC}).

The differences between the spectroscopic and photometric effective temperature are relatively small, but
the spectroscopically determined gravity (and the FWG) from \citet{Romaniello08,Romaniello22} and \citet{Lemasle17}
are systematically and significantly smaller than that derived from $L, T_{\rm eff}$, and stellar mass.
Like the effective temperature the gravity also changes over the pulsation cycle.
There is a change in radius, but there is also a dynamical term.
What is thus determined from HR spectroscopy is the effective gravity:
\begin{equation}
    g_{\rm eff} =  \frac{G\;M}{R(t)^{2}} + \frac{\partial^2{R}}{\partial{t}^2} = \frac{G\;M}{R(t)^{2}} - p \;\frac{\partial V_{\rm r}(t)}{\partial t}, 
 \end{equation}
where $M$ is the mass of the CC, $R(t)$ is the radius as a function of time (or pulsation phase),
$p$ is the projection factor (see \citealt{Nardetto04} and references therein)
and $V_{\rm r}(t)$ is the radial velocity at the time $t$.

Recently, \citet{daSilva22} presented time series of HR spectroscopy for 20 calibrating MW CCs and the analysis of these spectra in terms of
radial velocities, metallicities, micro turbulent velocities, gravities and effective temperatures (see their Appendix~B).
This unique dataset also allows to study the effect of the effective gravity.
The dynamical term $\frac{\partial V_{\rm r}(t)}{\partial t}$ at each phase point $\phi_{\rm i}$ was approximated
as $\frac{V_{\rm r}(\phi_{\rm i+1})-V_{\rm r}(\phi_{\rm i-1})}{\Delta t}$ and a typical $p$ factor of 1.25 was adopted.

\begin{figure*}
  
\begin{minipage}{0.48\textwidth}
\resizebox{\hsize}{!}{\includegraphics[angle=-0]{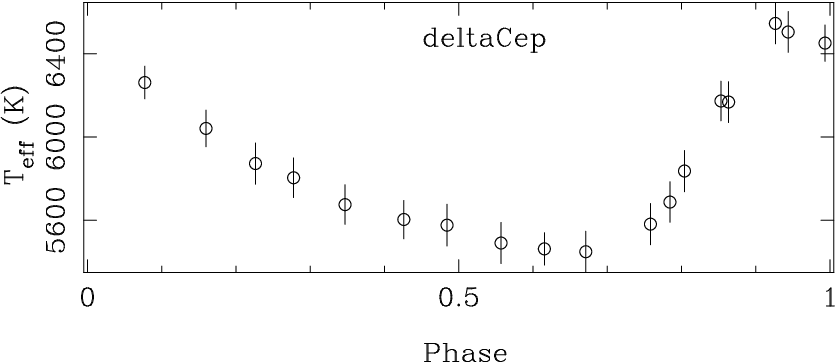}} 
\end{minipage}
\begin{minipage}{0.48\textwidth}
\resizebox{\hsize}{!}{\includegraphics[angle=-0]{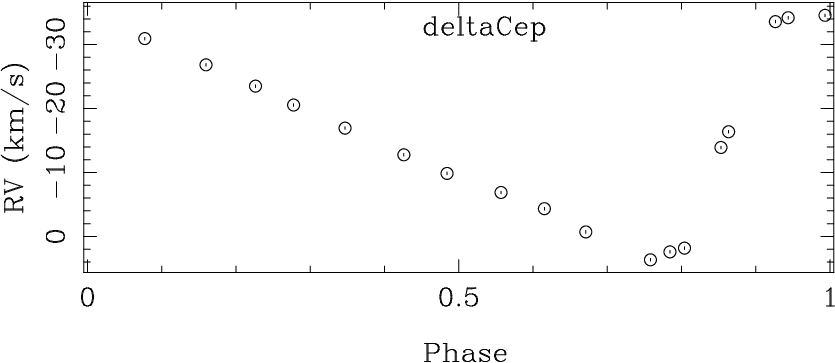}} 
\end{minipage}
 
\begin{minipage}{0.48\textwidth}
\resizebox{\hsize}{!}{\includegraphics[angle=-0]{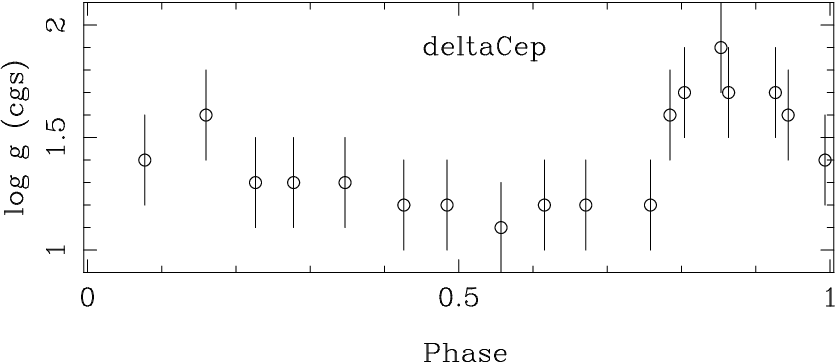}} 
\end{minipage}
\begin{minipage}{0.48\textwidth}
\resizebox{\hsize}{!}{\includegraphics[angle=-0]{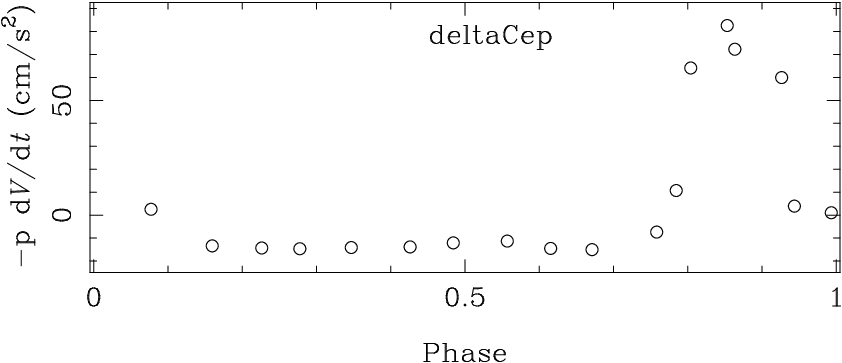}} 
\end{minipage}
 
\begin{minipage}{0.48\textwidth}
\resizebox{\hsize}{!}{\includegraphics[angle=-0]{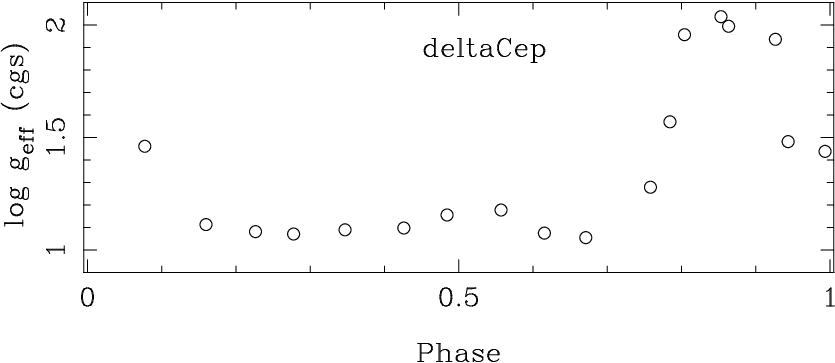}} 
\end{minipage}
\caption{Time series of effective temperature, radial velocity and gravity for $\delta$ Cep from \citet{daSilva22}.
  Calculated in the presented paper are the dynamical term and the effective gravity based on the mean gravity and the dynamical term.
  The ordinate for the effective gravity is set to that of the observed gravity.
}
\label{Fig:delCep}
\end{figure*}

The effective gravity is calculated as $\log g_{\rm eff} = 10^{\rm log\, g_{mean}} +$ (dynamical term), where $g_{\rm mean}$ 
is the weighted mean gravity over all available phase points (recalculated in the present paper and identical to the values
in Table~6 of \citealt{daSilva22}). This ignores the variation in radius with phase but this effect is smaller than the dynamical
term (Appendix~\ref{App:Vas}).
Figure~\ref{Fig:delCep} shows the result for $\delta$ Cep. The figure is ordered in such a way that the dynamical term appears below
the RV curve and the calculated effective gravity appears below the observed gravity.
Contrary to the convention in \citet{daSilva22} phase 0 is taken at maximum light.
Plots for some of the stars with the best phase coverage are shown in Appendix~\ref{App:AF} (Figs.~\ref{Fig:TVul}-\ref{Fig:zetaGem}).
Overall there is reasonable to good correspondence between the effective gravity and the observed gravity, and the expected rise in gravity
due to the dynamical terms occurs at the correct phase.
The results show that if observations are taken at phases that avoid the sharp rise in radial velocity the effective temperature and gravity
will be systematically lower than the average over the light curve.
However, the effect should be a few 0.1~dex, and cannot explain the large difference between the spectroscopic and evolutionary gravity
noted in Fig.~\ref{Fig:logg}.
The behaviour of the effective gravity is also confirmed by a theoretical model, see Appendix~\ref{App:Vas}.
It is beyond the scope of the present paper, but the analysis of time series data, as presented in \citet{daSilva22}, allows one 
to put constrains or derive the $p$-factor for individual stars, as the dynamical term is proportional to $p$ and the effective gravity should
match the observed gravity.

\begin{figure}
  \centering

\begin{minipage}{0.495\textwidth}
\resizebox{\hsize}{!}{\includegraphics{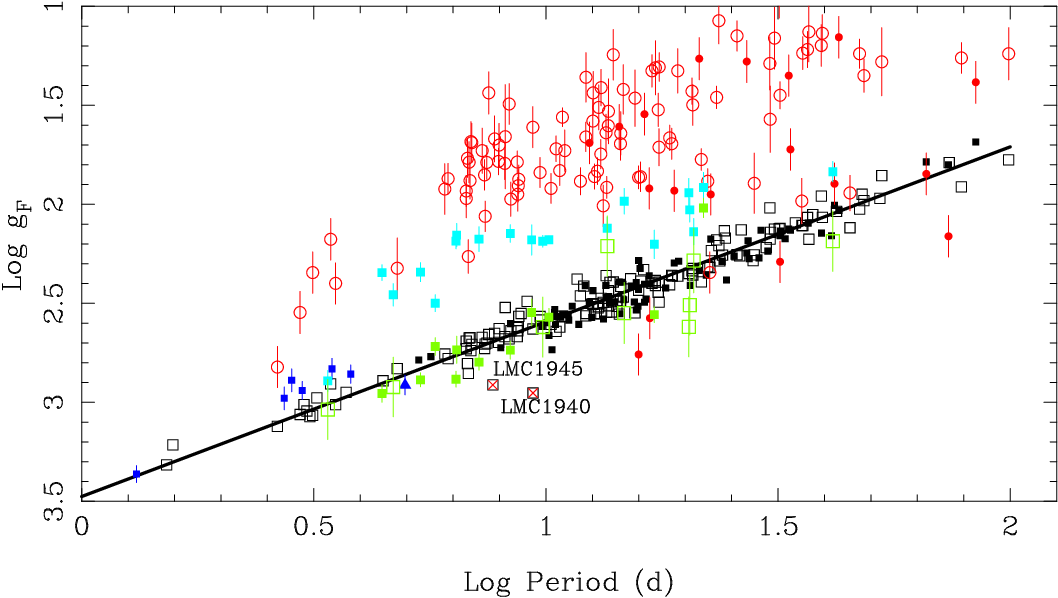}}
\end{minipage}

\caption{FWG versus  $\log P$. Black points indicate the FWGs derived from $L$ and $T_{\rm eff}$ (from the SED fitting) and stellar mass
in the present paper.
Black open squares are LMC objects, filled squares indicate SMC objects.
Blue symbols indicate the CCs (filled squares) and one T2C (filled triangle) in EB systems.
The black line is a fit to the black points, minus the two marked outliers.
The red circles indicate FWGs from HR spectroscopy (open circles indicate LMC, filled circles indicate SMC)
Light blue filled squares indicate FWGs of 20 MC CCs from \citet{daSilva22}, while
the green points indicate the FWGs for the same 20 objects from \citet{Luck18}
(open squares indicate that only one spectrum is available in \citet{Luck18}, filled squares
indicate that five or more spectra are available).
}
\label{Fig:gFPer}
\end{figure}

\begin{figure}
  \centering

\begin{minipage}{0.495\textwidth}
\resizebox{\hsize}{!}{\includegraphics{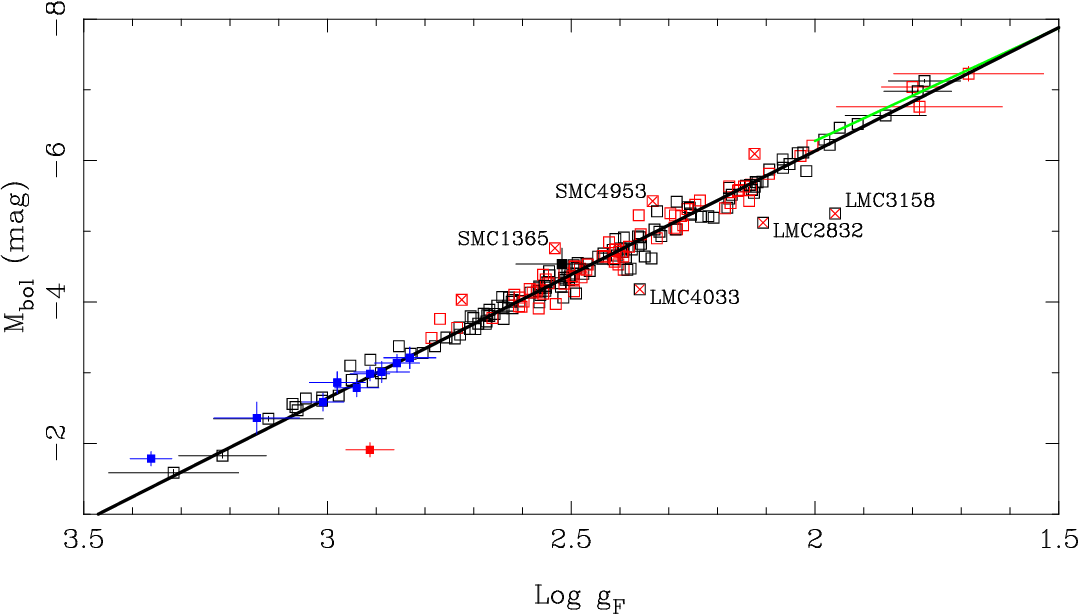}}
\end{minipage}

\caption{FWGLR: 
LMC and SMC objects are identified by black and red open squares.
To avoid cluttering the plot, representative error bars are plotted
for the smallest and largest values of $\log g_{\rm F}$ only.
Some outliers are marked, and stars excluded from the fit are marked by a red cross.
Blue filled squares indicate the six EB CCs and their companions, the filled red square indicates
the T2C in the LMC-T2CEP-098 system and the filled black square its companion.
The black line gives the best fit to the objects (Eq.~\ref{Eq:FWGPer}).
The green line at small FWGs is the fit to LMC BSGs \citep{Urbaneja17, Bresolin22}.
}
\label{Fig:gFLum}
\end{figure}

\subsection{FWG-period and the FWGLR} 
\label{S-FWG}

Figure~\ref{Fig:gFPer} shows the relation between the FWG and the pulsation period based on the analysis in the present paper as well as
on the available HR observations for the sample and two identical samples of MW CCs.
The best fit to the LMC objects is: 
\begin{equation}
  \log g_{\rm F} =  (-0.856 \pm 0.016) \;\log P + (3.442  \pm 0.019)
\label{Eq:FWGPerL}
\end{equation}
with an rms of 0.057~dex, and is indistinguishable from the best fit to the SMC objects:  
\begin{equation}
  \log g_{\rm F} =  (-0.854 \pm 0.033) \;\log P + (3.442  \pm 0.042)
\label{Eq:FWGPerS}
\end{equation}
with an rms of 0.062~dex. The preferred solution combines the SMC and LMC objects as follows: 
\begin{equation}
  \log g_{\rm F} =  (-0.853 \pm 0.014) \;\log P + (3.442  \pm 0.017)
\label{Eq:FWGPer}
\end{equation}
with an rms of 0.059~dex using 212 stars.
This relation is in good agreement both with the theoretical prediction
$\log g_{\rm F} =  (-0.834 \pm 0.011) \;\log P + (3.402  \pm 0.011),$ derived in \citealt{GrFWG} based on the models in
\citet{Anderson16} and the relation 
$\log g_{\rm F} =  (-0.80 \pm 0.03) \;\log P + (3.43  \pm 0.03)$ derived for MW CCs \citep{GrFWG}.

Of interest are the location of all the coloured points in this plot.
The blue points indicate the CCs and one T2C in EBs and these agree well with the observed relation.
That some appear to be slightly above the relation could be related to an overestimate of the effective temperature, as argued before.
The red circles indicate the objects from the sample where a FWG is available from HR spectroscopy.
Almost all lie clearly above the relation, and those that do appear to be on a line parallel to the derived relation
with on offset of about 0.8~dex (also see Fig.~\ref{Fig:loggFSPC}).
To investigate this further we again used the data from \citet{daSilva22} on 20 calibrating MW CCs.
These points are the light blue squares\footnote{FF Aql is an overtone pulsator and the star is plotted at its fundamentalised
  period of 6.401 days.}. Except for the shortest period value (R TrA at $P=$ 3.4 days), which agrees well with the mean relation
all others form a sequence that lies above and is inclined to the mean relation.
All these 20 CCs also have (in part multi-epoch) HR spectroscopy data that is analysed in \citet{Luck18}, which was the main source of data
used in our previous study on MW cepheids \citep{GrFWG}.
These points are the green squares and those fit Eq.~\ref{Eq:FWGPer} very well.
If we demand that the absolute difference between the observed FWG and that predicted by Eq.~\ref{Eq:FWGPer} is less than
$\left( \delta \cdot \sqrt{0.059^2 + \sigma_{\rm FWG}^2}\right)$ with $\sigma_{\rm FWG}$ the observed error in the FWG than
all 20 objects from \citet{Luck18} obey this relation for $\delta= 1.5$.
For this value of $\delta$ only 4 out of the same 20 objects obey this relation using the FWGs from \citet{daSilva22}
-- and even these four stars all lie above the relation -- and
only 9\footnote{These are LMC 0079, 0461, 2019, 2832, and 3724, and SMC 0431, 0574, 0921, and 4444.} out of 104 stars using
the FWGs as derived from HR spectroscopy in the MCs.
It is beyond the scope of the present paper to investigate this further (see also the discussion and appendix
in \citealt{GrFWG}, where a similar effect was noticed).
Nevertheless, Fig.~\ref{Fig:LCKdaS} shows the comparison between temperature, gravity, and the differential between \citet{Luck18} and
\citet{daSilva22} for the same 20 objects. Most interesting is the result that there is a correlation
between the two data sets. When for an object $T_{\rm eff}$ is larger in \citet{Luck18} then in \citet{daSilva22}, then also
the gravity is larger.

Figure~\ref{Fig:gFLum} shows the FWGLR.
Using a least-squares fit taking into account errors in both axes gives the following fit
to the LMC objects:
\begin{equation}
  M_{\rm bol} =  (3.479 \pm 0.032) (\log g_{\rm F} - 2.5) - (4.390 \pm 0.010),
\end{equation}
with an rms of 0.16, and a fit to the SMC CCs of: 
\begin{equation}
  M_{\rm bol} =  (3.577 \pm 0.097) (\log g_{\rm F} - 2.5) - (4.390 \pm 0.021), 
\end{equation}
with an rms of 0.12. 
The combined fit is the preferred solution and is expressed as:
\begin{equation}
  M_{\rm bol} =  (3.492 \pm 0.028) (\log g_{\rm F} - 2.5) - (4.388 \pm 0.009),
\label{Eq:FWGLR}
\end{equation}
with an rms of 0.12~mag using 207 objects.
Blue filled squares in the plot indicate the six EB CCs and their companions, the filled red square indicates
the T2C in the LMC-T2CEP-098 system and the filled black square its companion. Except for the T2C itself, the CCs, and the companions
agree very well with Eq.~\ref{Eq:FWGLR}.
Where there is overlap at small FWGs, there is also good agreement with the FWGLR derived for BSGs in the LMC \citep{Urbaneja17, Bresolin22}.

  This demonstrates the power of the FWGLR as the BSGs have masses in the range 12-40~\msol\ (Fig.~5 in \citealt{Urbaneja17}), while the cepheids
  in the present sample have lower masses that are estimated to be in the range 2.8-13.5~\msol\ (median of about 6~\msol).
  Using evolutionary tracks \citet{K20} demonstrated that, what they named an "extended," FWGLR is expected over 17~magnitudes
  in $M_{\rm bol}$ (with a scatter of 0.17~mag below $M_{\rm bol}= -3.0$~mag) and for masses in the range 0.8-40~\msol, which they verfied
  using a sample of RGB stars with a typical mass of 1.1~\msol.

\section{Discussion and summary}
\label{S:Dis}

This paper is a follow-up of  G20 and 
\citet{GrFWG}, where the SEDs of 477 MW cepheids were fitted. All stars had metallicities based on HR spectroscopy from the literature.
Excluding non-CCs and overtone pulsators the $PL$, the $PR$ and other relations were typically based on about 370 FM CCs.
Some of the relations have been redetermined in the present paper using a restricted range in metallicity to have a sample of MW CCs with
near solar metallicities and these relations are typically based on 190 CCs.

The present study covers 142 LMC and 77 SMC FM CCs. All known (FM) CCs in the MCs with metallicities based on HR spectroscopy are included and
those constitute about half of the sample. Other CCs are included because they were studied otherwise (for example a Baade-Wesselink analyses was conducted)
or may be of interest in future work (ongoing spectroscopic or {\it HST} observations).
The advantage of the current sample compared to the MW sample is that the reddening is better established and, in particular, the distance is well known for the MCs.
This means that the $PL$, $PR,$ and other relations have better determined slopes and smaller residuals compared to the MW relations.

One interesting result is that the zero point of the bolometric $PL$ relation (when fixing the slope to that of the LMC) does not seem to depend on metallicity, contrary
to the recent result that in photometric filters covering a large range in wavelength there is a significant metallicity terms that is essentially constant with 
wavelength (\citealt{Breuval22} and references therein). A new study of MW CCs with improved distances from \G\ DR4 could strengthen this conclusion.

The power of the FWG is again demonstrated. Both the relation of the FWG with period and with luminosity are very tight.
The relation based on the present analysis (gravity derived from the radius, that follows from $T_{\rm eff}$ and $L$, and the stellar mass as derived from several relations)
is in excellent agreement with theory and (where it overlaps) with the relation derived for BSGs.
However, a large fraction of the stars in the sample for which gravities and effective temperatures have been derived from HR spectroscopy show gravities
and FWGs that are smaller than expected by about 0.8~dex.
For the MW sample two recent studies that both analyse time series of HR spectra show strikingly different results in this respect.
The FWGs based on \citet{daSilva22} lie mostly above the expected FWG-period relation (but less so than for the MCs), while the
FWGs based on \citet{Luck18} for the identical sample of 20 stars are in very good agreement with this relation.
Its is beyond this paper to try to resolve this discrepancy as it must be related to the details of the spectroscopic analysis approach.
Of note is that the effective temperature and gravity differences between \citet{daSilva22} and \citet{Luck18} appear correlated.
Since gravity, effective temperature, micro turbulent velocity, and metallicity are determined simultaneously in a spectroscopic analysis it is of interest
to investigate whether these correlations also lead to different metallicity estimates. This appears not to be the case (bottom right panel in
Figure~\ref{Fig:LCKdaS}). Restricting oneself to the 11 stars, where there are 5 or more available spectra per star in \citet{Luck18} the difference in
metallicity between \citet{Luck18} and \citet{daSilva22} is $0.00 \pm 0.05$~dex. Thus,  at least at solar metallicities the (correlated) differences
between temperature and gravity determinations do not lead to differences in metallicity. It remains to be seen whether this is also the case
at lower metallicities where the differences in gravity are much larger.

\begin{acknowledgements}
This research was supported by the International Space Science Institute (ISSI) in Bern,
through ISSI International Team project \#490,  SHoT: The Stellar Path to the Ho Tension in the Gaia, TESS, LSST and JWST Era".
The paper benefitted from the interesting talks and discussions at the memorable "Large-scale surveys as bridges between spectroscopy and photometry"
conference at La Palma, Spain in September 2022.
MG would like to thank
Dr. Valeriy Vasilyev for making the data in \citet{Vasilyev17} available in electronic form,
Dr. Bertrand Lemasle for comments on the spectroscopic analysis in the literature, and
Dr. Lucas Marci for pointing out the webpage with the HST programs and target lists.
This work has made use of data from the European Space Agency (ESA) mission {\it Gaia} 
(\url{http://www.cosmos.esa.int/gaia}), processed by the {\it Gaia} Data Processing and Analysis Consortium 
(DPAC, \url{http://www.cosmos.esa.int/web/gaia/dpac/consortium}). 
Funding for the DPAC has been provided by national institutions, in particular
the institutions participating in the {\it Gaia} Multilateral Agreement.
This research has made use of the SIMBAD database and the VizieR catalogue access tool 
operated at CDS, Strasbourg, France.
\end{acknowledgements}

\bibliographystyle{aa.bst}
\bibliography{references.bib}

\begin{appendix}

\section{Walraven photometry}
\label{WApp}

\citet{vanGenderen1983} published Walraven data taken between 1971 and 1978 of CCs in the MCs 
at the Leiden southern station in Hartebeespoordam in South Africa. The telescope and photometer were
moved to La Silla observatory at the end of 1978 where the data taking continued from January 1979 onwards.
The move was also used to make several improvements to the system \citep{PelvGL81, deRuiter86}.

Table~\ref{Tab:WalPhot} collects these, as of yet, largely unpublished photometry for CCs (some initial results were presented in \citealt{PelvGL81}
and \citealt{vanGenderen1989} for HDE 270100).
The HV number, the Heliocentric Julian date (HJD) of the observations, and the $V$, $V-B$, $B-U$, $U-W$,
and $B-L$ colours on the Walraven system are listed\footnote{It is recalled that Walraven photometry is given on a $\log$ intensity ($I$)
  scale, not on a magnitude scale.}.
The last column lists a quality flag, that ranges between 0 and 9 (in fact, it can be a '*' for
extremely poor observations, but these have been filtered out).
However, photon statistics also play a role for fainter objects.
Light curves were inspected and fitted to determine what typical rms values can be achieved as
a function of the quality flags for the magnitude range of these Cepheids.
In the end an uncertainty of 0.015 is adopted for
quality flags 0-5, and 0.019, and 0.037 in $\log I$  
for quality flags 6 and 7, respectively. Points with a quality flags 8 and 9 are excluded from the fitting of the light curves.

For stars already observed by \citet{vanGenderen1983} the new data were 
added to the published data, after applying the following corrections to the data in \citet{vanGenderen1983}
that reflect the slightly different photometric system and set-up between the 1971-1978 and the later observations.
Referring to these as the "70" and the "80" system, respectively, these corrections are:
\begin{eqnarray}
  V80  &=& V70  + 0.0417 \; VB70 - 0.0007 \;  ({\rm std. dev.= 0.0055}) \nonumber  \\
  B80  &=& B70  - 0.0494 \; VB70 - 0.0003 \;  ({\rm std. dev.= 0.0067}) \nonumber  \\
  L80  &=& L70  - 0.0569 \; BL70 - 0.0007 \;  ({\rm std. dev.= 0.0075}) \nonumber \\
  U80  &=& U70  - 0.0151 \; BU70 - 0.0007 \;  ({\rm std. dev.= 0.0077}) \nonumber  \\
  W80  &=& W70  - 0.2085 \; UW70 + 0.0007 \;  ({\rm std. dev.= 0.0128}) \nonumber \\
\end{eqnarray}
which are based on a set of about 1000 stars measured at both sites.
The quality flag is not given by \citet{vanGenderen1983} and an error of 0.015 has been
adopted in VBL, 0.02 in U, and 0.025 in W.
Points marked by a ':' in that paper were excluded.

The procedure in MoD is to use photometric zero points that are determined independently based on a model of Vega using
the respective filter curves
(see \citealt{Gr_MOD} for details, and the link to the latest available version of MoD given before).
The calibration constants derived in this way are
$-11.184, -10.923,  -10.831,  -10.808$, and $-10.684$ (in units of $\log$ ergs/cm$^2$/s/\AA) in VBLUW, respectively,
and that differ on average by 0.008 from the empirically determined values of
$-11.176, -10.914, -10.818, -10.800$, and $-10.681$ (J.~Lub's unpublished determination in 2019), that supersede the values of
$-11.172, -10.910, -10.818, -10.793$, and $-10.673$ as published in \citet{deRuiter86}.
Although not used in this paper the updated conversions to Johnson $V_{\rm J}$ and $(B-V)_{\rm J}$ are
\begin{eqnarray}
V_{\rm J}     & = & 6.8819 -2.5 \cdot (V80 + 0.0280 \; (V80-B80))  \nonumber \\
  & & \hspace{42mm} ({\rm std. dev.= 0.017}), \nonumber \\
(B-V)_{\rm J} & = & 2.528 \cdot (B80-V80) -0.817 \cdot (B80-V80)^2 + \nonumber  \\
             &   &                        0.336 \cdot (B80-V80)^3 -0.0133 \; ({\rm std. dev.= 0.016}). \nonumber  \\
\end{eqnarray}

\begin{figure}
  \centering
  
\begin{minipage}{0.40\textwidth}
\resizebox{\hsize}{!}{\includegraphics[angle=-0]{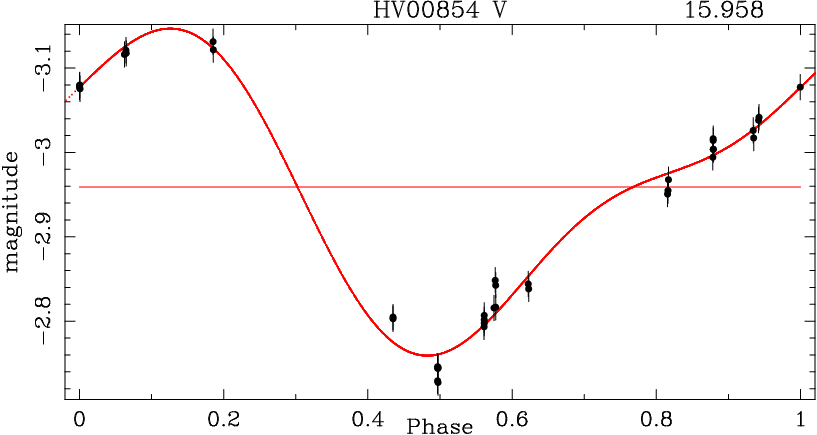}} 
\end{minipage}
\begin{minipage}{0.40\textwidth}
\resizebox{\hsize}{!}{\includegraphics[angle=-0]{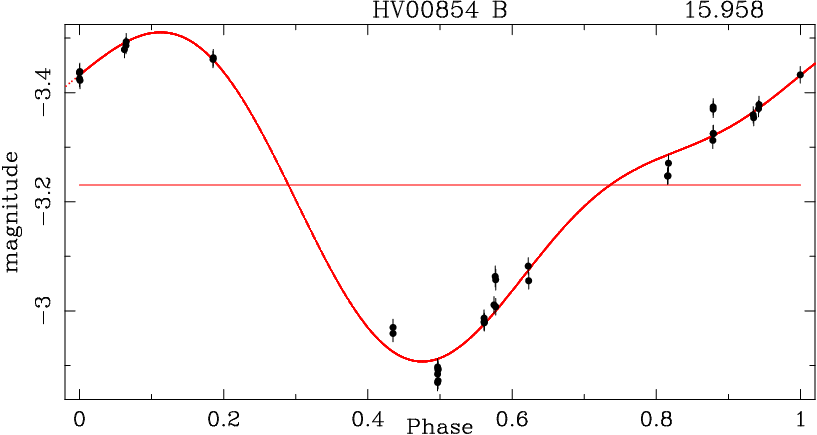}} 
\end{minipage}
 
\begin{minipage}{0.40\textwidth}
\resizebox{\hsize}{!}{\includegraphics[angle=-0]{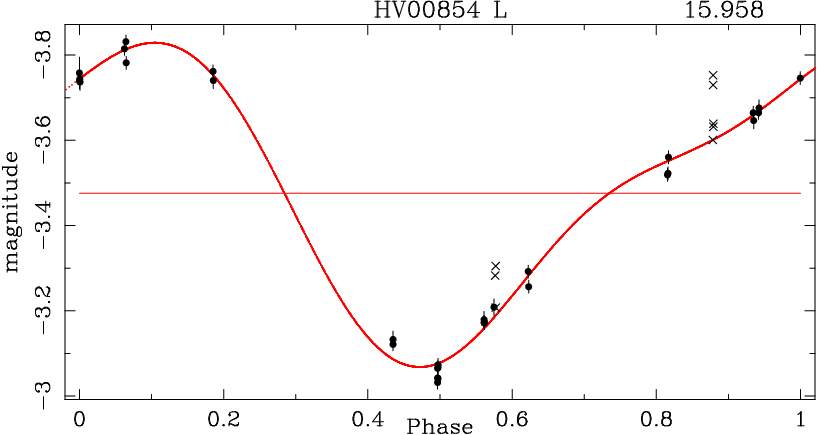}} 
\end{minipage}
\begin{minipage}{0.40\textwidth}
\resizebox{\hsize}{!}{\includegraphics[angle=-0]{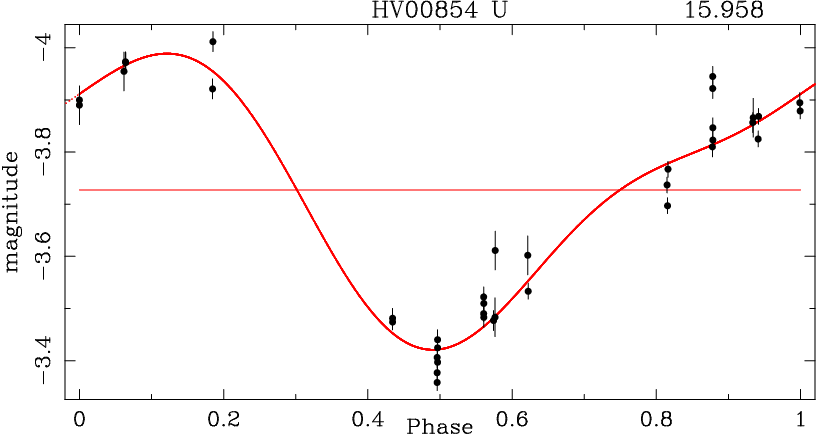}} 
\end{minipage}

\begin{minipage}{0.40\textwidth}
\resizebox{\hsize}{!}{\includegraphics[angle=-0]{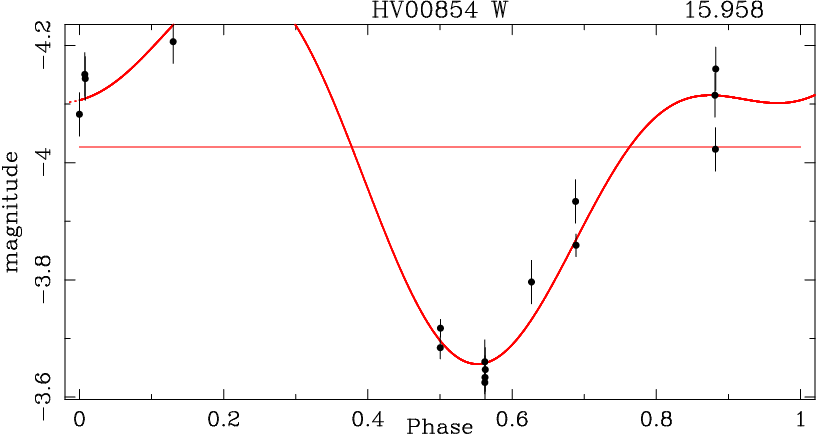}} 
\end{minipage}

\caption{Example of the fit to the light curves for HV 854.
  The name filter is listed at the top, as well as the period to the right at the top.
  The red lines indicate the model fit and the mean value.
  Crosses indicate points that were flagged as outliers and excluded from the fit.
}
\label{Fig-LC}
\end{figure}

The data were fitted with the code described and used in \citet{Groenewegen04,Gr20VMC,Gr22WISE} tailored to the Walraven data.
The light curves are analysed using a fixed period, fitting for the mean and the amplitude.
Depending on the number of available  data points the first harmonic period was added in the fit, solving for its amplitude as well.
The mean and total amplitude are reported in Table~\ref{Tab:WalPhotMean}.
To the error in the mean a value of 0.015 is added in quadrature.
The first entries (HV 824 - 5655) are the stars with new observations,
the latter part (HDE 270100 - HV 12815) are the stars from \citet{vanGenderen1983} (and \citealt{vanGenderen1989} for HDE 270100)
with any new data added in the analysis.
An example of the fit to the light curves is shown in Fig.~\ref{Fig-LC}.

\begin{table*}[h]
  \centering

\caption{Walraven photometry (first entries).}
\begin{tabular}{cccccccc} \hline \hline 
  HV   & HJD & $V$     & $V-B$  & $B-U$ & $U-W$ & $B-L$ &  quality flag  \\
\hline 
824 & 44888.71073 & -2.1162 & 0.3236 & 0.5142 & 0.3871 & 0.2958 & 23567 \\
824 & 44888.71293 & -2.1132 & 0.3251 & 0.5184 & 0.3475 & 0.2882 & 31455 \\
824 & 44888.72042 & -2.1097 & 0.3235 & 0.5253 & 0.4162 & 0.2944 & 22327 \\
824 & 44888.74559 & -2.1149 & 0.3195 & 0.4963 & 0.2780 & 0.2896 & 23465 \\
824 & 44889.66239 & -2.1098 & 0.3336 & 0.5410 & 0.2926 & 0.2862 & 23446 \\
824 & 44889.66896 & -2.1111 & 0.3293 & 0.5147 & 0.3777 & 0.3058 & 23557 \\
824 & 44923.60442 & -2.3975 & 0.5251 & 0.5336 & 0.3109 & 0.4165 & 35567 \\
824 & 44923.60992 & -2.3974 & 0.5199 & 0.5632 & 0.3065 & 0.4213 & 45558 \\
829 & 44889.55646 & -2.0115 & 0.3132 & 0.5072 & 0.4083 & 0.2822 & 24455 \\
829 & 44889.56281 & -2.0095 & 0.3140 & 0.5197 & 0.3526 & 0.2860 & 22456 \\
829 & 44889.58245 & -2.0132 & 0.3208 & 0.5008 & 0.3641 & 0.2753 & 23546 \\
829 & 44889.59266 & -1.9977 & 0.3145 & 0.4774 & 0.3383 & 0.2802 & 23455 \\
829 & 44923.62044 & -1.9865 & 0.3854 & 0.5519 & 0.4616 & 0.3439 & 23357 \\
829 & 44923.62577 & -1.9864 & 0.3862 & 0.5640 & 0.4396 & 0.3506 & 22446 \\
843 & 44130.82462 & -3.1981 & 0.3675 & 0.7497 & -0.2837 & 0.3028 & 55777 \\
843 & 44171.58065 & -3.0746 & 0.2458 & 0.5169 & 0.3916 & 0.2061 & 34578 \\
843 & 44171.59016 & -3.0740 & 0.2341 & 0.5149 & 0.4214 & 0.2391 & 33667 \\
843 & 44936.54307 & -2.3005 & 0.2043 & 0.3194 & 0.1822 & 0.2102 & 22344 \\
843 & 44936.65037 & -3.0870 & 0.2432 & 0.5052 & 0.2616 & 0.2453 & 34577 \\
843 & 44936.65920 & -3.0772 & 0.2414 & 0.5205 & 0.3958 & 0.2399 & 43467 \\
847 & 44187.59606 & -2.7581 & 0.3987 & 0.4901 & 0.3969 & 0.3502 & 34557 \\
847 & 44187.60540 & -2.7651 & 0.3930 & 0.5186 & 0.3545 & 0.3331 & 34666 \\
847 & 44189.55048 & -2.7952 & 0.4225 & 0.4982 & 0.3248 & 0.3698 & 33668 \\
847 & 44189.56003 & -2.7988 & 0.4291 & 0.5096 & 0.2621 & 0.3851 & 23569 \\
847 & 44191.54832 & -2.8459 & 0.4411 & 0.4812 & 0.3590 & 0.4065 & 33669 \\
847 & 44191.55783 & -2.8389 & 0.4527 & 0.5121 & 0.4444 & 0.3958 & 33679 \\
847 & 44201.53999 & -2.9831 & 0.4450 & 0.4870 & 0.2504 & 0.3559 & 24558 \\
847 & 44201.54938 & -2.9890 & 0.4314 & 0.5376 & 0.1662 & 0.3896 & 44779 \\
847 & 44545.56849 & -2.8976 & 0.4821 & 0.6089 & 0.1574 & 0.3579 & 25578 \\
847 & 44545.57784 & -2.9020 & 0.4738 & 0.5385 & 0.3119 & 0.4230 & 35688 \\
847 & 44547.51972 & -1.7594 & 0.5009 & 0.5153 & 0.3512 & 0.4070 & 35678 \\
847 & 44547.52874 & -2.9328 & 0.5034 & 0.5868 & 0.4855 & 0.3927 & 35579 \\
847 & 44550.52146 & -2.9967 & 0.5188 & 0.5078 & 0.4073 & 0.4608 & 45679 \\
847 & 44550.53026 & -2.9921 & 0.4897 & 0.5453 & 0.3483 & 0.4340 & 35668 \\
847 & 44554.52225 & -3.0104 & 0.4771 & 0.5549 & 0.0930 & 0.3693 & 45678 \\
847 & 44554.53139 & -3.0000 & 0.4389 & 0.4742 & 0.2362 & 0.3245 & 24657 \\
847 & 44557.52307 & -2.8514 & 0.3373 & 0.4617 & 0.2849 & 0.2399 & 35556 \\
847 & 44557.53159 & -2.8423 & 0.3261 & 0.3934 & 0.3693 & 0.2416 & 13647 \\
847 & 45019.53295 & -2.6549 & 0.2285 & 0.5019 & 0.2985 & 0.1970 & 44469 \\
847 & 45019.53546 & -2.6527 & 0.2372 & 0.5040 & 0.4039 & 0.2373 & 15569 \\
847 & 45019.54418 & -2.6619 & 0.2339 & 0.5134 & 0.3240 & 0.1964 & 13687 \\
847 & 45019.54615 & -2.6574 & 0.2334 & 0.5154 & 0.3289 & 0.2223 & 32277 \\
847 & 45019.54863 & -2.6756 & 0.2423 & 0.5389 & 0.4643 & 0.2087 & 35689 \\
847 & 45314.58503 & -2.3392 & 0.8868 & 0.6791 & -0.0731 & 0.4027 & 24799 \\
847 & 45314.59289 & -2.3392 & 0.8412 & 0.5456 & 0.1085 & 0.4604 & 24657 \\
847 & 45314.60175 & -2.3432 & 0.8718 & 0.5697 & 0.1327 & 0.5204 & 24656 \\
847 & 45316.61214 & -2.6810 & 0.7457 & 0.6714 & 0.2676 & 0.5462 & 25678 \\
847 & 45316.62416 & -2.6831 & 0.7617 & 0.6834 & 0.1953 & 0.5135 & 44677 \\
847 & 45669.59889 & -2.6415 & 0.2320 & 0.4901 & 0.2764 & 0.2263 & 23557 \\
847 & 45669.59893 & -2.6476 & 0.2265 & 0.4889 & 0.3010 & 0.2239 & 43556 \\
847 & 45669.60066 & -2.6488 & 0.2362 & 0.4874 & 0.1476 & 0.2240 & 32668 \\

\hline
\end{tabular} 
\tablefoot{
  Column~1: HV number.
  Column~2: HJD ($-2400000.0$).
  Column~3: Walraven $V$.
  Column~4: Walraven $V-B$.
  Column~5: Walraven $B-U$.
  Column~6: Walraven $U-W$.
  Column~7: Walraven $B-L$.
  Column~8: quality flag in the VBLUW bands, respectively.
This flag indicates the internal dispersion in the photometer signal in each channel, and
0, 1, 2, ..., 9 implies, $<$1, 1-2, 2-4, ..., 256-512 promille variation, respectively.
}
\label{Tab:WalPhot}
\end{table*}

\begin{sidewaystable*}
  \scriptsize
\setlength{\tabcolsep}{1.5mm}
  \caption{\label{Tab:WalPhotMean} Results of the light curve fitting. Mean VBLUW photometry (first entries). }  
\begin{tabular}{crrccccccccrrrrlllll}
\hline  \hline
  HV   &  $V$    & Amp$_{\rm V}$ & $N$ &  $B$    & Amp$_{\rm B}$ & $N$ & $L$     & Amp$_{\rm L}$ & $N$ & $U$   & Amp$_{\rm U}$ & $N$ & $W$  & Amp$_{\rm W}$ & $N$  \\
\hline
%
%
 824 & -2.253 $\pm$ 0.015 &  0.145 $\pm$ 0.021 & 8 & -2.685 $\pm$ 0.015 &  0.253 $\pm$ 0.030 & 8 & -3.044 $\pm$ 0.016 &  0.329 $\pm$ 0.063 & 8 & -3.230 $\pm$ 0.019 &  0.333 $\pm$ 0.143 & 8 & &   & \\ 
 847 & -2.860 $\pm$ 0.015 &  0.224 $\pm$ 0.009 & 30 & -3.239 $\pm$ 0.025 &  0.293 $\pm$ 0.049 & 35 & -3.563 $\pm$ 0.025 &  0.366 $\pm$ 0.052 & 32 & -3.735 $\pm$ 0.021 &  0.290 $\pm$ 0.041 & 35 & -4.140 $\pm$ 0.054 &  0.399 $\pm$ 0.078 & 15  \\ 
 854 & -2.959 $\pm$ 0.016 &  0.175 $\pm$ 0.010 & 40 & -3.231 $\pm$ 0.016 &  0.277 $\pm$ 0.015 & 40 & -3.476 $\pm$ 0.016 &  0.351 $\pm$ 0.014 & 37 & -3.727 $\pm$ 0.019 &  0.263 $\pm$ 0.028 & 38 & -4.027 $\pm$ 0.031 &  0.286 $\pm$ 0.058 & 16  \\ 
 872 & -2.585 $\pm$ 0.024 &  0.380 $\pm$ 0.020 & 49 & -2.924 $\pm$ 0.029 &  0.669 $\pm$ 0.027 & 49 & &   && &   && &   & \\ 
 876 & -2.667 $\pm$ 0.018 &  0.144 $\pm$ 0.018 & 37 & -3.107 $\pm$ 0.018 &  0.368 $\pm$ 0.023 & 25 & -3.518 $\pm$ 0.025 &  0.532 $\pm$ 0.041 & 24 & -3.620 $\pm$ 0.021 &  0.322 $\pm$ 0.026 & 30 & -3.979 $\pm$ 0.035 &  0.379 $\pm$ 0.060 & 30  \\ 
 880 & -2.849 $\pm$ 0.020 &  0.276 $\pm$ 0.018 & 48 & -3.063 $\pm$ 0.025 &  0.420 $\pm$ 0.026 & 48 & -3.296 $\pm$ 0.030 &  0.518 $\pm$ 0.034 & 47 & -3.554 $\pm$ 0.027 &  0.397 $\pm$ 0.028 & 45 & -3.866 $\pm$ 0.042 &  0.427 $\pm$ 0.048 & 34  \\ 
 899 & -2.736 $\pm$ 0.015 &  0.137 $\pm$ 0.007 & 6 & -3.262 $\pm$ 0.016 &  0.194 $\pm$ 0.009 & 6 & -3.761 $\pm$ 0.021 &  0.228 $\pm$ 0.028 & 6 & -3.880 $\pm$ 0.023 &  0.165 $\pm$ 0.037 & 6 & &    & \\ 
 955 & -2.883 $\pm$ 0.017 &  0.156 $\pm$ 0.022 & 32 & -3.253 $\pm$ 0.018 &  0.312 $\pm$ 0.025 & 30 & -3.576 $\pm$ 0.025 &  0.394 $\pm$ 0.045 & 22 & -3.716 $\pm$ 0.034 &  0.474 $\pm$ 0.070 & 21 & -4.103 $\pm$ 0.019 &  0.574 $\pm$ 0.036 & 12  \\ 
 969 & -3.108 $\pm$ 0.015 &  0.110 $\pm$ 0.006 & 76 & -3.466 $\pm$ 0.015 &  0.182 $\pm$ 0.010 & 76 & -3.819 $\pm$ 0.016 &  0.308 $\pm$ 0.016 & 73 & -3.980 $\pm$ 0.017 &  0.193 $\pm$ 0.021 & 69 & -4.144 $\pm$ 0.094 &  0.608 $\pm$ 0.102 & 24  \\ 
1013 & -2.790 $\pm$ 0.019 &  0.105 $\pm$ 0.013 & 19 & -3.276 $\pm$ 0.030 &  0.193 $\pm$ 0.029 & 19 & -3.735 $\pm$ 0.122 &  0.282 $\pm$ 0.138 & 18 & -3.838 $\pm$ 0.110 &  0.254 $\pm$ 0.125 & 16 & &   & \\ 
1345 & -3.133 $\pm$ 0.022 &  0.142 $\pm$ 0.022 & 45 & -3.445 $\pm$ 0.029 &  0.217 $\pm$ 0.034 & 45 & -3.753 $\pm$ 0.035 &  0.301 $\pm$ 0.043 & 43 & -3.945 $\pm$ 0.047 &  0.221 $\pm$ 0.055 & 33 & -4.207 $\pm$ 0.245 &  0.161 $\pm$ 0.276 & 15  \\ 
1374 & -3.393 $\pm$ 0.037 &  0.169 $\pm$ 0.051 & 13 & -3.695 $\pm$ 0.040 &  0.260 $\pm$ 0.058 & 12 & -4.015 $\pm$ 0.045 &  0.378 $\pm$ 0.066 & 13 & -4.141 $\pm$ 0.031 &  0.230 $\pm$ 0.037 & 10 & &    & \\ 
1610 & -2.960 $\pm$ 0.027 &  0.492 $\pm$ 0.025 & 13 & &   && &   && -3.671 $\pm$ 0.093 &  0.346 $\pm$ 0.098 & 13 & &   & \\ 
1618 & -3.393 $\pm$ 0.016 &  0.164 $\pm$ 0.012 & 51 & -3.622 $\pm$ 0.016 &  0.243 $\pm$ 0.017 & 51 & -3.819 $\pm$ 0.019 &  0.271 $\pm$ 0.025 & 47 & -4.034 $\pm$ 0.024 &  0.175 $\pm$ 0.045 & 21 & &   & \\ 
1705 & -3.337 $\pm$ 0.016 &  0.092 $\pm$ 0.014 & 13 & -3.680 $\pm$ 0.016 &  0.134 $\pm$ 0.015 & 12 & -4.026 $\pm$ 0.030 &  0.217 $\pm$ 0.065 & 12 & -4.104 $\pm$ 0.039 &  0.315 $\pm$ 0.043 & 7 & &    & \\ 
1744 & -3.077 $\pm$ 0.015 &  0.148 $\pm$ 0.010 & 80 & -3.347 $\pm$ 0.016 &  0.225 $\pm$ 0.013 & 80 & -3.590 $\pm$ 0.017 &  0.273 $\pm$ 0.021 & 79 & -3.814 $\pm$ 0.020 &  0.232 $\pm$ 0.033 & 49 & -4.120 $\pm$ 0.056 &  0.253 $\pm$ 0.076 & 17  \\ 
1768 & -3.265 $\pm$ 0.042 &  0.162 $\pm$ 0.047 & 16 & -3.543 $\pm$ 0.064 &  0.189 $\pm$ 0.074 & 16 & -3.809 $\pm$ 0.101 &  0.384 $\pm$ 0.118 & 16 & -4.032 $\pm$ 0.109 &  0.183 $\pm$ 0.130 & 15 & &    & \\ 
1884 & -2.975 $\pm$ 0.027 &  0.218 $\pm$ 0.029 & 7 & -3.302 $\pm$ 0.025 &  0.317 $\pm$ 0.026 & 7 & -3.597 $\pm$ 0.044 &  0.421 $\pm$ 0.053 & 7 & -3.796 $\pm$ 0.044 &  0.323 $\pm$ 0.052 & 6 & &    & \\ 
1967 & -2.633 $\pm$ 0.023 &  0.283 $\pm$ 0.040 & 22 & -3.009 $\pm$ 0.021 &  0.265 $\pm$ 0.038 & 21 & -3.331 $\pm$ 0.023 &  0.321 $\pm$ 0.046 & 22 & -3.531 $\pm$ 0.026 &  0.269 $\pm$ 0.057 & 22 & -3.845 $\pm$ 0.028 &  0.302 $\pm$ 0.065 & 8  \\ 
2063 & -3.227 $\pm$ 0.018 &  0.179 $\pm$ 0.013 & 40 & -3.540 $\pm$ 0.016 &  0.225 $\pm$ 0.009 & 32 & -3.902 $\pm$ 0.026 &  0.360 $\pm$ 0.027 & 21 & -3.688 $\pm$ 0.241 &  0.609 $\pm$ 0.250 & 15 & &   & \\ 
2205 & -2.909 $\pm$ 0.023 &  0.198 $\pm$ 0.033 & 28 & -3.328 $\pm$ 0.021 &  0.380 $\pm$ 0.031 & 24 & -3.747 $\pm$ 0.021 &  0.583 $\pm$ 0.026 & 26 & -3.891 $\pm$ 0.026 &  0.429 $\pm$ 0.033 & 28 & &   & \\ 
2209 & -2.679 $\pm$ 0.015 &  0.124 $\pm$ 0.006 & 29 & -2.958 $\pm$ 0.015 &  0.197 $\pm$ 0.011 & 29 & -3.217 $\pm$ 0.016 &  0.254 $\pm$ 0.015 & 29 & -3.451 $\pm$ 0.016 &  0.180 $\pm$ 0.016 & 27 & -3.788 $\pm$ 0.019 &  0.195 $\pm$ 0.032 & 18  \\ 
2249 & -2.825 $\pm$ 0.017 &  0.269 $\pm$ 0.023 & 58 & -3.153 $\pm$ 0.020 &  0.415 $\pm$ 0.038 & 58 & -3.471 $\pm$ 0.021 &  0.493 $\pm$ 0.045 & 55 & -3.715 $\pm$ 0.021 &  0.350 $\pm$ 0.040 & 52 & -4.001 $\pm$ 0.028 &  0.370 $\pm$ 0.051 & 35  \\ 
2260 & -3.211 $\pm$ 0.015 &  0.163 $\pm$ 0.009 & 52 & -3.623 $\pm$ 0.016 &  0.253 $\pm$ 0.016 & 52 & -4.026 $\pm$ 0.018 &  0.384 $\pm$ 0.024 & 44 & -4.166 $\pm$ 0.019 &  0.302 $\pm$ 0.031 & 38 & -4.194 $\pm$ 0.032 &  0.635 $\pm$ 0.031 & 6  \\ 
2299 & -2.932 $\pm$ 0.017 &  0.161 $\pm$ 0.011 & 92 & -3.267 $\pm$ 0.020 &  0.264 $\pm$ 0.018 & 93 & -3.578 $\pm$ 0.025 &  0.335 $\pm$ 0.025 & 92 & -3.774 $\pm$ 0.025 &  0.263 $\pm$ 0.025 & 91 & -3.983 $\pm$ 0.041 &  0.163 $\pm$ 0.048 & 50  \\ 
2454 & -3.014 $\pm$ 0.018 &  0.152 $\pm$ 0.026 & 56 & -3.533 $\pm$ 0.020 &  0.273 $\pm$ 0.040 & 56 & -4.020 $\pm$ 0.021 &  0.474 $\pm$ 0.059 & 51 & -4.182 $\pm$ 0.027 &  0.416 $\pm$ 0.087 & 48 & -4.363 $\pm$ 0.032 &  0.307 $\pm$ 0.053 & 24  \\ 
2680 & -3.163 $\pm$ 0.016 &  0.133 $\pm$ 0.016 & 39 & -3.505 $\pm$ 0.016 &  0.218 $\pm$ 0.021 & 39 & -3.868 $\pm$ 0.021 &  0.338 $\pm$ 0.035 & 37 & -3.998 $\pm$ 0.020 &  0.245 $\pm$ 0.048 & 33 & &   & \\ 
2686 & -3.422 $\pm$ 0.017 &  0.140 $\pm$ 0.015 & 25 & -3.751 $\pm$ 0.018 &  0.233 $\pm$ 0.024 & 22 & -4.202 $\pm$ 0.048 &  0.450 $\pm$ 0.058 & 22 & -4.230 $\pm$ 0.039 &  0.114 $\pm$ 0.088 & 19 & &    & \\ 
5655 & -3.090 $\pm$ 0.020 &  0.169 $\pm$ 0.045 & 46 & -3.532 $\pm$ 0.017 &  0.345 $\pm$ 0.028 & 42 & -3.930 $\pm$ 0.022 &  0.535 $\pm$ 0.074 & 41 & -4.139 $\pm$ 0.021 &  0.415 $\pm$ 0.060 & 37 & &   & \\ 
\\
270100 & -1.951 $\pm$ 0.015 & 0.091 $\pm$ 0.003 & 47 & -2.426 $\pm$ 0.015 & 0.170 $\pm$ 0.007 & 42 & -2.852 $\pm$ 0.015 &  0.251 $\pm$ 0.010 & 37 & -2.977 $\pm$ 0.015 &  0.186 $\pm$ 0.010 & 30 & -3.491 $\pm$ 0.018 &  0.220 $\pm$ 0.027 & 26  \\ 
 817 & -2.802 $\pm$ 0.017 &  0.164 $\pm$ 0.019 & 34 & -3.080 $\pm$ 0.017 &  0.251 $\pm$ 0.022 & 34 & -3.322 $\pm$ 0.018 &  0.292 $\pm$ 0.031 & 29 & -3.440 $\pm$ 0.051 &  0.158 $\pm$ 0.098 & 22 & &    &  \\ 
 821 & -2.041 $\pm$ 0.017 &  0.131 $\pm$ 0.020 & 34 & -2.514 $\pm$ 0.017 &  0.259 $\pm$ 0.020 & 29 & -2.945 $\pm$ 0.022 &  0.341 $\pm$ 0.040 & 23 & -3.079 $\pm$ 0.018 &  0.285 $\pm$ 0.030 & 25 & &    &  \\ 
 822 & -3.110 $\pm$ 0.020 &  0.228 $\pm$ 0.041 & 15 & -3.469 $\pm$ 0.018 &  0.289 $\pm$ 0.029 & 14 & -3.785 $\pm$ 0.236 &  0.481 $\pm$ 0.451 & 9 & -3.848 $\pm$ 0.034 &  0.346 $\pm$ 0.088 & 10 & &    &  \\ 
 823 & -2.755 $\pm$ 0.017 &  0.169 $\pm$ 0.022 & 21 & -3.176 $\pm$ 0.019 &  0.277 $\pm$ 0.031 & 20 & -3.455 $\pm$ 0.022 &  0.307 $\pm$ 0.044 & 18 & -3.637 $\pm$ 0.042 &  0.261 $\pm$ 0.094 & 13 & &    &  \\ 
 827 & -3.046 $\pm$ 0.017 &  0.139 $\pm$ 0.020 & 33 & -3.308 $\pm$ 0.016 &  0.220 $\pm$ 0.016 & 32 & -3.532 $\pm$ 0.020 &  0.236 $\pm$ 0.036 & 26 & -3.747 $\pm$ 0.021 &  0.210 $\pm$ 0.044 & 22 & &    &  \\ 
 829 & -2.029 $\pm$ 0.015 &  0.130 $\pm$ 0.011 & 33 & -2.406 $\pm$ 0.016 &  0.215 $\pm$ 0.012 & 31 & -2.734 $\pm$ 0.018 &  0.259 $\pm$ 0.029 & 28 & -2.909 $\pm$ 0.019 &  0.235 $\pm$ 0.035 & 26 & &    &  \\ 
 834 & -2.138 $\pm$ 0.016 &  0.111 $\pm$ 0.015 & 43 & -2.529 $\pm$ 0.017 &  0.217 $\pm$ 0.024 & 41 & -2.864 $\pm$ 0.017 &  0.319 $\pm$ 0.025 & 33 & -3.024 $\pm$ 0.018 &  0.248 $\pm$ 0.030 & 37 & &    &  \\ 
 837 & -2.568 $\pm$ 0.016 &  0.171 $\pm$ 0.013 & 36 & -2.979 $\pm$ 0.016 &  0.271 $\pm$ 0.020 & 37 & -3.315 $\pm$ 0.017 &  0.323 $\pm$ 0.024 & 25 & -3.468 $\pm$ 0.019 &  0.284 $\pm$ 0.032 & 23 & &    &  \\ 
 877 & -2.606 $\pm$ 0.016 &  0.127 $\pm$ 0.016 & 22 & -3.183 $\pm$ 0.016 &  0.218 $\pm$ 0.017 & 22 & -3.640 $\pm$ 0.021 &  0.256 $\pm$ 0.041 & 20 & -3.720 $\pm$ 0.021 &  0.281 $\pm$ 0.041 & 11 & &    &  \\ 
 883 & -2.130 $\pm$ 0.016 &  0.206 $\pm$ 0.015 & 28 & -2.735 $\pm$ 0.017 &  0.337 $\pm$ 0.019 & 28 & -3.188 $\pm$ 0.021 &  0.374 $\pm$ 0.044 & 18 & -3.282 $\pm$ 0.018 &  0.343 $\pm$ 0.028 & 21 & &    &  \\ 
 886 & -2.607 $\pm$ 0.018 &  0.219 $\pm$ 0.028 & 20 & -3.014 $\pm$ 0.018 &  0.396 $\pm$ 0.028 & 17 & -3.166 $\pm$ 0.025 &  0.566 $\pm$ 0.039 & 8 & -3.566 $\pm$ 0.023 &  0.458 $\pm$ 0.049 & 13 & &    &  \\ 
 900 & -2.391 $\pm$ 0.017 &  0.183 $\pm$ 0.024 & 22 & -2.855 $\pm$ 0.020 &  0.306 $\pm$ 0.037 & 21 & -3.246 $\pm$ 0.022 &  0.413 $\pm$ 0.048 & 15 & -3.399 $\pm$ 0.022 &  0.380 $\pm$ 0.045 & 17 & &    &  \\ 
 902 & -2.596 $\pm$ 0.017 &  0.189 $\pm$ 0.022 & 21 & -2.990 $\pm$ 0.021 &  0.317 $\pm$ 0.039 & 19 & -3.425 $\pm$ 0.019 &  0.490 $\pm$ 0.035 & 14 & -3.506 $\pm$ 0.043 &  0.381 $\pm$ 0.096 & 13 & &    &  \\ 
 909 & -2.369 $\pm$ 0.017 &  0.207 $\pm$ 0.023 & 29 & -2.767 $\pm$ 0.018 &  0.357 $\pm$ 0.025 & 26 & -3.138 $\pm$ 0.019 &  0.469 $\pm$ 0.030 & 22 & -3.303 $\pm$ 0.021 &  0.405 $\pm$ 0.038 & 26 & &    &  \\ 
 953 & -2.187 $\pm$ 0.017 &  0.178 $\pm$ 0.022 & 22 & -2.617 $\pm$ 0.018 &  0.301 $\pm$ 0.030 & 22 & -2.986 $\pm$ 0.018 &  0.369 $\pm$ 0.026 & 17 & -3.155 $\pm$ 0.032 &  0.347 $\pm$ 0.074 & 6 & &    &  \\ 
1002 & -2.438 $\pm$ 0.018 &  0.201 $\pm$ 0.031 & 30 & -2.855 $\pm$ 0.019 &  0.332 $\pm$ 0.032 & 24 & -3.195 $\pm$ 0.019 &  0.496 $\pm$ 0.041 & 15 & -3.392 $\pm$ 0.021 &  0.377 $\pm$ 0.043 & 19 & &    &  \\ 
1003 & -2.569 $\pm$ 0.018 &  0.209 $\pm$ 0.028 & 25 & -2.932 $\pm$ 0.018 &  0.331 $\pm$ 0.028 & 22 & -3.274 $\pm$ 0.019 &  0.424 $\pm$ 0.031 & 18 & -3.415 $\pm$ 0.021 &  0.316 $\pm$ 0.040 & 21 & &    &  \\ 
1365 & -3.291 $\pm$ 0.017 &  0.138 $\pm$ 0.019 & 42 & -3.584 $\pm$ 0.017 &  0.212 $\pm$ 0.020 & 39 & -3.808 $\pm$ 0.018 &  0.214 $\pm$ 0.028 & 31 & -4.022 $\pm$ 0.025 &  0.164 $\pm$ 0.060 & 23 & &    &  \\ 
\hline
%
\end{tabular}
\tablefoot{
  Column~1: HV number; 270100 refers to HDE 270100 \citep{vanGenderen1989}.
  Column~2-11: Mean values and amplitudes with error bars and the number of data points in the VBLUW filters.
}
\end{sidewaystable*}

\section{Mass estimates}
\label{MApp}

Table~\ref{Tab:AppMass} compiles the mass estimates using the different methods outlined in the main text (also see the table footnote).
The adopted mass is the median among the five estimates.
To estimate the error bar the error in the mass estimate of the median value is added in quadrature to
the median-absolute-deviation times 1.48 (to get the equivalent of one sigma in a Gaussian distribution)
among the five estimates.

\begin{table*}[h]
  \centering

\caption{Stellar mass estimates (selected entries only).}
\begin{tabular}{crrrrrrrrr} \hline \hline 
  Identifier  &  Period & $M_1$ & $M_2$ & $M_3$ & $M_4$ & $M_5$ &  $M_{\rm adopted}$  \\
              & (d)    &  (\msol)  & (\msol)  & (\msol)  & (\msol)  & (\msol)  & (\msol) \\
\hline 
LMC3004 &   1.52 &   2.50 $\pm$   0.01 &   2.81 $\pm$   0.01 &   2.94 $\pm$   0.02 &   2.84 $\pm$   0.08 &   2.17 $\pm$   0.01 &   2.81 $\pm$   0.13 \\ 
LMC1523 &   1.57 &   2.64 $\pm$   0.01 &   2.81 $\pm$   0.01 &   3.03 $\pm$   0.02 &   2.91 $\pm$   0.09 &   2.45 $\pm$   0.01 &   2.81 $\pm$   0.17 \\ 
LMC3724 &   2.64 &   3.41 $\pm$   0.01 &   3.91 $\pm$   0.01 &   3.96 $\pm$   0.03 &   3.63 $\pm$   0.14 &   3.21 $\pm$   0.01 &   3.63 $\pm$   0.34 \\ 
LMC3750 &   2.95 &   3.37 $\pm$   0.01 &   3.81 $\pm$   0.01 &   3.86 $\pm$   0.04 &   3.55 $\pm$   0.13 &   3.32 $\pm$   0.01 &   3.55 $\pm$   0.30 \\ 
LMC2138 &   3.01 &   3.48 $\pm$   0.01 &   3.82 $\pm$   0.01 &   3.94 $\pm$   0.03 &   3.61 $\pm$   0.13 &   3.25 $\pm$   0.01 &   3.61 $\pm$   0.29 \\ 
LMC3744 &   3.05 &   3.86 $\pm$   0.01 &   4.41 $\pm$   0.01 &   4.46 $\pm$   0.01 &   4.01 $\pm$   0.17 &   2.08 $\pm$   0.01 &   4.01 $\pm$   0.47 \\ 
LMC4646 &   3.10 &   3.82 $\pm$   0.01 &   4.47 $\pm$   0.01 &   4.45 $\pm$   0.04 &   4.00 $\pm$   0.17 &   3.68 $\pm$   0.01 &   4.00 $\pm$   0.41 \\ 
LMC3723 &   3.14 &   3.58 $\pm$   0.01 &   4.16 $\pm$   0.01 &   4.15 $\pm$   0.04 &   3.77 $\pm$   0.15 &   3.33 $\pm$   0.01 &   3.77 $\pm$   0.44 \\ 
LMC3756 &   3.21 &   3.47 $\pm$   0.01 &   3.78 $\pm$   0.01 &   3.90 $\pm$   0.04 &   3.57 $\pm$   0.13 &   3.46 $\pm$   0.01 &   3.57 $\pm$   0.23 \\ 
LMC3752 &   3.44 &   3.59 $\pm$   0.01 &   3.79 $\pm$   0.01 &   3.96 $\pm$   0.04 &   3.62 $\pm$   0.14 &   3.63 $\pm$   0.01 &   3.63 $\pm$   0.07 \\ 
LMC3726 &   3.52 &   3.59 $\pm$   0.01 &   4.11 $\pm$   0.01 &   4.10 $\pm$   0.04 &   3.73 $\pm$   0.14 &   3.50 $\pm$   0.01 &   3.73 $\pm$   0.31 \\ 
LMC0961 &   3.71 &   4.01 $\pm$   0.01 &   4.49 $\pm$   0.01 &   4.55 $\pm$   0.03 &   4.06 $\pm$   0.17 &   3.26 $\pm$   0.01 &   4.06 $\pm$   0.50 \\ 
LMC1124 &   4.46 &   3.88 $\pm$   0.01 &   4.35 $\pm$   0.01 &   4.34 $\pm$   0.05 &   3.90 $\pm$   0.16 &   3.89 $\pm$   0.01 &   3.90 $\pm$   0.23 \\ 
LMC3320 &   4.79 &   4.35 $\pm$   0.01 &   4.75 $\pm$   0.01 &   4.81 $\pm$   0.06 &   4.25 $\pm$   0.19 &   4.27 $\pm$   0.01 &   4.35 $\pm$   0.10 \\ 
SMC3588 &   5.32 &   4.85 $\pm$   0.02 &   5.29 $\pm$   0.01 &   5.35 $\pm$   0.07 &   4.64 $\pm$   0.22 &   4.57 $\pm$   0.02 &   4.85 $\pm$   0.27 \\ 
SMC2254 &   5.65 &   6.00 $\pm$   0.02 &   6.64 $\pm$   0.02 &   6.69 $\pm$   0.10 &   5.61 $\pm$   0.33 &   5.58 $\pm$   0.03 &   6.00 $\pm$   0.42 \\ 
LMC1939 &   6.06 &   4.53 $\pm$   0.02 &   4.98 $\pm$   0.01 &   4.97 $\pm$   0.06 &   4.36 $\pm$   0.20 &   4.31 $\pm$   0.02 &   4.53 $\pm$   0.22 \\ 
LMC1466 &   6.15 &   4.27 $\pm$   0.01 &   4.80 $\pm$   0.01 &   4.72 $\pm$   0.06 &   4.17 $\pm$   0.18 &   4.21 $\pm$   0.02 &   4.27 $\pm$   0.11 \\ 
LMC0594 &   6.73 &   4.62 $\pm$   0.02 &   5.02 $\pm$   0.01 &   5.01 $\pm$   0.08 &   4.38 $\pm$   0.20 &   4.90 $\pm$   0.02 &   4.90 $\pm$   0.17 \\ 
LMC1526 &   6.73 &   4.47 $\pm$   0.02 &   4.71 $\pm$   0.01 &   4.77 $\pm$   0.07 &   4.21 $\pm$   0.18 &   4.67 $\pm$   0.01 &   4.67 $\pm$   0.15 \\ 
LMC1424 &   6.78 &   4.13 $\pm$   0.01 &   4.86 $\pm$   0.01 &   4.62 $\pm$   0.05 &   4.09 $\pm$   0.17 &   3.93 $\pm$   0.02 &   4.13 $\pm$   0.19 \\ 
LMC1941 &   6.81 &   5.04 $\pm$   0.02 &   6.25 $\pm$   0.01 &   5.82 $\pm$   0.07 &   4.97 $\pm$   0.26 &   4.67 $\pm$   0.03 &   5.04 $\pm$   0.37 \\ 
LMC1313 &   6.83 &   4.39 $\pm$   0.02 &   4.60 $\pm$   0.01 &   4.67 $\pm$   0.07 &   4.13 $\pm$   0.18 &   4.50 $\pm$   0.01 &   4.50 $\pm$   0.15 \\ 
LMC1128 &   6.86 &   4.53 $\pm$   0.02 &   4.76 $\pm$   0.01 &   4.83 $\pm$   0.07 &   4.25 $\pm$   0.19 &   4.60 $\pm$   0.01 &   4.60 $\pm$   0.19 \\ 
LMC1374 &   6.89 &   4.17 $\pm$   0.01 &   4.61 $\pm$   0.01 &   4.53 $\pm$   0.06 &   4.02 $\pm$   0.17 &   4.27 $\pm$   0.01 &   4.27 $\pm$   0.26 \\ 
LMC1327 &   6.92 &   4.37 $\pm$   0.01 &   4.70 $\pm$   0.01 &   4.70 $\pm$   0.07 &   4.15 $\pm$   0.18 &   4.55 $\pm$   0.01 &   4.55 $\pm$   0.18 \\ 
$\cdots$ & \\
LMC1113 &  37.56 &   8.57 $\pm$   0.04 &   8.83 $\pm$   0.02 &   8.39 $\pm$   0.33 &   6.63 $\pm$   0.45 &  10.14 $\pm$   0.05 &   8.57 $\pm$   0.27 \\ 
LMC3158 &  39.19 &   3.83 $\pm$   0.01 &   3.62 $\pm$   0.01 &   3.48 $\pm$   0.08 &   3.15 $\pm$   0.10 &   4.88 $\pm$   0.01 &   3.62 $\pm$   0.20 \\ 
SMC2706 &  39.20 &   7.81 $\pm$   0.04 &   8.76 $\pm$   0.02 &   7.90 $\pm$   0.24 &   6.30 $\pm$   0.41 &   8.64 $\pm$   0.05 &   7.90 $\pm$   0.75 \\ 
LMC0512 &  39.40 &   8.40 $\pm$   0.04 &   8.88 $\pm$   0.02 &   8.29 $\pm$   0.33 &   6.56 $\pm$   0.44 &  10.17 $\pm$   0.05 &   8.40 $\pm$   0.48 \\ 
SMC1797 &  41.21 &   7.70 $\pm$   0.04 &   8.87 $\pm$   0.02 &   7.85 $\pm$   0.22 &   6.26 $\pm$   0.41 &   8.31 $\pm$   0.05 &   7.85 $\pm$   0.46 \\ 
SMC4444 &  41.83 &   9.90 $\pm$   0.06 &  10.22 $\pm$   0.03 &   9.69 $\pm$   0.43 &   7.49 $\pm$   0.58 &  11.60 $\pm$   0.07 &   9.90 $\pm$   0.33 \\ 
LMC0943 &  42.24 &   9.64 $\pm$   0.06 &  10.24 $\pm$   0.03 &   9.54 $\pm$   0.41 &   7.39 $\pm$   0.56 &  11.24 $\pm$   0.07 &   9.64 $\pm$   0.60 \\ 
SMC2470 &  42.75 &   9.09 $\pm$   0.05 &   9.59 $\pm$   0.02 &   8.95 $\pm$   0.34 &   6.99 $\pm$   0.51 &  10.27 $\pm$   0.06 &   9.09 $\pm$   0.50 \\ 
LMC0461 &  45.17 &   7.77 $\pm$   0.04 &   8.86 $\pm$   0.02 &   7.85 $\pm$   0.24 &   6.25 $\pm$   0.40 &   8.55 $\pm$   0.05 &   7.85 $\pm$   0.70 \\ 
LMC0966 &  47.38 &   9.33 $\pm$   0.05 &  10.06 $\pm$   0.03 &   9.22 $\pm$   0.33 &   7.17 $\pm$   0.53 &  10.13 $\pm$   0.07 &   9.33 $\pm$   0.74 \\ 
LMC4663 &  47.96 &  10.96 $\pm$   0.07 &  11.24 $\pm$   0.03 &  10.65 $\pm$   0.50 &   8.10 $\pm$   0.68 &  12.51 $\pm$   0.08 &  10.96 $\pm$   0.33 \\ 
LMC1290 &  48.38 &  10.00 $\pm$   0.06 &  10.48 $\pm$   0.03 &   9.77 $\pm$   0.70 &   7.53 $\pm$   0.58 &  14.69 $\pm$   0.07 &  10.00 $\pm$   0.49 \\ 
LMC2253 &  52.37 &   9.24 $\pm$   0.05 &   9.85 $\pm$   0.03 &   9.03 $\pm$   0.36 &   7.03 $\pm$   0.51 &  10.62 $\pm$   0.06 &   9.24 $\pm$   0.61 \\ 
LMC0992 &  52.87 &  10.61 $\pm$   0.07 &  10.39 $\pm$   0.03 &  10.02 $\pm$   0.47 &   7.68 $\pm$   0.61 &  12.04 $\pm$   0.07 &  10.39 $\pm$   0.38 \\ 
SMC0921 &  65.94 &  10.12 $\pm$   0.06 &   9.91 $\pm$   0.03 &   9.42 $\pm$   0.62 &   7.27 $\pm$   0.55 &  13.83 $\pm$   0.06 &   9.91 $\pm$   0.50 \\ 
SMC2099 &  73.62 &  13.14 $\pm$   0.10 &  13.46 $\pm$   0.05 &  12.53 $\pm$   0.66 &   9.24 $\pm$   0.88 &  14.30 $\pm$   0.12 &  13.14 $\pm$   0.63 \\ 
LMC4691 &  73.90 &  12.24 $\pm$   0.09 &  12.45 $\pm$   0.04 &  11.60 $\pm$   0.61 &   8.66 $\pm$   0.77 &  13.72 $\pm$   0.10 &  12.24 $\pm$   0.66 \\ 
LMC4689 &  78.51 &  10.54 $\pm$   0.07 &  12.02 $\pm$   0.03 &  10.42 $\pm$   0.46 &   7.90 $\pm$   0.64 &  11.92 $\pm$   0.10 &  10.54 $\pm$   1.39 \\ 
SMC1502 &  84.30 &  12.27 $\pm$   0.09 &  11.90 $\pm$   0.04 &  11.29 $\pm$   0.76 &   8.45 $\pm$   0.74 &  15.35 $\pm$   0.09 &  11.90 $\pm$   0.62 \\ 
LMC4628 &  99.20 &  13.56 $\pm$   0.11 &  14.88 $\pm$   0.05 &  13.12 $\pm$   0.83 &   9.57 $\pm$   0.95 &  16.03 $\pm$   0.15 &  13.56 $\pm$   1.33 \\ 
LMC4629 & 108.70 &   6.77 $\pm$   0.03 &   6.86 $\pm$   0.01 &   6.12 $\pm$   0.33 &   5.01 $\pm$   0.26 &  10.05 $\pm$   0.03 &   (6.77 $\pm$   0.66) \\ 
LMC1591 & 118.62 &   9.50 $\pm$   0.05 &   9.84 $\pm$   0.02 &   8.74 $\pm$   0.39 &   6.77 $\pm$   0.47 &  11.02 $\pm$   0.06 &   (9.50 $\pm$   0.76) \\ 
SMC0417 & 128.20 &   6.71 $\pm$   0.03 &   6.02 $\pm$   0.01 &   5.68 $\pm$   0.62 &   4.69 $\pm$   0.23 &  13.83 $\pm$   0.02 &   (6.02 $\pm$   0.70) \\ 
LMC0619 & 133.78 &   8.74 $\pm$   0.05 &   9.18 $\pm$   0.02 &   8.00 $\pm$   0.21 &   6.28 $\pm$   0.41 &   8.03 $\pm$   0.06 &   (8.03 $\pm$   0.77) \\ 
SMC3611 & 208.80 &   9.60 $\pm$   0.05 &  10.39 $\pm$   0.02 &   8.71 $\pm$   0.33 &   6.71 $\pm$   0.46 &  10.14 $\pm$   0.07 &   (9.60 $\pm$   0.79) \\ 
\hline
\end{tabular} 
\tablefoot{
  Column~1: Identifier (sources are sorted by period).
  Column~2: Period.
  Column~3: Mass estimate from the period-luminosity-mass-effective temperature-metallicity relation derived in \citet{GrJu17b} based on the
  the models in \cite{Bono_2000}.
  Column~4: Mass estimate from Eq.~\ref{EqPCEP}, based on the  models of \citet{Anderson16}.
  Column~5: Mass estimate from the relation in \citet{Ragosta19}.
  Column~6: Mass estimate from the relation in \citet{Pilecki2018}.
  Column~7: Mass estimate from the relation in \citet{Marconi20}.
  Column~8: Adopted mass based on the median and the MAD (see text).
  Values between parenthesis are deemed unreliable and have not been used in the analysis of the mass-luminosity relation
  and in the calculation of the evolutionary $\log g$ and FWG values.
}
\label{Tab:AppMass}
\end{table*}

\section{2D Cepheid model}
\label{App:Vas}

\begin{figure}[b]
  \centering

\begin{minipage}{0.49\textwidth}
\resizebox{\hsize}{!}{\includegraphics{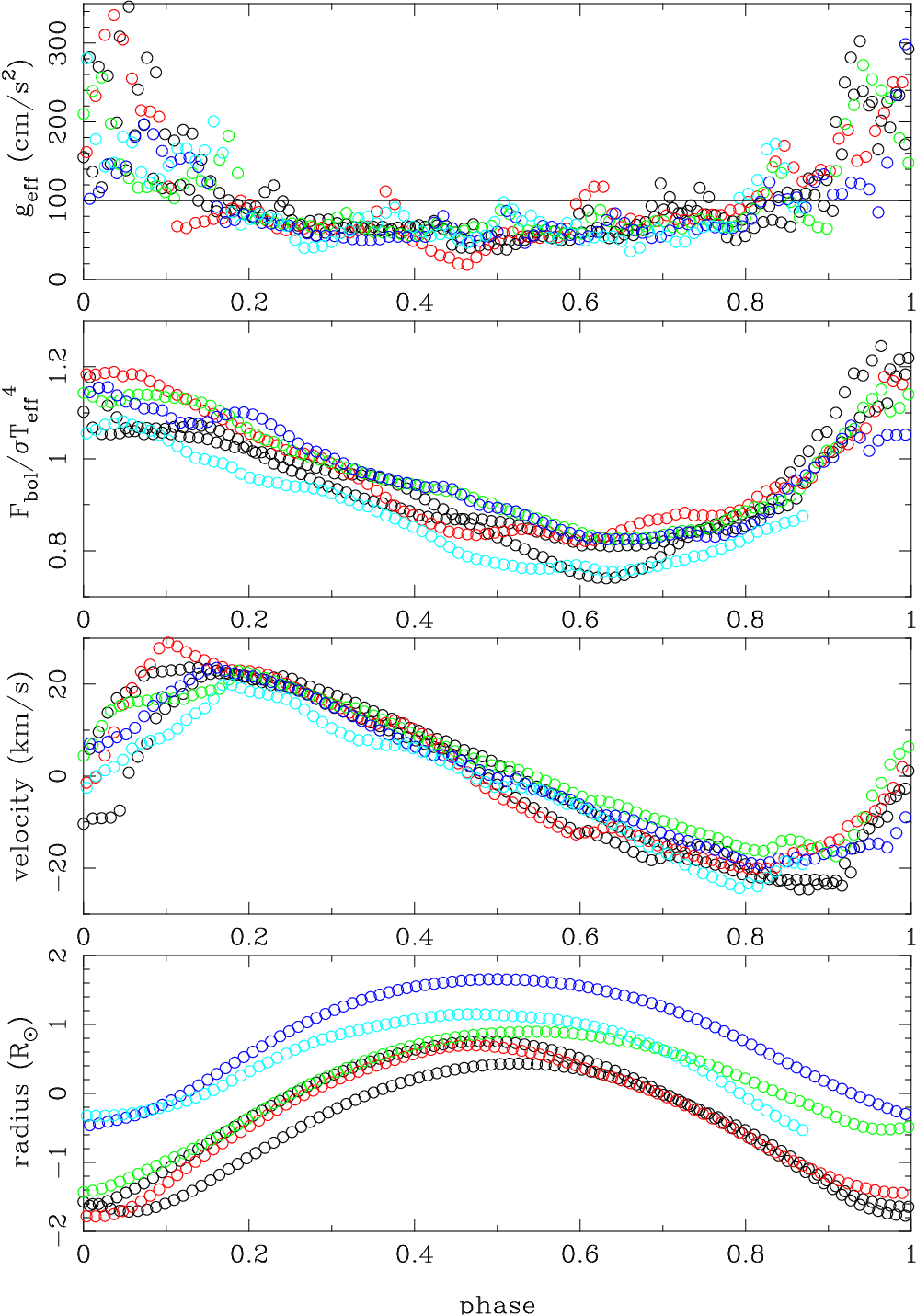}}
\end{minipage}

\caption{Various quantities plotted against phase based on the 2D models of \citet{Vasilyev17,Vasilyev18}.
  Different pulsation cycles are plotted in different colours.
From top to bottom, the effective gravity, normalised flux, velocity and change in radius. See text for details.
}
\label{Fig:Vas}
\end{figure}

\citet{Vasilyev17,Vasilyev18} present the results of a two-dimensional time-dependent envelope model of
a CC with $T_{\rm eff}$ = 5600~K and $\log g_0 = 2.0$~dex. In Figure~5 of \citet{Vasilyev17}, the term
$g_0 + \frac{\partial{v}}{\partial{t}}$ is plotted against time. The time series of various quantities were kindly made
available, and are plotted in a slightly different way in Fig.~\ref{Fig:Vas}.
A Fourier analysis of the velocity time series showed a periodicity of 2.6426 days (and a mean of $-7.86$ \ks), which is
used to phase the data. Phase zero is taken at the instance in time when the normalised flux reaches a maximum
for the first time. Consecutive pulsation cycles are plotted with different colours.
The cycles are not very smooth. This is explained by convection, which adds statistical fluctuations
to the velocity and thermal structure of their model (Sect.~2.1 in \citealt{Vasilyev17}).

The integral of the velocity curve is used to calculate the change in radius.
For a mass of 3 \msol,  $\log g = 2.0$~dex implies a radius of about 29~\rsol.
The top panel shows how $g_{\rm eff}$ changes over the pulsation cycle.
The effective gravity is below 100 for 68\% of the time, with an average value of 66~cm/s$^2$ ($-0.18$~dex),
and above 100 cm/s$^2$ 32\% of the time for an average of 166 ($+0.22$~dex).
The 5 and 95\% percentiles correspond to values of $\pm 0.35$~dex.
The effect of the change in radius is almost negligible (of order 5-10 cm/s$^2$, or $\pm 0.04$~dex at most) compared to the
derivative of the velocity in determining $g_{\rm eff}$.

\section{Additional figures}
\label{App:AF}

\begin{figure*}

\begin{minipage}{0.46\textwidth}
\resizebox{\hsize}{!}{\includegraphics{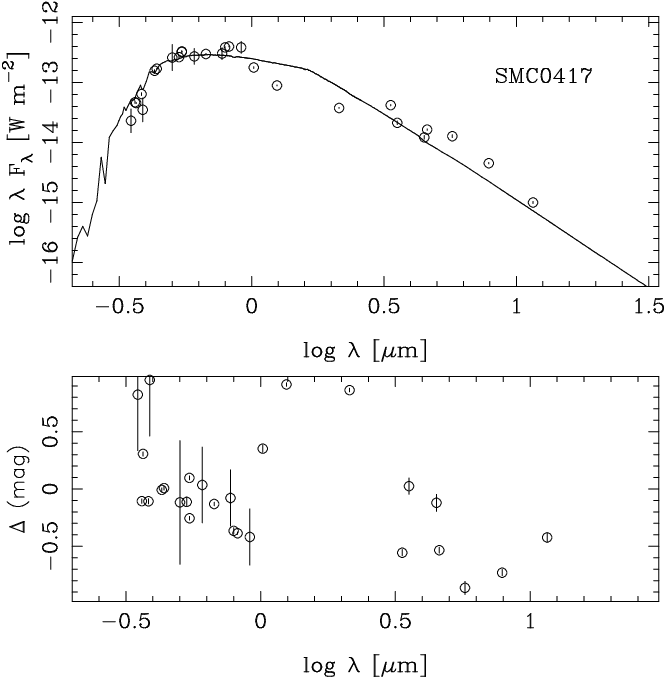}}
\end{minipage}
\begin{minipage}{0.46\textwidth}
\resizebox{\hsize}{!}{\includegraphics{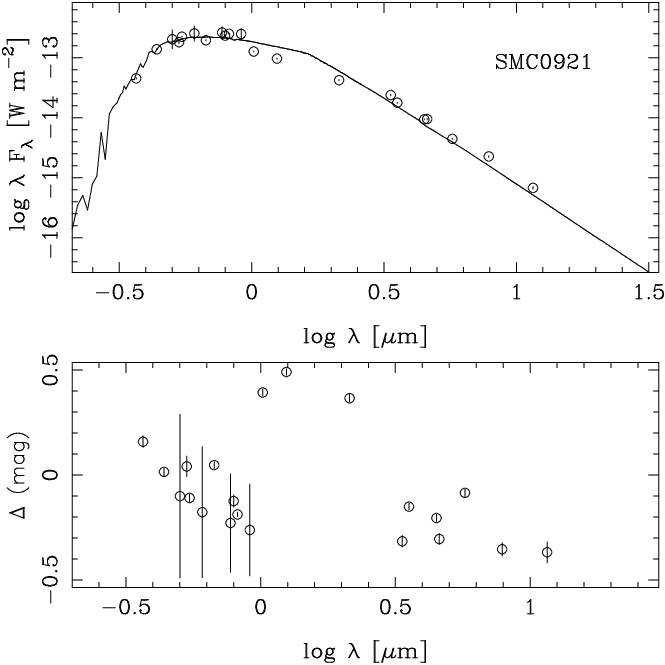}}
\end{minipage}

\begin{minipage}{0.46\textwidth}
\resizebox{\hsize}{!}{\includegraphics{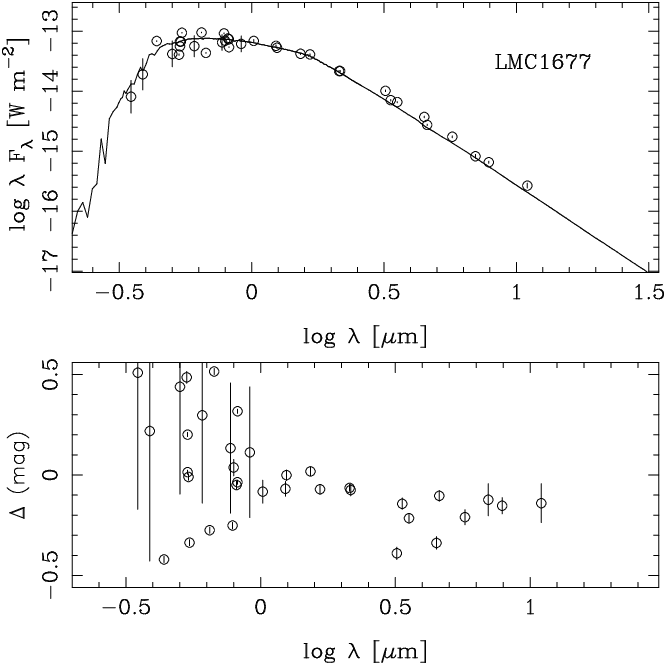}}
\end{minipage}
\begin{minipage}{0.46\textwidth}
\resizebox{\hsize}{!}{\includegraphics{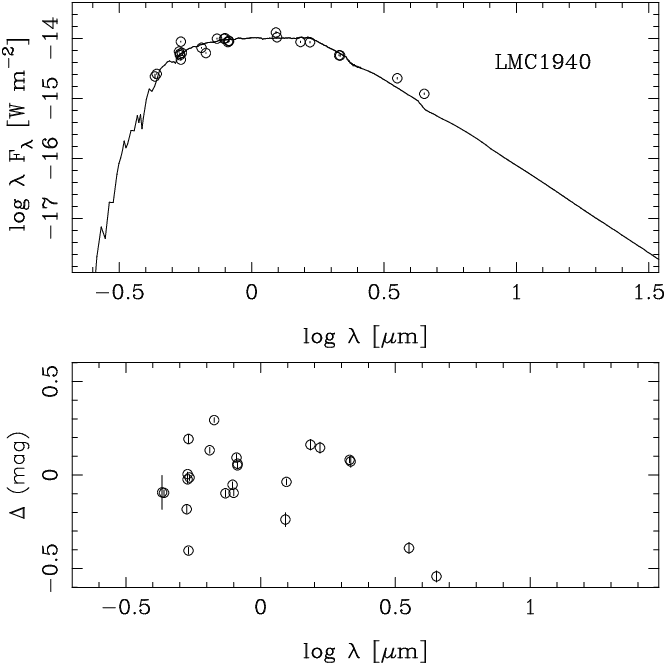}}
\end{minipage}

\caption{Details as in Fig.~\ref{Fig:sed1} for the four poorest-fitting models.
}
\label{Fig:sed2}
\end{figure*}

\begin{figure}

\begin{minipage}{0.242\textwidth}
\resizebox{\hsize}{!}{\includegraphics{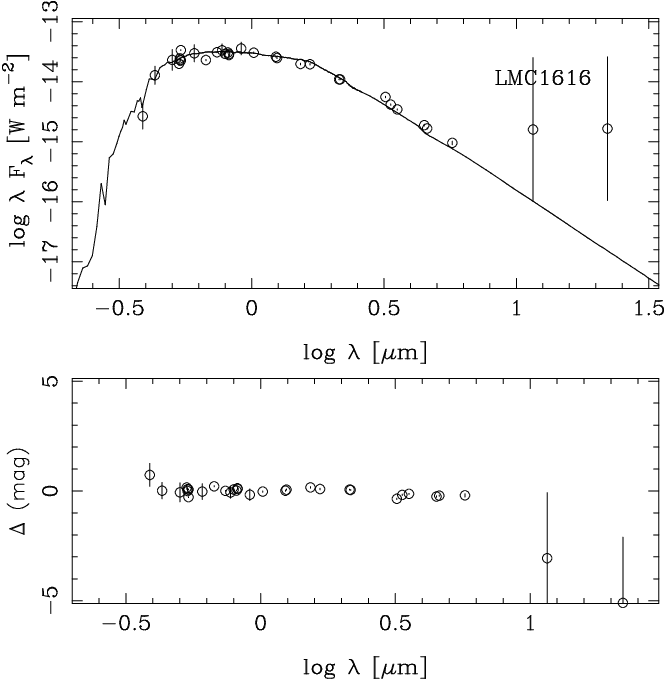}}
\end{minipage}
\begin{minipage}{0.242\textwidth}
\resizebox{\hsize}{!}{\includegraphics{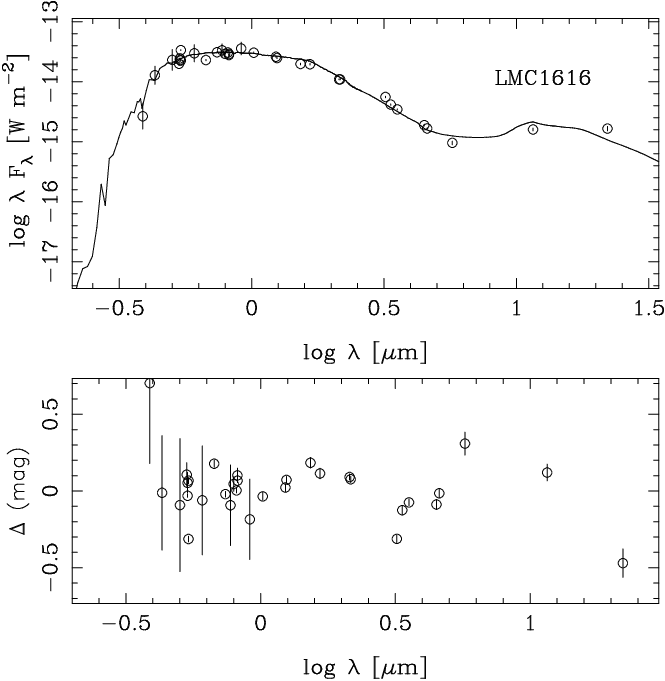}}
\end{minipage}

\begin{minipage}{0.242\textwidth}
\resizebox{\hsize}{!}{\includegraphics{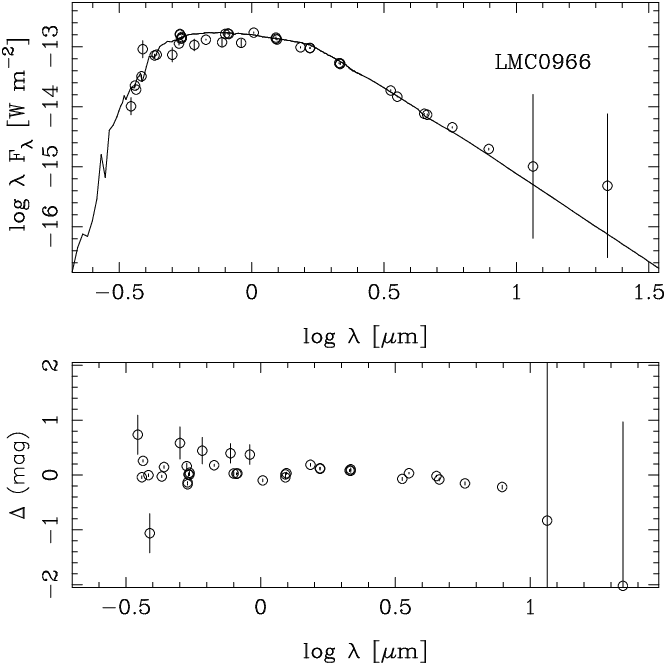}}
\end{minipage}
\begin{minipage}{0.242\textwidth}
\resizebox{\hsize}{!}{\includegraphics{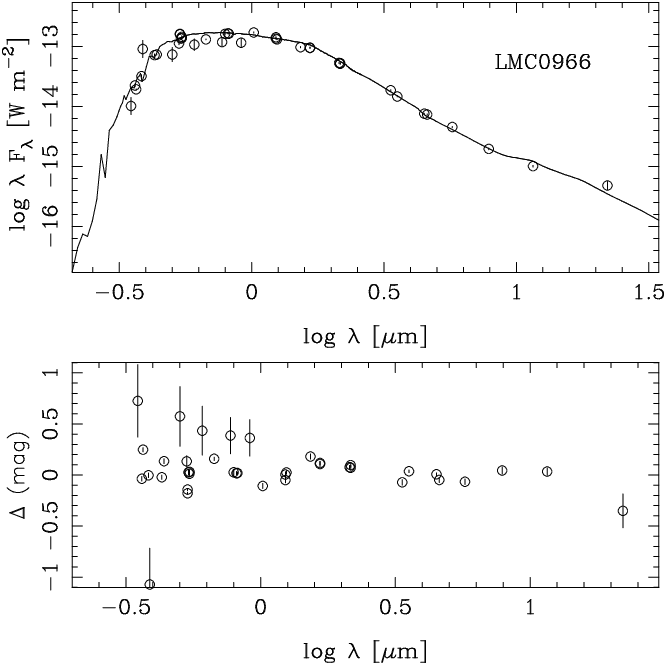}}
\end{minipage}

\begin{minipage}{0.242\textwidth}
\resizebox{\hsize}{!}{\includegraphics{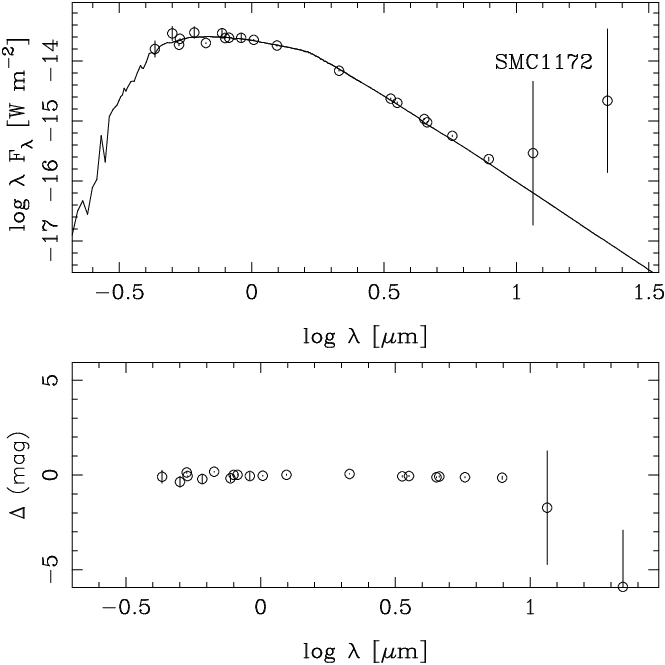}}
\end{minipage}
\begin{minipage}{0.242\textwidth}
\resizebox{\hsize}{!}{\includegraphics{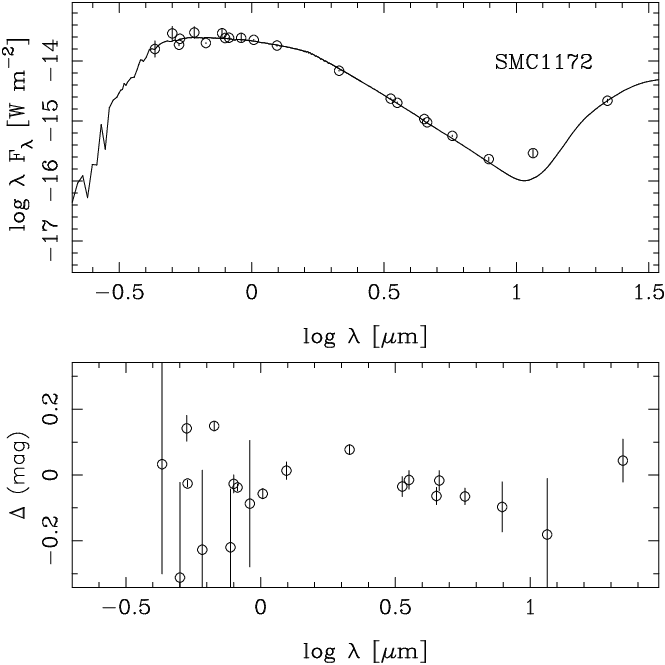}}
\end{minipage}

\begin{minipage}{0.242\textwidth}
\resizebox{\hsize}{!}{\includegraphics{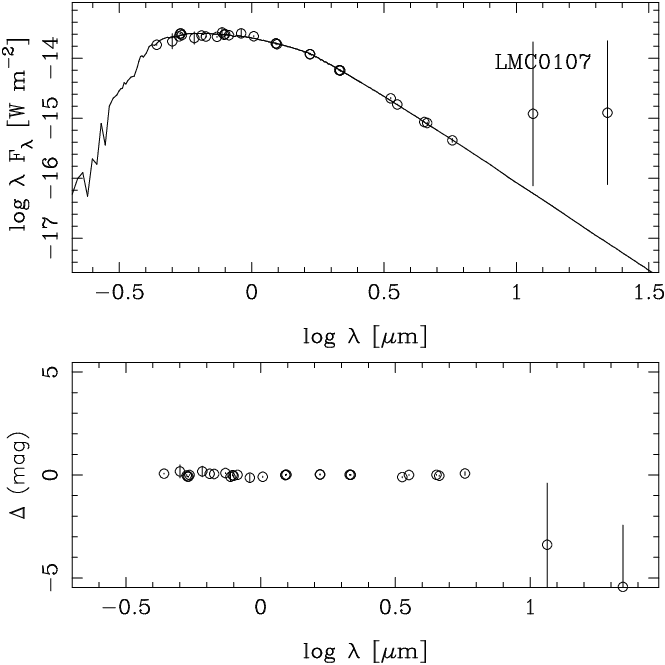}}
\end{minipage}
\begin{minipage}{0.242\textwidth}
\resizebox{\hsize}{!}{\includegraphics{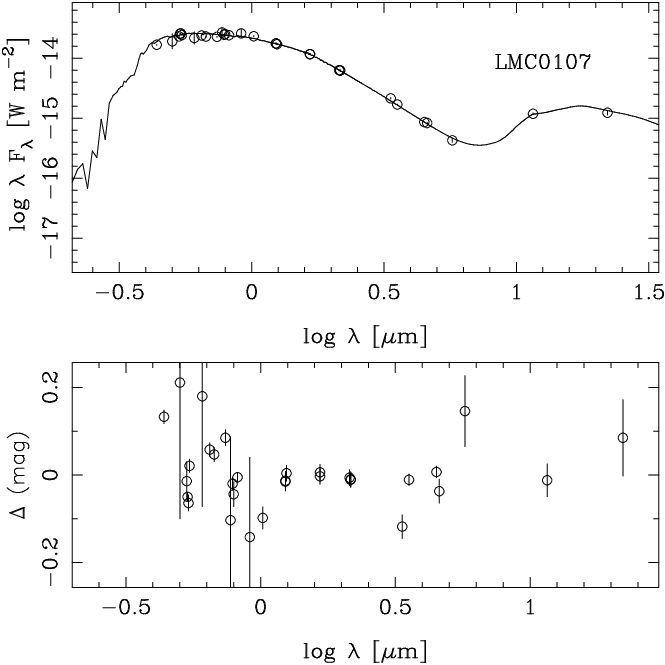}}
\end{minipage}

\begin{minipage}{0.242\textwidth}
\resizebox{\hsize}{!}{\includegraphics{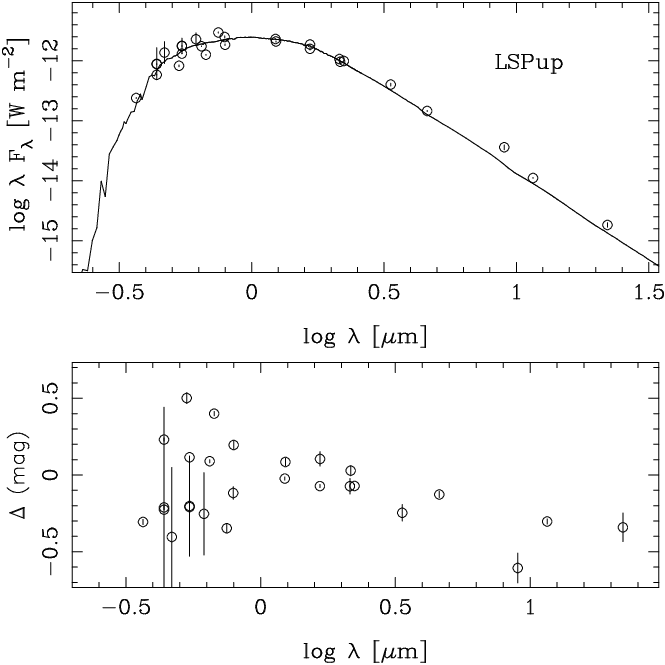}}
\end{minipage}
\begin{minipage}{0.242\textwidth}
\resizebox{\hsize}{!}{\includegraphics{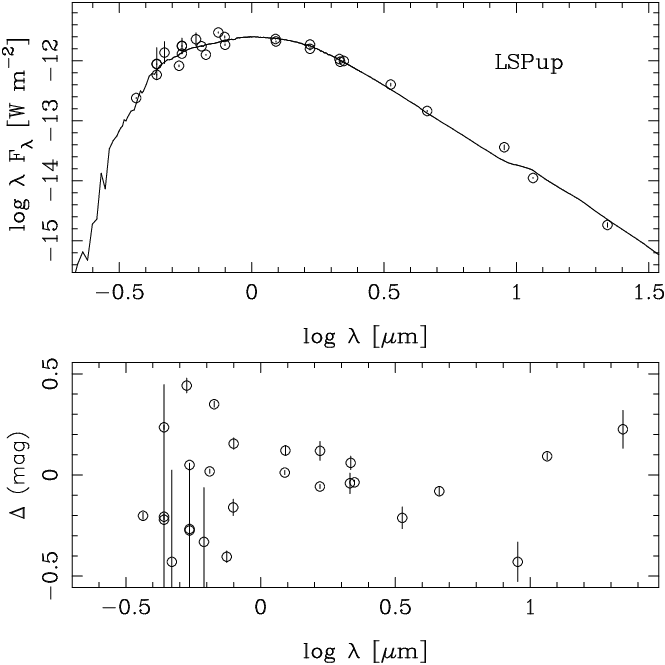}}
\end{minipage}

\caption{Models without dust (left-hand side) with dust (right-hand side) for the five MC stars where the dust
excess is unlikely to be related to the star, and the model for the MW star LS Pup for comparison.
}
\label{Fig:dust2}
\end{figure}

\begin{figure}
  \centering

\begin{minipage}{0.49\textwidth}
\resizebox{\hsize}{!}{\includegraphics{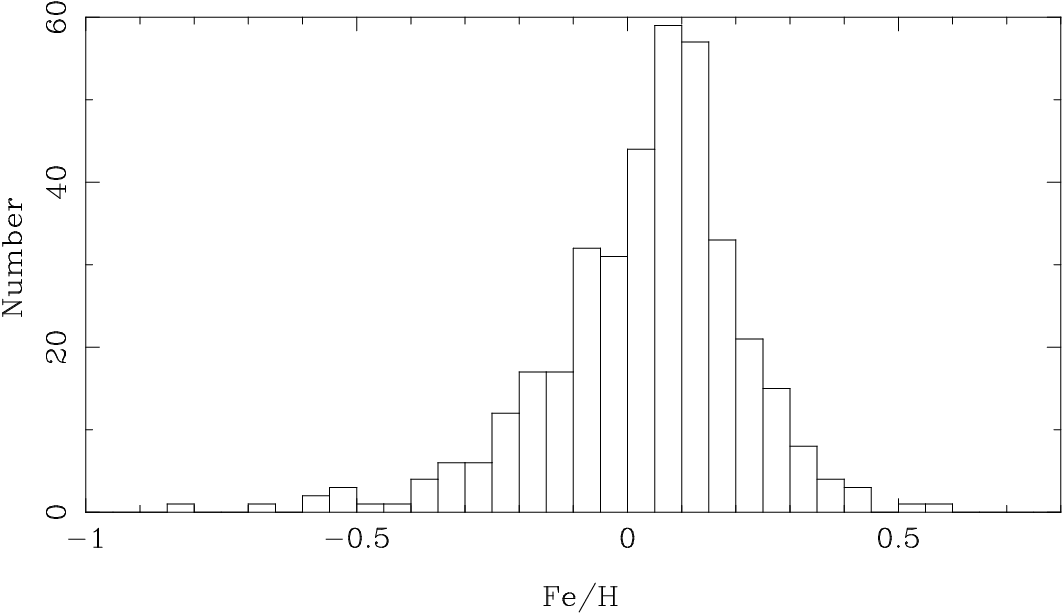}}
\end{minipage}

\caption{Distribution in [Fe/H] of the 380 galactic CC used in G20 to determine the $PL$ and $PR$ relations.
}
\label{Fig:fehhist}
\end{figure}

\begin{figure}
  \centering

\begin{minipage}{0.49\textwidth}
\resizebox{\hsize}{!}{\includegraphics{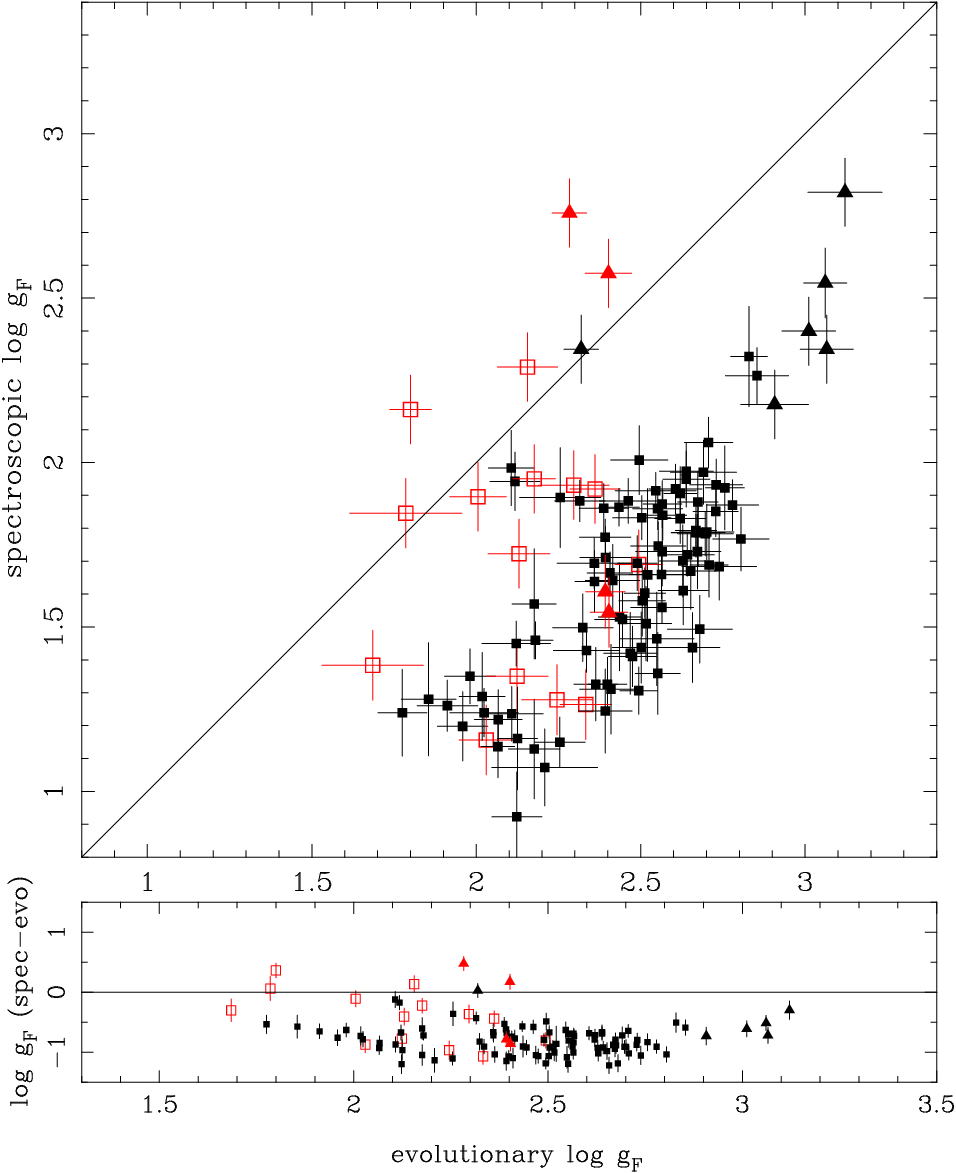}}
\end{minipage}

\caption{FWG determined from the SED fitting and the derived stellar mass compared to that from
  spectroscopic temperature and gravity determination from the literature.
Symbols as in Fig.~\ref{Fig:TeffTeff}.  
The median offset is $-0.79 \pm 0.24$~dex.  
}
\label{Fig:loggFSPC}
\end{figure}

\begin{figure*}
\begin{minipage}{0.45\textwidth}
\resizebox{\hsize}{!}{\includegraphics[angle=-0]{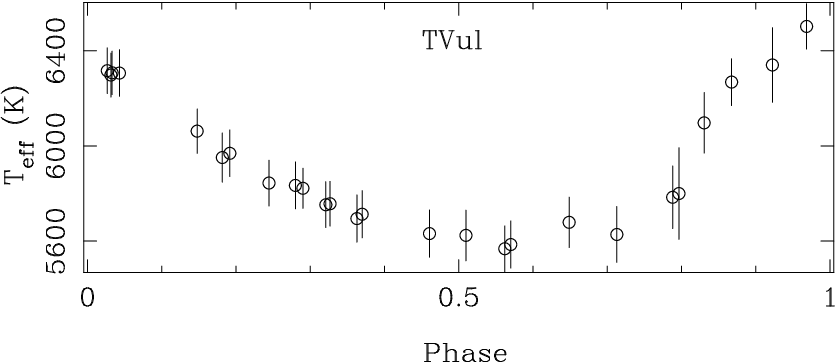}} 
\end{minipage}
\begin{minipage}{0.45\textwidth}
\resizebox{\hsize}{!}{\includegraphics[angle=-0]{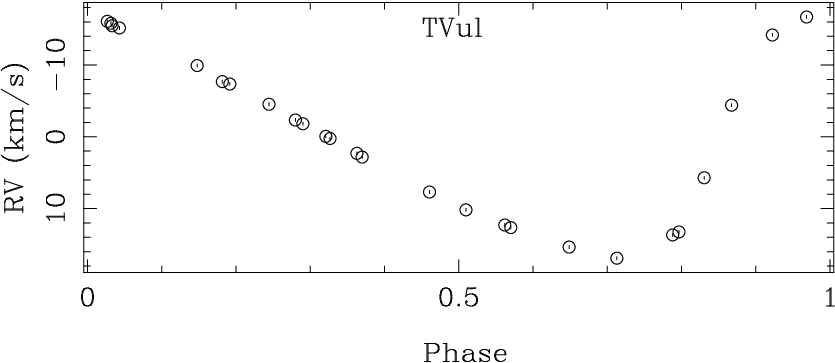}} 
\end{minipage}
 
\begin{minipage}{0.45\textwidth}
\resizebox{\hsize}{!}{\includegraphics[angle=-0]{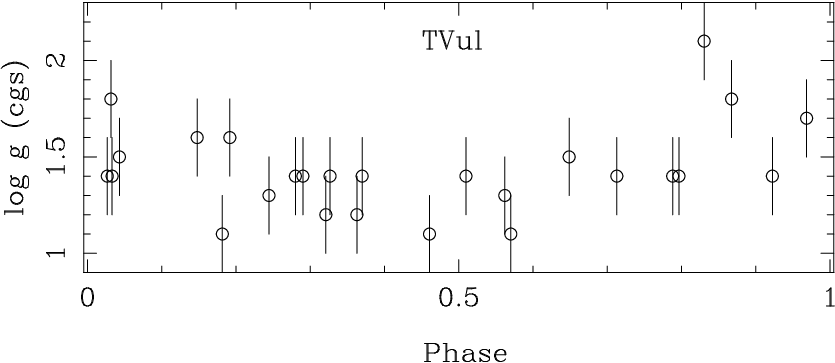}} 
\end{minipage}
\begin{minipage}{0.45\textwidth}
\resizebox{\hsize}{!}{\includegraphics[angle=-0]{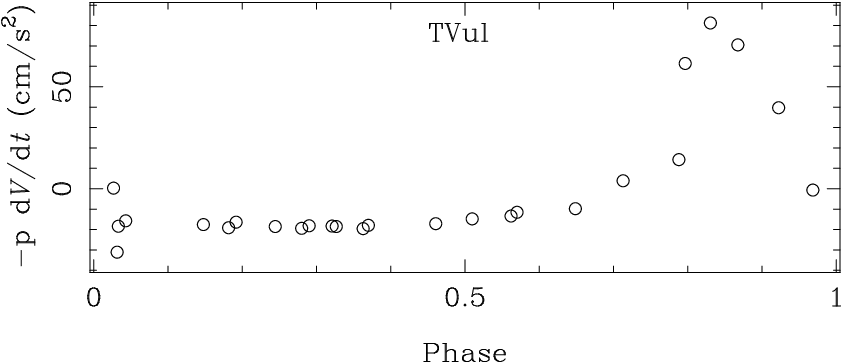}} 
\end{minipage}
 
\begin{minipage}{0.45\textwidth}
\resizebox{\hsize}{!}{\includegraphics[angle=-0]{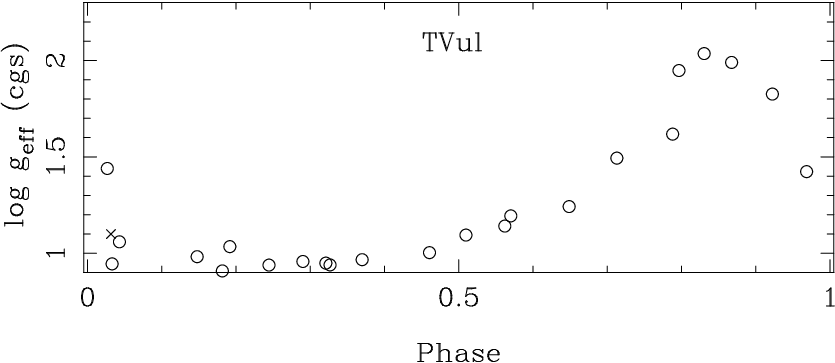}} 
\end{minipage}
\caption{Same as Fig.~\ref{Fig:delCep}, but for T Vul.
}
\label{Fig:TVul}
\end{figure*}
 
\begin{figure*}
\begin{minipage}{0.45\textwidth}
\resizebox{\hsize}{!}{\includegraphics[angle=-0]{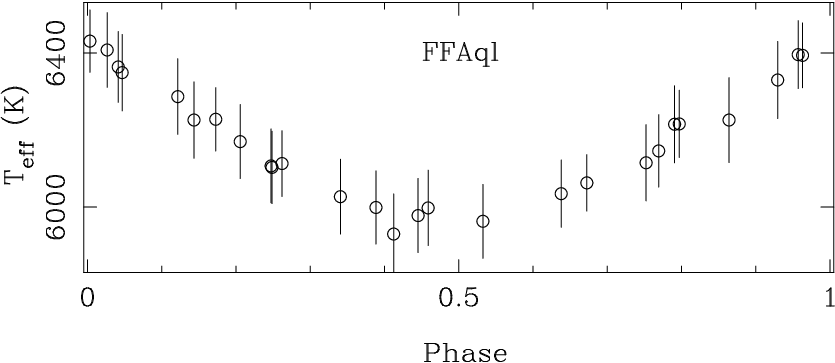}} 
\end{minipage}
\begin{minipage}{0.45\textwidth}
\resizebox{\hsize}{!}{\includegraphics[angle=-0]{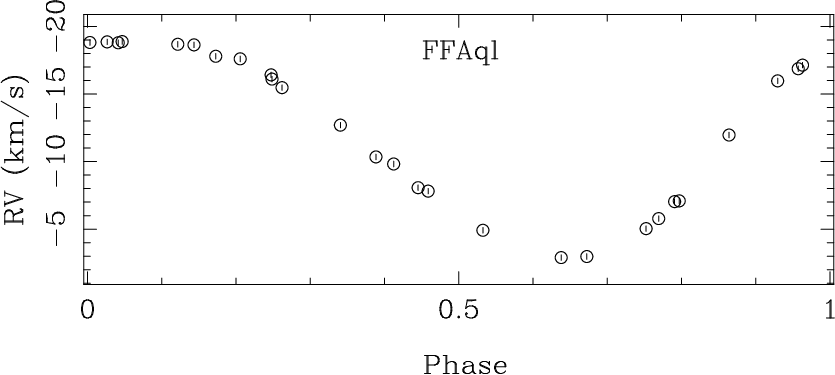}} 
\end{minipage}
 
\begin{minipage}{0.45\textwidth}
\resizebox{\hsize}{!}{\includegraphics[angle=-0]{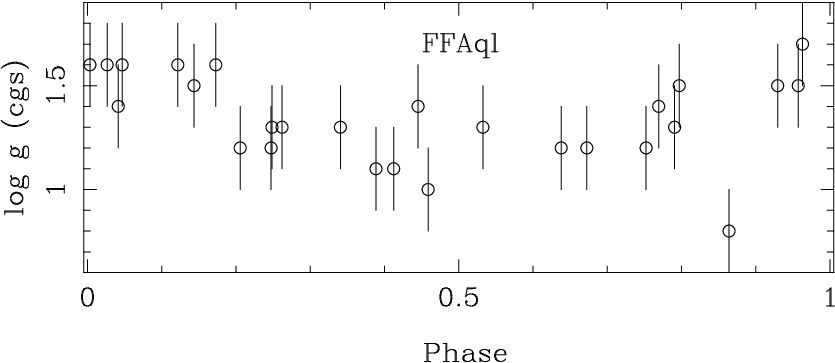}} 
\end{minipage}
\begin{minipage}{0.45\textwidth}
\resizebox{\hsize}{!}{\includegraphics[angle=-0]{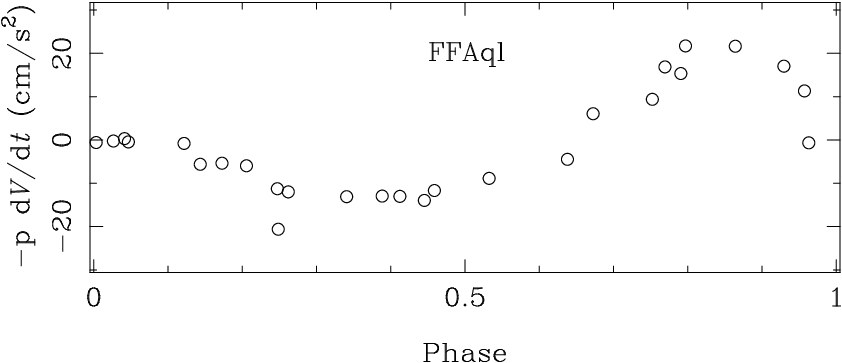}} 
\end{minipage}
 
\begin{minipage}{0.45\textwidth}
\resizebox{\hsize}{!}{\includegraphics[angle=-0]{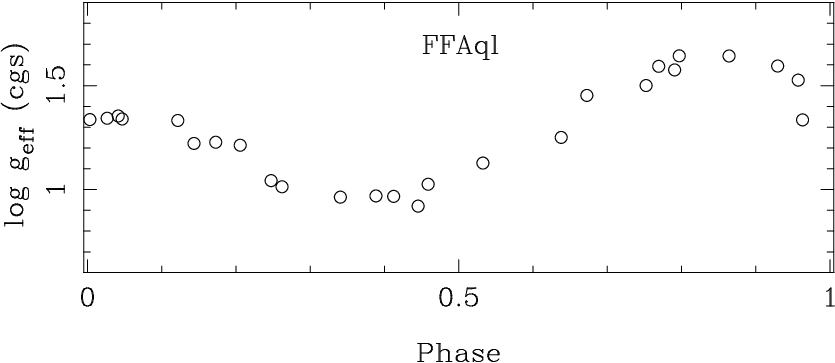}} 
\end{minipage}
\caption{Same as Fig.~\ref{Fig:delCep}, but for FF Aql.
}
\label{Fig:FFAql}
\end{figure*}
 
\begin{figure*}
\begin{minipage}{0.45\textwidth}
\resizebox{\hsize}{!}{\includegraphics[angle=-0]{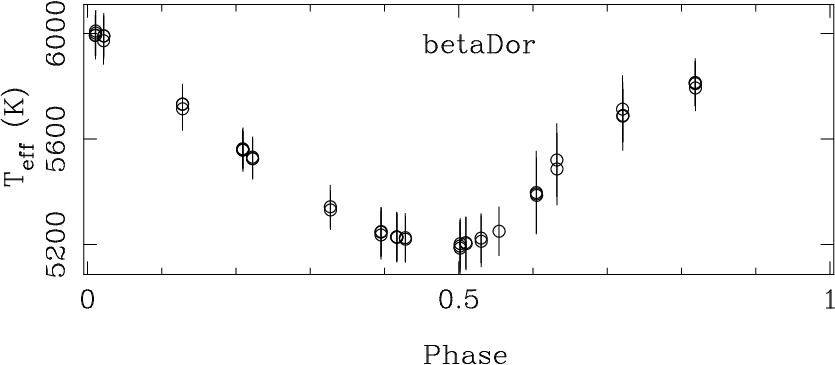}} 
\end{minipage}
\begin{minipage}{0.45\textwidth}
\resizebox{\hsize}{!}{\includegraphics[angle=-0]{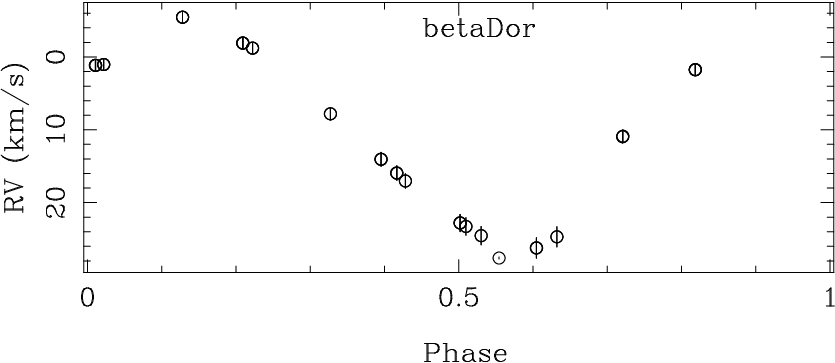}} 
\end{minipage}
 
\begin{minipage}{0.45\textwidth}
\resizebox{\hsize}{!}{\includegraphics[angle=-0]{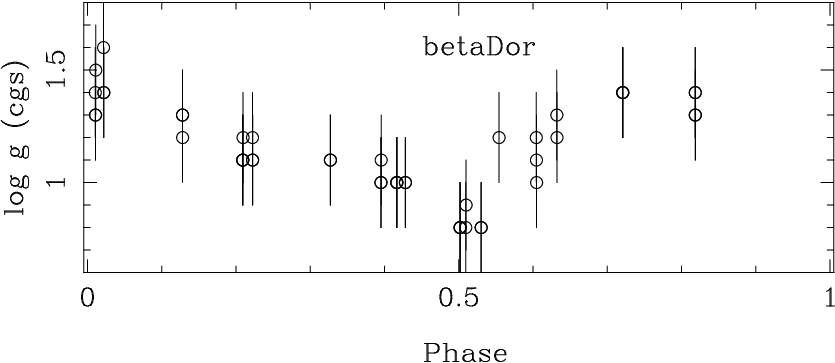}} 
\end{minipage}
\begin{minipage}{0.45\textwidth}
\resizebox{\hsize}{!}{\includegraphics[angle=-0]{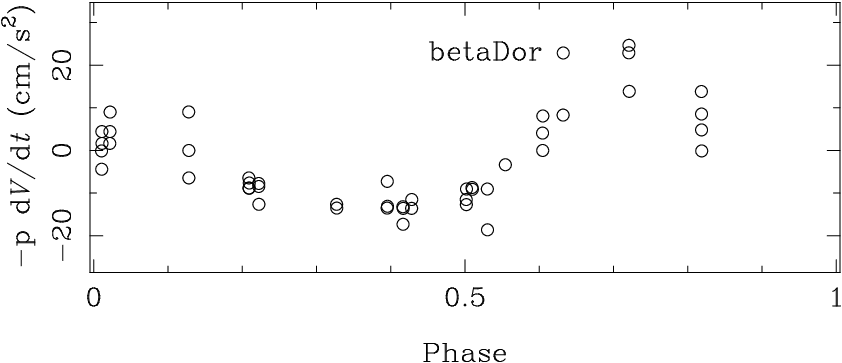}} 
\end{minipage}
 
\begin{minipage}{0.45\textwidth}
\resizebox{\hsize}{!}{\includegraphics[angle=-0]{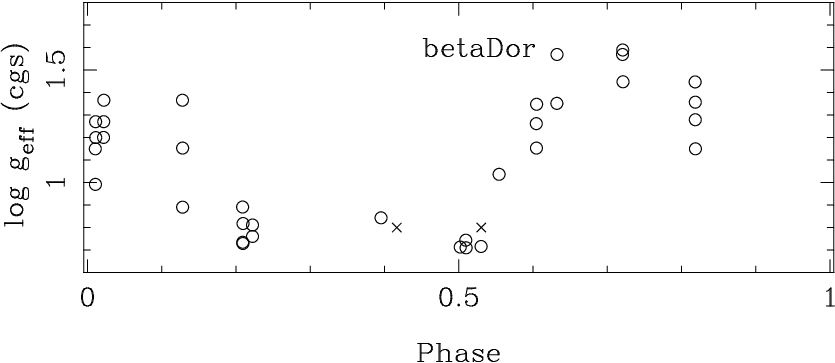}} 
\end{minipage}
\caption{Same as Fig.~\ref{Fig:delCep}, but for $\beta$ Dor.
  Effective gravities below zero are unphysical and are plotted with a cross at the lowest observed $\log$ g.
}
\label{Fig:betaDor}
\end{figure*}

\begin{figure*}
\begin{minipage}{0.45\textwidth}
\resizebox{\hsize}{!}{\includegraphics[angle=-0]{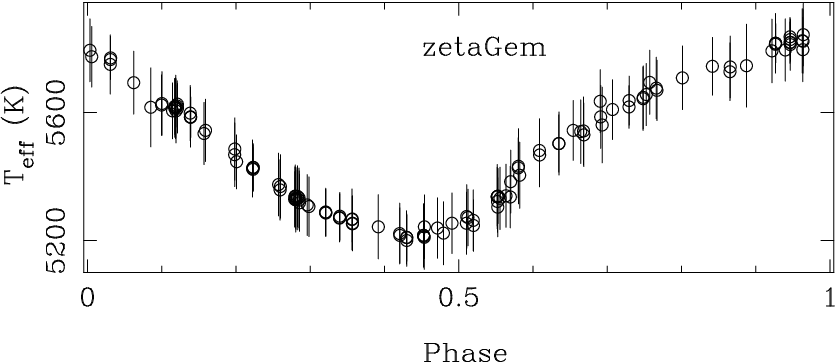}} 
\end{minipage}
\begin{minipage}{0.45\textwidth}
\resizebox{\hsize}{!}{\includegraphics[angle=-0]{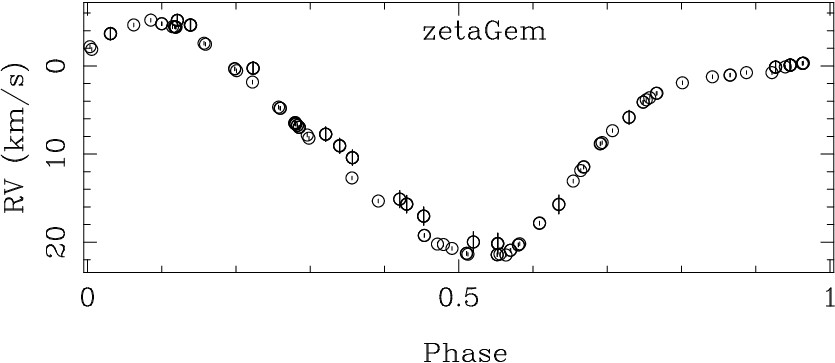}} 
\end{minipage}
 
\begin{minipage}{0.45\textwidth}
\resizebox{\hsize}{!}{\includegraphics[angle=-0]{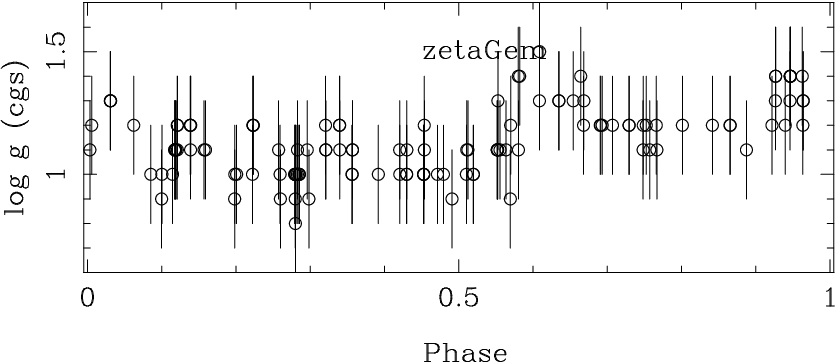}} 
\end{minipage}
\begin{minipage}{0.45\textwidth}
\resizebox{\hsize}{!}{\includegraphics[angle=-0]{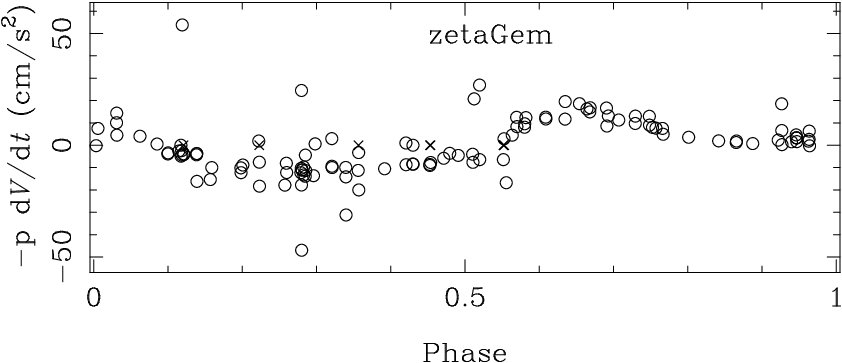}} 
\end{minipage}
 
\begin{minipage}{0.45\textwidth}
\resizebox{\hsize}{!}{\includegraphics[angle=-0]{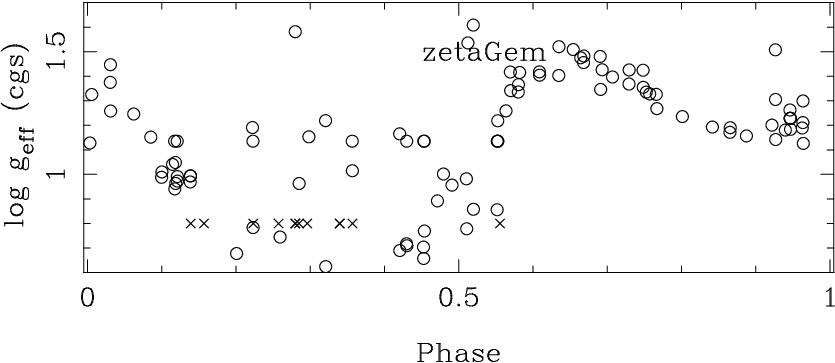}} 
\end{minipage}
\caption{Same as Fig.~\ref{Fig:delCep}, but for $\zeta$ Gem. Dynamical terms larger than 50 cm/s$^2$ in absolute values are discarded and plotted
  with a cross. Effective gravities below zero are unphysical and are plotted with a cross at the lowest observed $\log$ g.
}
\label{Fig:zetaGem}
\end{figure*}

\begin{figure}

\begin{minipage}{0.49\textwidth}
\resizebox{\hsize}{!}{\includegraphics{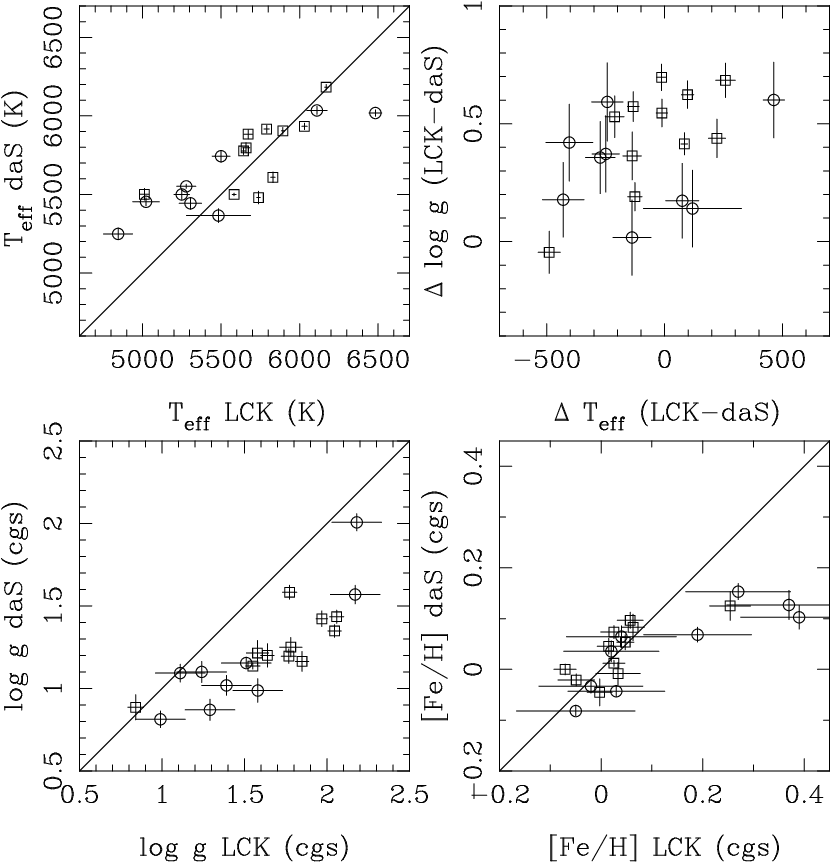}}
\end{minipage}

\caption{Comparison between \citet{Luck18} (LCK) and \citet{daSilva22} (daS) in terms of effective temperature, $\log g$, the
  differential between the two, and iron abundance.
  The stars with only one spectrum in \citet{Luck18} are marked by circles, objects plotted with squares have 5 or more spectra.
  Abundances in  \citet{Luck18} are given on a scale where $\log$ Hydrogen abundance = 12 and have been converted to the scale
  in \citet{daSilva22} using their adopted iron abundance of $\log$ Fe = 7.50.
}
\label{Fig:LCKdaS}
\end{figure}

\end{appendix}

\end{document}